\documentstyle[aps,prl]{revtex}

\input psfig.tex
\begin{document}

\catcode`@=11
\def\seceqaa{\@addtoreset{equation}{section}
           \def\theequation{A.\arabic{equation}}}
\def\seceqbb{\@addtoreset{equation}{section}
           \def\theequation{B.\arabic{equation}}}
\def\seceqcc{\@addtoreset{equation}{section}
           \def\theequation{C.\arabic{equation}}}
\def\seceqdd{\@addtoreset{equation}{section}
           \def\theequation{D.\arabic{equation}}}
\catcode`@=12

\def\slash#1{\setbox0=\hbox{$#1$}#1\hskip-\wd0\hbox to\wd0{\hss\sl/\/\hss}}

\title{A new approach to study energy transfer in magnetohydrodynamic 
turbulence}
\author{Gaurav Dar\thanks{e-mail: gdar@iitk.ac.in} and
Mahendra K. Verma  \\ 
Department of Physics \\ 
Indian Institute of Technology, 
Kanpur 208016, India \\
V. Eswaran  \\ 
Department of Mechanical Engineering \\ 
Indian Institute of Technology, 
Kanpur 208016, India}
\maketitle

\bibliographystyle{unsrt}

\begin{abstract}
The unit of nonlinear interaction in Navier-Stokes and the 
Magnetohydrodynamic (MHD) equations is a wavenumber triad ({\bf k,p,q})
satisfying ${\bf k+p+q=0}$. The expression for the combined energy
transfer from two of these wavenumbers to the third wavenumber is
known. In this paper we introduce the idea of an effective energy
transfer between a pair of modes through the mediation of the third mode
and then find an expression for it. In fluid turbulence, energy transfer
takes place between a pair of velocity modes, whereas in MHD turbulence
energy transfer takes places between (1) a pair of velocity modes, 
(2) a pair of magnetic modes, and (3) between a velocity
and a magnetic mode in a triad. In this paper we have obtained the
expression for each of these transfers. 
We also show how the effective mode-to-mode energy
transfer rate can be utilised to study energy cascades and shell-to-shell
energy transfer rates. 
\end{abstract}

\newpage

\section{Introduction}
\label{s:intro}

In fluid and MHD turbulence, eddies of various
sizes interact amongst themselves; energy is exchanged among them in this
process. These interactions arise due to the nonlinearity
present in these systems. 
The fundamental interactions involve wavenumber triads ({\bf k,p,q}) with
${\bf k+p+q=0}$. 

The {\it combined energy transfer} computation to a mode from the other
two modes of a triad has generally been considered to be fundamental. 
This formalism has played
an important role in furthering our understanding of
locality in fluid turbulence ~\cite{Domaradzki90,Domaradzki88,Zhou93a,Zhou93b},
and also in analysing subgrid scale eddy viscosity 
~\cite{Kraichnan76,Domaradzki87}.  Energy interactions are
 considered to be local if the triads with $k \sim p \sim q$ contribute most
to the energy transfer to the wavenumber $k$. Numerical simulations of
Domaradzki and Rogallo ~\cite{Domaradzki88,Domaradzki90} showed that the
contribution of the nonlocal triads ($k \sim p >> q$) to the interactions is
not small, but in such triads the dominant exchange of energy occurs
between wavenumbers of similar magnitudes, i.e., between $k$ and $p$.
Domaradzki {\it et al.} ~\cite{Domaradzki87} 
computed the sub-grid scale viscosity by 
summing up the energy lost by a wavenumber $k<k_c$ (cut off wavenumber)
by triad interactions with atleast one wavenumber larger than $k_c$.
Batchelor ~\cite{Batchelor50} first suggested a formula for the energy transfer
in fluid turbulence between two regions in Fourier space, and
Domaradzki and Rogallo ~\cite{Domaradzki90} calculated shell-to-shell energy
transfer rates. Taking energy transfer between shells
as an example, we will show in Section~\ref{s:shell_intro}, that these
formalisms have certain limitations. 

In this paper we present a scheme to calculate the energy transfer rates
between two modes in a triad interaction. Using our scheme we can
calculate the energy transfer rates between two shells, cascade rates,
and also study locality of energy transfer.
The formulae for the combined energy transfer rate were used to study
energy transfer rates in MHD turbulence
~\cite{MKV96,Pouquet76,Pouquet78,Ishizawa298,Ishizawa198}. Ishizawa and
Hattori ~\cite{Ishizawa198} numerically computed the kinetic energy 
flux to the velocity modes 
outside the sphere and the magnetic energy fluxes to the magnetic modes outside
the sphere. They also computed the kinetic energy fluxes due to transfer of
energy  to all magnetic modes and a similar magnetic energy flux due to
transfer of energy to all velocity modes ~\cite{Ishizawa198}. 
Pouquet {\it et al.} ~\cite{Pouquet76}, Pouquet ~\cite{Pouquet78}, and
Ishizawa and Hattori ~\cite{Ishizawa298} used the combined energy transfer
formalism in an Eddy damped quasi-normal Markovian (EDQNM) closure
calculation to study non-local energy transfer. However, several
energy transfers cannot be calculated using these formalisms. 
Our formalism of ``mode-to-mode'' transfer enables us to calculate all 
these energy transfer rates between the spheres and the shells. 

This paper is organized into the following sections. 
In Section~\ref{s:triad_intro} we will state the known results regarding
the combined energy transfer between velocity modes of a triad. In Section 
\ref{s:shell_intro} we will present a discussion on the formulae which
are used for studying energy transfer between shells in fluid turbulence.
In Section~\ref{s:mode_to_mode_fluids} we will formulate a way of studying
energy transfer between a pair of modes within a triad, with the third
mode of the triad meditating the transfer. This new method is used to
re-define shell-to-shell transfer in Section~\ref{s:shell_redefn}. 
The known results of energy transfer in a MHD triad are discussed in
Section~\ref{s:mhdtriad_intro} and the ``mode-to-mode'' transfers in MHD 
in Section~\ref{s:mode_to_mode_mhd}.
The results of Section~\ref{s:mode_to_mode_mhd}
are used in Section~\ref{s:shell_to_shell_mhd}
to define shell-to-shell transfer rates and the energy cascade rates
in MHD turbulence. The conclusions of this paper follows in
Section~\ref{s:conclusion_chap3}.

\section{Energy transfer in a triad: Known results}
\label{s:triad_intro}

  The Navier Stokes equation in real space are written as

\begin{equation}
\frac{\partial {\bf u}}{\partial t} + ({\bf u.\nabla}){\bf u} =
 - {\bf \nabla}p + \nu {\bf \nabla^{2} u} ,
\label{eq:u_nseq}
\end{equation}
where ${\bf u}$ is the velocity field, and $\nu$ is the
fluid kinematic viscosity.
In Fourier space, the kinetic energy  equations
for a Fourier mode is \\
\begin{equation}
\frac{\partial E^{u}(\bf k)}{\partial t} + 2 \nu k^{2} E^{u}({\bf k}) =
\sum_{{\bf k+p+q}=0} \frac{1}{2}S^{uu}({\bf k|p,q}),
\label{eq:eu_nseq_chap3}
\end{equation}
where $E^{u}({\bf k})\equiv{\bf |u(k)|}^{2}/2$ is the kinetic energy in
Fourier space for
mode {\bf k}. The nonlinear term  $S^{uu}({\bf k|p,q})$ is
\begin{equation}
      S^{uu}({\bf k|p,q}) \equiv
                         -  Re \left( i{\bf \left [k.u(q)\right]} 
                                     {\bf \left [u(k).u(p)\right ]} + 
                                   i{\bf \left [k.u(p)\right]} 
                                    {\bf \left [u(k).u(q)\right]} \right)
\label{eq:ukpq_def}
\end{equation}
This term represents the combined
transfer of kinetic energy from mode ${\bf p}$ and ${\bf q}$ to mode ${\bf k}$ 
~\cite{Lesieur}.
Note that the wavenumber triad $\bf k$, $\bf p$, and $\bf q$ should satisfy
the condition ${\bf k+p+q}=0$. 
This nonlinear term satisfies the following conservation
property
\begin{equation}
S^{uu}({\bf k|p,q})+ S^{uu}({\bf p|k,q}) +
S^{uu}({\bf q|k,p}) = 0 .
\label{eq:ukpq_triad}
\end{equation}
While the quantity $S^{uu}({\bf k|p,q})$ represents the 
nonlinear energy transfer from the {\it two modes} $\bf p$ and $\bf q$ 
to mode $\bf k$, it would be useful to know the exact energy
transfer between {\it any two modes}, say from ${\bf p}$ to ${\bf k}$.
In Section~\ref{s:mode_to_mode_fluids} we will explore this idea.

\section{Shell-to-Shell energy transfer in fluid turbulence}
\label{s:shell_intro}
Using the combined energy transfer
$S^{uu}({\bf k|p,q})$, Domaradzki and Rogallo ~\cite{Domaradzki90}
have discussed energy transfer between two
shells.  They interpret the quantity
\begin{equation}
T^{uu}_{mn} = \frac{1}{2}\sum_{{\bf k} {\large \epsilon}{\it m}} \sum_{{\bf p} 
              {\large \epsilon {\it n}}} S^{uu}({\bf k|p,q}).
\label{eq:shell_old_defn}
\end{equation}
as the rate of energy transfer from shell
{\it n} to shell {\it m} ~\cite{Domaradzki90,Batchelor50}. 
Note that {\bf k}-sum is over shell {\it m}, {\bf p}-sum 
over shell {\it n}, and  ${\bf q = -k-p}$. However, 
Domaradzki and Rogallo ~\cite{Domaradzki90} themselves
points out that it may not be entirely correct to interpret the 
formula~(\ref{eq:shell_old_defn}) as the shell-to-shell
energy transfer. The reason for this is as follows:

In the energy transfer between two shells {\it m} and {\it n}, two types
of wavenumber triads are involved (see Fig. 1 shown below).
In a triad of type I, the wavenumbers
{\bf p, q} are located in one shell, and {\bf k}
is located in the other. In a type II triad, wavenumber {\bf k}
is in one shell, {\bf p} {\it or} {\bf q} in the other shell, and the
third wavenumber is located outside the two shells.
In Eq.~(\ref{eq:shell_old_defn}) the 
summation is carried over both kinds of triads. However, it is easy 
to see that the energy transfer from shell {\it n} to shell {\it m}
takes place through both the {\bf k-p} and {\bf k-q} legs of triad I  but 
only through the {\bf k-p} leg of triad II. 
Hence Batchelor's and Domaradzki's formalism
do not yield correct shell-to-shell energy transfers (as was pointed out by
Batchelor and Domaradzki themselves).

\section{Mode-to-Mode energy transfer in a triad}
\label{s:mode_to_mode_fluids}
The nonlinear interaction in the Navier-Stokes and the MHD equations are
intrinsically three-mode interactions. The expression for the energy
transfer to one mode of the triad from the other two was presented
in Section \ref{s:triad_intro}. In this section we shall 
explore the possibility of 
obtaining an expression for the energy transfer between any two modes
{\it within} a triad --- we will call it the {\it mode-to-mode transfer}.
We emphasize that this approach is still within the framework of the triad
interaction --- that is, the triad is still the fundamental interaction of 
which the mode-to-mode transfer is a part. The energy transfer between 
two modes of a triad by the mediation of the third mode is sought here.

We shall first consider only the Navier-Stokes equation. In a later
section, the discussion will be generalised to the MHD equation.

\subsection{Definition of Mode-to-Mode transfer in a triad}
\label{subs:mode_to_mode_defn}

Consider a triad {\bf k, p, q}.
Let the quantity ${\cal{\slash{R}}}^{uu}({\bf k|p|q})$ 
denote the energy transferred from mode {\bf p} to
mode {\bf k} with mode {\bf q} playing the role of a mediator (see Fig. 2).
We wish to obtain an expression for ${\cal{\slash{R}}}^{uu}$.

The ${\cal{\slash{R}}}^{uu}$'s should satisfy the following relationships :
\begin{enumerate}
\item The sum of ${\cal{\slash{R}}}^{uu}({\bf k|p|q})$ and
 ${\cal{\slash{R}}}^{uu}({\bf k|q|p})$, which represent energy transfer
from mode {\bf p} to mode {\bf k} and from mode {\bf q} to mode {\bf k},
respectively, should be equal to the total energy transferred to mode
{\bf k} from modes {\bf p} and {\bf q}, i.e., $S^{uu}({\bf k|p,q})$
[see Eq.~(\ref{eq:ukpq_def})]. Thus,
\begin{equation}
{\cal{\slash{R}}}^{uu}({\bf k|p|q}) + {\cal{\slash{R}}}^{uu}({\bf k|q|p})
                                    = S^{uu}({\bf k|p,q}), 
\label{eq:Rukup_ukuq}
\end{equation}
\begin{equation}
{\cal{\slash{R}}}^{uu}({\bf p|k|q}) + {\cal{\slash{R}}}^{uu}({\bf p|q|k})
                                    = S^{uu}({\bf p|k,q}),
\label{eq:Rupuk_upuq}
\end{equation}
\begin{equation}
{\cal{\slash{R}}}^{uu}({\bf q|k|p}) + {\cal{\slash{R}}}^{uu}({\bf q|p|k})
                                    = S^{uu}({\bf q|k,p}).
\label{eq:Ruquk_uqup}
\end{equation}

\item The energy transferred from mode {\bf p} to mode {\bf k}, 
i.e., ${\cal{\slash{R}}}^{uu}({\bf k|p|q})$, will be equal and opposite
to the energy transferred from mode {\bf k} to mode {\bf p}
i.e., ${\cal{\slash{R}}}^{uu}({\bf p|k|q})$. Thus,
\begin{equation}
{\cal{\slash{R}}}^{uu}({\bf k|p|q})+ {\cal{\slash{R}}}^{uu}({\bf p|k|q})=0 ,
\label{eq:Rukup_upuk}
\end{equation}
\begin{equation}
{\cal{\slash{R}}}^{uu}({\bf k|q|p})+ {\cal{\slash{R}}}^{uu}({\bf q|k|p})=0 ,
\label{eq:Rukuq_uquk}
\end{equation}
\begin{equation}
{\cal{\slash{R}}}^{uu}({\bf p|q|k})+ {\cal{\slash{R}}}^{uu}({\bf q|p|k})=0 .
\label{eq:Rupuq_uqup}
\end{equation}
\end{enumerate}
These are six equations with six
unknowns. However, the value of the
determinant formed from the 
Eqs.~(\ref{eq:Rukup_ukuq})-(\ref{eq:Rupuq_uqup}) is zero. 
Therefore we cannot find unique
${\cal{\slash{R}}}^{uu}$'s given just these equations. However, it is
reasonable to expect that there is a {\it definite} amount of energy
transfer from one mode to the other mode, say from mode {\bf k} to
mode {\bf p} given {\bf u(k), u(p), u(q)}. Since 
Eqs.~(\ref{eq:Rukup_ukuq})-(\ref{eq:Rupuq_uqup}) do not yield a
unique ${\cal{\slash{R}}}^{uu}$, we need to use constraints based
on invariance, symmetries, etc. to get a definite
${\cal{\slash{R}}}^{uu}$ using
Eqs.~(\ref{eq:Rukup_ukuq})-(\ref{eq:Rupuq_uqup}).

\subsection{Solutions of equations of mode-to-mode transfer}
\label{subs:mode_to_mode_soln}

One solution of Eqs.~(\ref{eq:Rukup_ukuq})-(\ref{eq:Rupuq_uqup})
is
\begin{equation}
{\cal{\slash{S}}}^{uu}({\bf k|p|q}) \equiv - Re \left( i{\bf \left[k.u(q)\right]} %
{\bf \left[u(k).u(p)\right]} \right) ,
\label{eq:Sukup_def}
\end{equation}    
From the definition
of ${\cal{\slash{S}}}^{uu}({\bf k|p|q})$, it directly follows that
${\cal{\slash{S}}}^{uu}$'s  satisfy the following conditions.
\begin{equation}
{\cal{\slash{S}}}^{uu}({\bf k|p|q}) + {\cal{\slash{S}}}^{uu}({\bf k|q|p})
                                    = S^{uu}({\bf k|p,q}) ,  
\label{eq:Sukup_ukuq}
\end{equation}
\begin{equation}
{\cal{\slash{S}}}^{uu}({\bf p|k|q}) + {\cal{\slash{S}}}^{uu}({\bf p|q|k})
                                  = S^{uu}({\bf p|k,q}) ,  
\label{eq:Supuk_upuq}
\end{equation}
\begin{equation}
{\cal{\slash{S}}}^{uu}({\bf q|k|p}) + {\cal{\slash{S}}}^{uu}({\bf q|p|k})
                                    = S^{uu}({\bf q|k,p}) .  
\label{eq:Suquk_uqup}
\end{equation}
Using the triad relationship ${\bf k+p+q = 0}$, and the
incompressibility constraint [${\bf k.u(k)=0}$], it can be seen that
${\cal{\slash{S}}}^{uu}$'s, etc. also satisfy the following conditions
\begin{equation}
{\cal{\slash{S}}}^{uu}({\bf k|p|q})+ {\cal{\slash{S}}}^{uu}({\bf p|k|q})=0 ,  
\label{eq:Sukup_upuk}
\end{equation}
\begin{equation}
{\cal{\slash{S}}}^{uu}({\bf k|q|p})+ {\cal{\slash{S}}}^{uu}({\bf q|k|p})=0 ,
\label{eq:Sukuq_uquk}
\end{equation}
\begin{equation}
{\cal{\slash{S}}}^{uu}({\bf p|q|k})+ {\cal{\slash{S}}}^{uu}({\bf q|p|k})=0 . 
\label{eq:Supuq_uqup}
\end{equation}
Comparing Eqs.~(\ref{eq:Sukup_ukuq})-(\ref{eq:Supuq_uqup}) with
Eqs.~(\ref{eq:Rukup_ukuq})-(\ref{eq:Rupuq_uqup}), it is clear that
the set of ${\cal{\slash{S}}}^{uu}$'s is {\it one instance} of the
${\cal{\slash{R}}}^{uu}$'s, i.e., 
${\cal{\slash{R}}}^{uu}({\bf k|p|q})={\cal{\slash{S}}}^{uu}({\bf k|p|q})$.
However, this is not a unique solution.

If another solution ${\cal{\slash{R}}}^{uu}({\bf k|p|q})$ differs from
${\cal{\slash{S}}}^{uu}({\bf k|p|q})$ by an arbitrary function 
$X_\Delta$, i.e.,
${\cal{\slash{R}}}^{uu}({\bf k|p|q})=
{\cal{\slash{S}}}^{uu}({\bf k|p|q})+X_\Delta$, then by inspection we can
easily see that every possible solution of 
Eqs.~(\ref{eq:Rukup_ukuq})-(\ref{eq:Rupuq_uqup}) must be of the form
\begin{equation}
{\cal{\slash{R}}}^{uu}({\bf k|p|q}) = 
{\cal{\slash{S}}}^{uu}({\bf k|p|q})+X_{\Delta}
\label{eq:Skpq_X}
\end{equation}
\begin{equation}
{\cal{\slash{R}}}^{uu}({\bf k|q|p}) = 
{\cal{\slash{S}}}^{uu}({\bf k|q|p})-X_{\Delta}
\label{eq:Skqp_X}
\end{equation}
\begin{equation}
{\cal{\slash{R}}}^{uu}({\bf p|k|q}) = 
{\cal{\slash{S}}}^{uu}({\bf p|k|q})-X_{\Delta}
\label{eq:Spkq_X}
\end{equation}
\begin{equation}
{\cal{\slash{R}}}^{uu}({\bf p|q|k}) = 
{\cal{\slash{S}}}^{uu}({\bf p|q|k})+X_{\Delta}
\label{eq:Spqk_X}
\end{equation}
\begin{equation}
{\cal{\slash{R}}}^{uu}({\bf q|k|p}) = 
{\cal{\slash{S}}}^{uu}({\bf q|k|p})+X_{\Delta}
\label{eq:Sqkp_X}
\end{equation}
\begin{equation}
{\cal{\slash{R}}}^{uu}({\bf q|p|k}) = 
{\cal{\slash{S}}}^{uu}({\bf q|p|k})-X_{\Delta}
\label{eq:Sqpk_X}
\end{equation}
Note that $X_\Delta$ can depend upon the wavenumber triad
{\bf k, p, q,} and the Fourier components {\bf u(k), u(p), u(q)}. 
$X_\Delta$ needs to be determined from other symmetry 
and invariance arguments to
uniquely fix ${\cal{\slash{R}}}^{uu}$.

In Appendix A we have attempted to determine $X_\Delta$. We construct
$X_\Delta$ by observing that it can depend on {\bf k, p, q, u(k),
u(p), u(q)}; and it must satisfy rotational invariance,
galilean invariance, and it should be finite.  In our procedure we write down
a general scalar function of {\bf k, p, q, u(k), u(p)}, and {\bf u(q)} 
with undetermined coefficients. 
The coefficients can be written as a series in the scalars of the type
{\bf k.p}, {\bf k.u(p)}, {\bf u(k).u(p)} with unknown constants. By 
imposing invariance,and finiteness constraints we have shown that, upto 
the linear order in these scalars, the coefficients must vanish,
giving $X_{\Delta}=0$.
Unfortunately, we have not been able to determine the higher order
coefficients. 
Hence, from the analysis of Appendix A we cannot show 
that $X_\Delta$ will definitely vanish.
However, if $X_{\Delta}=0$ we get a simple relationship, 
\begin{equation}
{\cal{\slash{R}}}^{uu} = {\cal{\slash{S}}}^{uu},
\label{eq:ReqS}
\end{equation}
giving the energy transfer between a pair of modes in a triad with the
third mode mediating the transfer, as simply
${\cal{\slash{S}}}^{uu}$.
But if $X_{\Delta} \ne 0$ then the energy
transfer picture is more complex and the energy transfer between a pair
of modes will be given by Eqs.~(\ref{eq:Skpq_X})-(\ref{eq:Sqpk_X}) where
$X_\Delta$ is unknown.
However, even if $X_{\Delta}$ is unknown we can 
obtain very relevant information from the ${\cal{\slash{S}}}^{uu}$'s 
as is shown below.

We will now give a physical interpretation to the two parts 
of ${\cal{\slash{R}}}^{uu}$'s, i.e., ${\cal{\slash{S}}}^{uu}$'s 
and $X_\Delta$. We now see from Fig. 3 that
${\cal{\slash{S}}}^{uu}({\bf k|p|q}) + X_{\Delta}$ gets transferred
from {\bf p} to {\bf k}, 
${\cal{\slash{S}}}^{uu}({\bf q|k|p}) + X_{\Delta}$ gets transferred
from {\bf k} to {\bf q}, 
${\cal{\slash{S}}}^{uu}({\bf p|q|k}) + X_{\Delta}$ gets transferred
from {\bf q} to {\bf p}.  The quantity $X_\Delta$ flows along ${\bf p} 
\rightarrow {\bf k} \rightarrow {\bf q} \rightarrow {\bf p}$, 
circulating around the entire triad without changing the energy
of any of the modes. Therefore we will call it the 
{\it Circulating transfer}. 
Of the total energy transfer between
two modes, ${\cal{\slash{S}}}^{uu} + X_{\Delta}$,
only ${\cal{\slash{S}}}^{uu}$  can bring about  a change in modal energy. 
$X_{\Delta}$ transferred from mode {\bf p} to mode {\bf k} is transferred
back to mode {\bf p} via mode {\bf q}, i.e.,
the mode {\bf p} transfers $X_{\Delta}$ {\it directly}
to mode {\bf k}, and mode {\bf p} transfers $X_{\Delta}$ back to ${\bf k}$ 
{\it indirectly} through mode {\bf q}. 
Thus the energy that is effectively transferred from mode {\bf p} to 
mode {\bf k} is just ${\cal{\slash{S}}}^{uu}({\bf k|p|q})$. 
Therefore ${\cal{\slash{S}}}^{uu}({\bf k|p|q})$ can be
be termed as the {\it effective mode-to-mode transfer} from 
mode {\bf p} to mode {\bf k}. 

In summary, we have attempted to obtain an expression for the energy
transfer rate ${\cal{\slash{R}}}^{uu}$ between two modes in a triad.
To leading order in a series expansion, we can show that
${\cal{\slash{R}}}^{uu}={\cal{\slash{S}}}^{uu}$. But the ambiguity in
${\cal{\slash{R}}}^{uu}$ remains because of the lack of a complete
proof. However, the important conclusion is that the mode-to-mode 
energy transfer can be expressed as
${\cal{\slash{R}}}^{uu}={\cal{\slash{S}}}^{uu}+X_{\Delta}$ where
${\cal{\slash{S}}}^{uu}$ is the `effective mode-to-mode transfer' and
$X_{\Delta}$ is the `circulating transfer'.
In the next section we will further develop the notion of effective transfers 
by applying it to energy transfer between shells.

\section{Shell-to-Shell energy transfer and Cascade rates using ``mode-to-mode''
formalism}
\label{s:shell_redefn}
In Section~\ref{s:shell_intro}  we had pointed out the problem in treating
the expression in Eq.~(\ref{eq:shell_old_defn})
as the shell-to-shell transfer. This expression,
commonly taken to be the shell-to-shell transfer, actually gives
the energy transferred to shell {\it m} from both shell {\it n}
and the modes {\bf q} of type II triads (see Fig. 1).
In this section we will make use of the idea of ``circulating'' and
``effective mode-to-mode transfer'' to redefine shell-to-shell transfer. 

Consider energy transfer between shells {\it m} and
{\it n} in Fig. 4 . The mode {\bf k} is in shell
{\it m}, the mode {\bf p} is in shell {\it n}, and {\bf q} could be
inside or outside the shells. In terms of the mode-to-mode transfer
${\cal{\slash{R}}}^{uu}({\bf k|p|q})$ from mode {\bf p} to  mode {\bf k},
the energy transfer from shell {\it n} to shell {\it m} can be defined as
\begin{equation}
T^{uu}_{mn} = \sum_{{\bf k} {\large \epsilon}{\it m}} \sum_{{\bf p} 
              {\large \epsilon {\it n}}} 
                       {\cal{\slash{R}}}^{uu}({\bf k|p|q})
\label{eq:shell_new_defn}
\end{equation}
where the {\bf k}-sum is over shell {\it m}, {\bf p}-sum is over 
shell {\it n}, and {\bf k+p+q=0}.  The quantity
${\cal{\slash{R}}}^{uu}$ can be written
as a sum of an effective transfer 
${\cal{\slash{S}}}^{uu}({\bf k|p|q})$ and a circulating transfer  $X_\Delta$.
We know from the last section that the circulating transfer does
not contribute to the energy
change of modes. From Fig. 4 we can see that
$X_\Delta$ flows from shell {\it m} to shell {\it n} and
then flows back to {\it m} indirectly through the mode {\bf q}. Therefore
the {\it effective} energy transfer from shell {\it m} to shell {\it n}
is just ${\cal{\slash{S}}}^{uu}({\bf k|p|q})$ summed over all {\bf k}
in shell {\it m} and all {\bf p} in shell {\it n}, i.e.,
\begin{equation}
T^{uu}_{mn} = \sum_{{\bf k} {\large \epsilon}{\it m}} \sum_{{\bf p} 
              {\large \epsilon {\it n}}} 
                       {\cal{\slash{S}}}^{uu}({\bf k|p|q}),
\label{eq:shell_eff_defn}
\end{equation}
where ${\cal{\slash{S}}}^{uu}({\bf k|p|q})$ is the effective
mode-to-mode transfer.

If $X_{\Delta}=0$ then the effective energy transfer between two shells 
given by this equation will be the same as the
same as the total energy transfer between them given by
Eq.~(\ref{eq:shell_new_defn}). However, we have argued earlier that
the circulating transfer does not result in the energy change of any of
the modes.  Hence, from the physical point of view
the effective shell-to-shell transfer 
is a very relevant quantity to
study. In a subsequent paper we have studied the effective shell-to-shell
transfers in numerical simulations. 

\subsection{Energy cascade rates}
\label{subs:flux_ns_defn}
The kinetic energy cascade rate ($\Pi$) (or flux) in fluid turbulence
is defined as the rate of loss of kinetic energy by a sphere in
{\bf k}-space to the modes outside. 
In literature, the  energy flux in fluid turbulence has been computed using 
$S^{uu}({\bf k|p,q})$ ~\cite{Leslie,Kraichnan59}. 
Kraichnan ~\cite{Kraichnan59} shows by direct integration of
Eq.~(\ref{eq:eu_nseq_chap3}) that the kinetic energy lost from a sphere 
of radius $K$ by nonlinear convective transfers is
\begin{equation}
\Pi(K) = - \sum_{|{\bf k}|<K} \sum_{|{\bf p}|>K} 
\frac{1}{2}S^{uu}({\bf k|p,q}) .
\label{eq:deudt_fluid1}
\end{equation}
Although the energy cascade rate in fluid turbulence  can be found by 
the above formula, the mode-to-mode approach provides a  more natural way of 
looking at the energy transfers.
In later sections we will  obtain expressions for mode-to-mode transfer
in MHD turbulence and define various fluxes which are not
accessible to the conventional approach.
Here, we will obtain an
expression for the flux in terms of 
the effective mode-to-mode energy transfer.
Since ${\cal{\slash{R}}}^{uu}({\bf k|p|q})$ represents energy transfer
from ${\bf p}$ to ${\bf k}$ with {\bf q} as the mediator, we may 
alternatively write the energy loss from a sphere as
\begin{equation}
\Pi(K) =
  - \sum_{|{\bf k}|<K} \sum_{|{\bf p}|>K} 
            {\cal{\slash{R}}}^{uu}({\bf k|p|q}) .
\label{eq:deudt_fluid2}
\end{equation}
where mode {\bf k} is inside the sphere, {\bf p} is outside the sphere,
and ${\bf q=-k-p}$.
The mode-to-mode transfer ${\cal{\slash{R}}}^{uu}({\bf k|p|q})$
consists of a circulating part and an effective part. 
From Fig. 5 we see that the net amount of
circulating transfer leaving the sphere is zero. 
Thus the circulating transfer makes no contribution to the the
energy flux from the sphere.  Therefore, it can be
removed from Eq.~(\ref{eq:deudt_fluid2}) and the resultant expression 
for the flux can be written as
\begin{equation}
\Pi(K) = -\sum_{|{\bf k}|<K} \sum_{|{\bf p}|>K} 
            {\cal{\slash{S}}}^{uu}({\bf k|p|q}) .
\label{eq:deudt_fluid3}
\end{equation}
with a summation over the effective mode-to-mode transfer. 
The expressions in Eqs.~(\ref{eq:deudt_fluid1}) and
~(\ref{eq:deudt_fluid3}) are equivalent as is formally proved in Appendix B.


\section{Energy transfer in a MHD triad: known results}
\label{s:mhdtriad_intro}

  The MHD equations in real space are written as

\begin{equation}
\frac{\partial {\bf u}}{\partial t} + ({\bf u.\nabla}){\bf u} =
 - {\bf \nabla}p + ({\bf b.\nabla}){\bf b} + \nu {\bf \nabla^{2} u} ,
\label{eq:u_mhdeq_chap2}
\end{equation}
and
\begin{equation}
\frac{\partial {\bf b}}{\partial t} + ({\bf u.\nabla}){\bf b} =
  ({\bf b.\nabla}){\bf u} + \mu {\bf \nabla^{2} b},
\label{eq:b_mhdeq_chap2}
\end{equation}
where ${\bf u}$ and ${\bf b}$ are the velocity and 
magnetic fields respectively, and $\nu$ and $\mu$ are the
fluid kinematic viscosity and magnetic diffusivity, respectively.
In Fourier space, the kinetic energy and magnetic energy evolution equations
for a Fourier mode are 

\begin{equation}
\frac{\partial E^{u}(\bf k)}{\partial t} + 2 \nu k^{2} E^{u}({\bf k}) =
\sum_{{\bf k+p+q}=0} \frac{1}{2}S^{uu}({\bf k|p,q}) +
\sum_{{\bf k+p+q}=0} \frac{1}{2}S^{ub}({\bf k|p,q}) ,
\label{eq:eu_mhdeq}
\end{equation}

\begin{equation}
\frac{\partial E^{b}(\bf k)}{\partial t} + 2 \mu k^{2} E^{b}({\bf k}) =
\sum_{{\bf k+p+q}=0} \frac{1}{2}S^{bb}({\bf k|p,q}) +
\sum_{{\bf k+p+q}=0} \frac{1}{2}S^{bu}({\bf k|p,q}) ,
\label{eq:eb_mhdeq}
\end{equation}
where $E^{u}({\bf k})={\bf |u(k)|}^{2}/2$ is the kinetic energy,
and $E^{b}({\bf k})={\bf |b(k)|}^{2}/2$ is the magnetic energy.
The four nonlinear terms  $S^{uu}({\bf k|p,q})$, $S^{ub}({\bf k|p,q})$
$S^{bb}({\bf k|p,q})$ and $S^{bu}({\bf k|p,q})$ are 

\begin{equation}
      S^{uu}({\bf k|p,q}) \equiv
                         -  Re \left( i{\bf \left [k.u(q)\right]} 
                                     {\bf \left [u(k).u(p)\right ]} + 
                                   i{\bf \left [k.u(p)\right]} 
                                    {\bf \left [u(k).u(q)\right]} \right),
\label{eq:ukpq_mhd_def}
\end{equation}
\begin{equation}
      S^{bb}({\bf k|p,q}) \equiv
                        -  Re \left( i{\bf \left [k.u(q)\right]} 
                                     {\bf \left [b(k).b(p)\right ]} + 
                                   i{\bf \left [k.u(p)\right]} 
                                    {\bf \left [b(k).b(q)\right]} \right),
\label{eq:bkpq_def}
\end{equation}
\begin{equation}
      S^{ub}({\bf k|p,q}) \equiv
                          Re \left( i{\bf \left [k.b(q)\right]} 
                                     {\bf \left [u(k).b(p)\right ]} + 
                                   i{\bf \left [k.b(p)\right]} 
                                    {\bf \left [u(k).b(q)\right]} \right),
\label{eq:ubkpq_def}
\end{equation}
\begin{equation}
      S^{bu}({\bf k|p,q}) \equiv
                          Re \left( i{\bf \left [k.b(q)\right]} 
                                     {\bf \left [b(k).u(p)\right ]} + 
                                   i{\bf \left [k.b(p)\right]} 
                                    {\bf \left [b(k).u(q)\right]} \right).
\label{eq:bukpq_def}
\end{equation}
These terms are  conventionally taken to represent the nonlinear
transfer from modes ${\bf p}$ and ${\bf q}$ to mode ${\bf k}$ 
~\cite{Stanisic,Lesieur} of a triad such that ${\bf k+p+q}=0$. 
The term $S^{uu}(\bf k|p,q)$ represents the net transfer of kinetic energy
from modes ${\bf p}$ and ${\bf q}$ to mode ${\bf k}$. Likewise the term
$S^{ub}({\bf k|p,q})$ is the net magnetic energy transferred from modes
${\bf p}$ and ${\bf q}$ to the kinetic energy in mode ${\bf k}$,
whereas $S^{bu}({\bf k|p,q})$ is the net kinetic energy transferred from
modes ${\bf p}$ and ${\bf q}$ to the magnetic energy in mode ${\bf k}$.
The term $S^{bb}({\bf k|p,q})$ represents the transfer of magnetic 
energy from modes ${\bf p}$ and ${\bf q}$ to mode ${\bf k}$. All these
transfer terms are represented in the Fig. 6.

Stanisic ~\cite{Stanisic} 
showed that the nonlinear terms satisfy the following detailed conservation
properties:
  
\begin{equation}
S^{uu}({\bf k|p,q})+ S^{uu}({\bf p|k,q}) +
S^{uu}({\bf q|k,p}) = 0 ,
\label{eq:ukpq_triad_mhd}
\end{equation}

\begin{equation}
S^{bb}({\bf k|p,q})+ S^{bb}({\bf p|k,q}) +
S^{bb}({\bf q|k,p}) = 0 ,
\label{eq:bkpq_triad}
\end{equation}
and
\begin{equation}
S^{ub}({\bf k|p,q})+ S^{ub}({\bf p|k,q}) +
S^{ub}({\bf q|k,p}) +
S^{bu}({\bf k|p,q})+ S^{bu}({\bf p|k,q}) +
S^{bu}({\bf q|k,p}) = 0 .
\label{eq:ubkpq_triad}
\end{equation}
The Eq.~(\ref{eq:ukpq_triad_mhd}) implies that kinetic energy is transferred
conservatively between the velocity modes of a wavenumber triad, and
the Eq.~(\ref{eq:bkpq_triad}) implies that magnetic energy is also transferred
conservatively between the magnetic modes of a wavenumber triad.
The Eq.~(\ref{eq:ubkpq_triad})
implies that the cross transfers of 
kinetic and magnetic energy, $S^{ub}({\bf k|p,q})$ and 
$S^{bu}({\bf k|p,q})$, within a triad are also energy conserving. 

The quantities $S^{uu}({\bf k|p,q})$, $S^{ub}({\bf k|p,q})$, 
$S^{bb}({\bf k|p,q})$, and $S^{bu}({\bf k|p,q})$ represent the 
nonlinear energy transfer from the {\it two modes} $\bf p$ and $\bf q$ 
to mode $\bf k$. It would be useful to know the energy
transfer between {\it any two modes}, say from ${\bf p}$ to ${\bf k}$.
This issue will be explored in the next section.

\section{``Mode-to-Mode'' energy transfers in MHD Equations}
\label{s:mode_to_mode_mhd}
In Section~\ref{s:mode_to_mode_fluids} we had presented a new
approach to describe the energy transfer 
between two velocity modes of a triad mediated by the third velocity mode,
in the Navier-Stokes equation. 
In the same spirit we will now attempt to find the mode-to-mode
energy transfers in MHD. In MHD turbulence there will be three kinds of
mode-to-mode transfer within a wavenumber triad {\bf k, p, q}: 
kinetic energy transfer from
{\bf u(p)} to {\bf u(k)}; magnetic energy transfer from
{\bf b(p)} to {\bf b(k)}; and transfer of kinetic energy from
{\bf u(p)} to magnetic energy in {\bf b(k)}. We denote these three transfers
by 
${\cal{\slash{R}}}^{uu}({\bf k|p|q})$,
${\cal{\slash{R}}}^{bb}({\bf k|p|q})$, and
${\cal{\slash{R}}}^{bu}({\bf k|p|q})$ respectively, where the index
{\bf q} indicates that the energy transfer between modes {\bf p} and {\bf k}
is mediated by the mode {\bf q}. Since the nonlinear interactions 
fundamentally involve three-modes, the energy transfer between a pair of
modes should depend on the third mode.
The mode-to-mode transfers ${\cal{\slash{R}}}^{uu}$,
 ${\cal{\slash{R}}}^{bb}$, and ${\cal{\slash{R}}}^{bu}$ are schematically
illustrated in Fig. 7.

In this section we will obtain a description for each of these transfers.

\subsection{Velocity mode to velocity mode energy transfers}
\label{subs:umode_umode}
In Section~\ref{s:mode_to_mode_fluids} we discussed  the 
mode-to-mode transfer, ${\cal{\slash{R}}}^{uu}$, between velocity modes
in fluid flows. In this section we will find
${\cal{\slash{R}}}^{uu}$ for MHD flows. 
The transfer of kinetic energy between the velocity modes is brought about
by the term ${\bf (\bf u.\nabla)u}$, both in  the Navier Stokes and 
MHD equations. Hence, the expression for the combined kinetic energy transfer 
to a mode from the other two modes of the triad is also same for the two.
i.e., the combined transfer to {\bf u(k)} from {\bf u(p)} and {\bf u(q)} is
given by $S^{uu}({\bf k|p,q})$ 
[see Eqs.~(\ref{eq:ukpq_def}) and (\ref{eq:ukpq_mhd_def})]. 
Consequently, ${\cal{\slash{R}}}^{uu}$ 's for MHD  will satisfy the 
constraints given
in Eqs.~(\ref{eq:Rukup_ukuq})-(\ref{eq:Rupuq_uqup}) for the
corresponding ${\cal{\slash{R}}}^{uu}$ 's for fluids.
As a result, ${\cal{\slash{R}}}^{uu}({\bf k|p,q})$  in MHD can be expressed as
a sum of a circulating transfer $X_\Delta$ and the
effective transfer ${\cal{\slash{S}}}^{uu}({\bf k|p|q})$ given by
Eq.~(\ref{eq:Sukup_def}), i.e.,
\begin{equation}
{\cal{\slash{R}}}^{uu}({\bf k|p|q})=
{\cal{\slash{S}}}^{uu}({\bf k|p|q})+X_\Delta
\label{eq:mode_uu_mhd}
\end{equation}
The arguments are the same as those presented in 
Section ~\ref{subs:mode_to_mode_soln} for fluid turbulence.

\subsection{Magnetic mode to Magnetic mode energy transfers}
\label{subs:bmode_bmode}

Now we consider the magnetic  energy transfer 
from mode {\bf b(p)} to {\bf b(k)} in the triad ({\bf k,p,q}) 
[see Fig. 7]. This transfer which is denoted by
${\cal{\slash{R}}}^{bb}({\bf k|p|q})$, should satisfy the following 
relationships :

\begin{enumerate}
\item The sum of the mode-to-mode energy transfers from {\bf b(p)} to
{\bf b(k)} and from {\bf b(q)} to {\bf b(k)} should be equal to
$S^{bb}({\bf k|p,q})$ [Eq.~(\ref{eq:bkpq_def}) in 
Section~\ref{s:mhdtriad_intro}], 
which is the combined energy transfer to {\bf b(k)} from {\bf b(p)} 
and {\bf b(q)}. Thus,
\begin{equation}
{\cal{\slash{R}}}^{bb}({\bf k|p|q}) + {\cal{\slash{R}}}^{bb}({\bf k|q|p})
                                    = S^{bb}({\bf k|p,q}) , 
\label{eq:Rbkbp_bkbq}
\end{equation}
\begin{equation}
{\cal{\slash{R}}}^{bb}({\bf p|k|q}) + {\cal{\slash{R}}}^{bb}({\bf p|q|k})
                                    = S^{bb}({\bf p|k,q}) , 
\label{eq:Rbpbk_bpbq}
\end{equation}
\begin{equation}
{\cal{\slash{R}}}^{bb}({\bf q|k|p}) + {\cal{\slash{R}}}^{bb}({\bf q|p|k})
                                = S^{bb}({\bf q|k,p}).
\label{eq:Rbqbk_bqbp}
\end{equation}
\item the energy transfer from {\bf b(k}) to {\bf b(p)},
${\cal{\slash{R}}}^{bb}({\bf k|p|q})$,  
and the transfer from {\bf b(p}) to {\bf b(k)},
${\cal{\slash{R}}}^{bb}({\bf p|k|q})$, should be equal but opposite in
sign. That is,
\begin{equation}
{\cal{\slash{R}}}^{bb}({\bf k|p|q})+ {\cal{\slash{R}}}^{bb}({\bf p|k|q})=0 ,  
\label{eq:Rbkbp_bpbk}
\end{equation}
\begin{equation}
{\cal{\slash{R}}}^{bb}({\bf k|q|p})+ {\cal{\slash{R}}}^{bb}({\bf q|k|p})=0 ,
\label{eq:Rbkbq_bqbk}
\end{equation}
\begin{equation}
{\cal{\slash{R}}}^{bb}({\bf p|q|k})+ {\cal{\slash{R}}}^{bb}({\bf q|p|k})=0 .
\label{eq:Rbpbq_bqbp}
\end{equation}
\end{enumerate}

Once again, the above equations cannot uniquely determine 
${\cal{\slash{R}}}^{bb}({\bf k|p|q})$ since the value of the determinant
formed from these equations is zero. While discussing the mode-to-mode
energy transfers in Navier Stokes equation in 
Section~\ref{s:mode_to_mode_fluids} we  got the same
result for ${\cal{\slash{R}}}^{uu}({\bf k|p|q})$. Following the 
reasoning in Section ~\ref{s:mode_to_mode_fluids}, 
we can extract physically meaningful information from
Eqs.~(\ref{eq:Rbkbp_bkbq})-(\ref{eq:Rbpbq_bqbp}).

The combined energy transfer to {\bf b(k)} from {\bf b(p)} and {\bf b(q)}
is given by Eq.~(\ref{eq:bkpq_def}) of Section~\ref{s:mhdtriad_intro}.
We denote the first term on the right hand side of that equation by
${\cal{\slash{S}}}^{bb}({\bf k|p|q})$, i.e.,
\begin{equation}
{\cal{\slash{S}}}^{bb}({\bf k|p|q}) \equiv - Re \left( i{\bf \left[k.u(q)\right]} %
{\bf \left[b(k).b(p)\right]} \right) ,  
\label{eq:Sbkbp_def}
\end{equation}    
By replacing ${\cal{\slash{S}}}^{bb}({\bf k|p|q})$ for 
${\cal{\slash{R}}}^{bb}({\bf k|p|q})$ in 
Eqs.~(\ref{eq:Rbkbp_bkbq})-(\ref{eq:Rbpbq_bqbp}) we find that 
${\cal{\slash{S}}}^{bb}({\bf k|p|q})$ is a solution of these equations, i.e.,
\begin{equation}
{\cal{\slash{S}}}^{bb}({\bf k|p|q}) + {\cal{\slash{S}}}^{bb}({\bf k|q|p})
                                    = S^{bb}({\bf k|p,q}) , 
\label{eq:Sbkbp_bkbq}
\end{equation}
\begin{equation}
{\cal{\slash{S}}}^{bb}({\bf p|k|q}) + {\cal{\slash{S}}}^{bb}({\bf p|q|k})
                                    = S^{bb}({\bf p|k,q}) , 
\label{eq:Sbpbk_bpbq}
\end{equation}
\begin{equation}
{\cal{\slash{S}}}^{bb}({\bf q|k|p}) + {\cal{\slash{S}}}^{bb}({\bf q|p|k})
                                = S^{bb}({\bf q|k,p}) .  
\label{eq:Sbqbk_bqbp}
\end{equation}
and
\begin{equation}
{\cal{\slash{S}}}^{bb}({\bf k|p|q})+ {\cal{\slash{S}}}^{bb}({\bf p|k|q})=0 ,  
\label{eq:Sbkbp_bpbk}
\end{equation}
\begin{equation}
{\cal{\slash{S}}}^{bb}({\bf k|q|p})+ {\cal{\slash{S}}}^{bb}({\bf q|k|p})=0 ,
\label{eq:Sbkbq_bqbk}
\end{equation}
\begin{equation}
{\cal{\slash{S}}}^{bb}({\bf p|q|k})+ {\cal{\slash{S}}}^{bb}({\bf q|p|k})=0 .
\label{eq:Sbpbq_bqbp}
\end{equation}
Thus, ${\cal{\slash{S}}}^{bb}({\bf k|p|q})$ is a solution of the equations %
~(\ref{eq:Rbkbp_bkbq})---(\ref{eq:Rbpbq_bqbp}). By inspection it can be
seen that {\it all} solutions of the equations
can be expressed as
\begin{equation}
{\cal{\slash{R}}}^{bb}({\bf k|p|q}) = 
{\cal{\slash{S}}}^{bb}({\bf k|p|q})+Y_{\Delta},
\label{eq:Skpq_Ybb}
\end{equation}
\begin{equation}
{\cal{\slash{R}}}^{bb}({\bf k|q|p}) = 
{\cal{\slash{S}}}^{bb}({\bf k|q|p})-Y_{\Delta},
\label{eq:Skqp_Ybb}
\end{equation}
\begin{equation}
{\cal{\slash{R}}}^{bb}({\bf p|k|q}) = 
{\cal{\slash{S}}}^{bb}({\bf p|k|q})-Y_{\Delta},
\label{eq:Spkq_Ybb}
\end{equation}
\begin{equation}
{\cal{\slash{R}}}^{bb}({\bf p|q|k}) = 
{\cal{\slash{S}}}^{bb}({\bf p|q|k})+Y_{\Delta},
\label{eq:Spqk_Ybb}
\end{equation}
\begin{equation}
{\cal{\slash{R}}}^{bb}({\bf q|k|p}) = 
{\cal{\slash{S}}}^{bb}({\bf q|k|p})+Y_{\Delta},
\label{eq:Sqkp_Ybb}
\end{equation}
\begin{equation}
{\cal{\slash{R}}}^{bb}({\bf q|p|k}) = 
{\cal{\slash{S}}}^{bb}({\bf q|p|k})-Y_{\Delta},
\label{eq:Sqpk_Ybb}
\end{equation}
where $Y_\Delta$ is an arbitrary scalar function dependent on
the wavenumber triad {\bf k, p, q} and the Fourier components 
{\bf u(k), u(p), u(q), b(k), b(p), b(q)} at those wavenumbers. To determine
the unique form of this function we need to impose additional constraints,
similar to those imposed on $X_{\Delta}$ in Appendix A for 
determining mode-to-mode energy transfer between 
velocity modes in fluid turbulence (see
Section~\ref{subs:mode_to_mode_soln}). 
The arguments for determination of  $Y_{\Delta}$ is given in Appendix B.

The solutions in Eqs.~(\ref{eq:Skpq_Ybb})-(\ref{eq:Sqpk_Ybb})
have been schematically illustrated in Fig. 8.
We see from the figure that $Y_\Delta$ is transferred from
${\bf b(p) \rightarrow b(k) \rightarrow b(q) \rightarrow}$ and back to
${\bf b(p)}$, i.e., it circulates around the triad 
without causing any change in modal energy --- it is a thus a 
circulating transfer.
The magnetic energy {\it effectively}
transferred from {\bf b(p)} to {\bf b(k)} is just
${\cal{\slash{S}}}^{bb}({\bf k|p|q})$\footnote{the idea of circulating
transfer and effective mode-to-mode transfer in a triad was introduced
in Section~\ref{subs:mode_to_mode_soln} ---  a detailed discussion 
can be found in that section.},i.e., 
\begin{equation}
{\cal{\slash{R}}}^{bb}_{eff}({\bf k|p|q})=
{\cal{\slash{S}}}^{bb}({\bf k|p|q}).
\end{equation}
The effective mode-to-mode transfer
${\cal{\slash{S}}}^{bb}({\bf k|p|q})$ from {\bf b(p)} to {\bf b(k)} is
mediated by the {\it velocity mode} {\bf u(q)}. 

In the next section we will use ${\cal{\slash{R}}}^{bb}_{eff}$ to
calculate effective energy transfer rate between the magnetic modes 
inside two shells and the effective cascade rate of magnetic energy.

\subsection{Velocity mode to Magnetic mode energy transfers}
\label{subs:umode_bmode}
                             
In Section~\ref{subs:umode_umode} and Section~\ref{subs:bmode_bmode}
we discussed mode-to-mode  energy transfer between
a pair of velocity modes and between a pair of magnetic modes in a
wavenumber triad, respectively. We now consider  the
energy transfer between {\bf u(p)} and {\bf b(k)}, 
${\cal{\slash{R}}}^{ub}({\bf k|p|q})$, 
within the triad ({\bf k, p, q}) illustrated in Fig. 7 .
We will follow the same sequence of steps as in the two
sections \ref{subs:umode_umode} and \ref{subs:bmode_bmode}.

${\cal{\slash{R}}}^{ub}$ 's will satisfy the following relationships:

\begin{enumerate}
\item Since ${\cal{\slash{R}}}^{ub}({\bf k|p|q})$ and 
${\cal{\slash{R}}}^{ub}({\bf k|q|p})$ are the mode-to-mode energy transfers
from {\bf b(p)} to {\bf u(k)} and from {\bf b(q)} to {\bf u(k)}, the sum of
the two should be equal to $S^{ub}({\bf k|p,q})$, the combined energy
transfer to {\bf u(k)} from {\bf b(p)} and {\bf b(q)}. Therefore, we get
\noindent 
\begin{equation}
{\cal{\slash{R}}}^{ub}({\bf k|p|q}) + {\cal{\slash{R}}}^{ub}({\bf k|q|p})
                                    = S^{ub}({\bf k|p,q}) ,
\label{eq:Rukbp_ukbq}
\end{equation}
\begin{equation}
{\cal{\slash{R}}}^{ub}({\bf p|k|q}) + {\cal{\slash{R}}}^{ub}({\bf p|q|k})
                                    = S^{ub}({\bf p|k,q}) ,
\label{eq:Rupbk_upbq}
\end{equation}
\begin{equation}
{\cal{\slash{R}}}^{ub}({\bf q|k|p}) + {\cal{\slash{R}}}^{ub}({\bf q|p|k})
                                    = S^{ub}({\bf q|k,p}) .
\label{eq:Ruqbk_uqbp}
\end{equation}
\begin{equation}
{\cal{\slash{R}}}^{bu}({\bf k|p|q}) + {\cal{\slash{R}}}^{bu}({\bf k|q|p})
                                    = S^{bu}({\bf k|p,q}) ,
\label{eq:Rbkup_bkuq}
\end{equation}
\begin{equation}
{\cal{\slash{R}}}^{bu}({\bf p|k|q}) + {\cal{\slash{R}}}^{bu}({\bf p|q|k})
                                    = S^{bu}({\bf p|k,q}) ,
\label{eq:Rbpuk_bpuq}
\end{equation}
\begin{equation}
{\cal{\slash{R}}}^{bu}({\bf q|k|p}) + {\cal{\slash{R}}}^{bu}({\bf q|p|k})
                                    = S^{bu}({\bf q|k,p}) .
\label{eq:Rbquk_bqup}
\end{equation}
\item ${\cal{\slash{R}}}^{ub}({\bf k|p|q})$ denotes the energy transfer
from {\bf b(p)} to {\bf u(k)}.  ${\cal{\slash{R}}}^{bu}({\bf p|k|q})$ is
physically the same transfer but seen as a transfer from 
{\bf u(k)} to {\bf b(p)}. Therefore,
\begin{equation}
{\cal{\slash{R}}}^{ub}({\bf k|p|q})+ {\cal{\slash{R}}}^{bu}({\bf p|k|q})=0 ,
\label{eq:Rukbp_bpuk}
\end{equation}
\begin{equation}
{\cal{\slash{R}}}^{ub}({\bf k|q|p})+ {\cal{\slash{R}}}^{bu}({\bf q|k|p})=0 ,
\label{eq:Rukbq_bquk}
\end{equation}
\begin{equation}
{\cal{\slash{R}}}^{ub}({\bf p|q|k})+ {\cal{\slash{R}}}^{bu}({\bf q|p|k})=0 .
\label{eq:Rupbq_bqup}
\end{equation}
\begin{equation}
{\cal{\slash{R}}}^{bu}({\bf k|p|q})+ {\cal{\slash{R}}}^{ub}({\bf p|k|q})=0 ,
\label{eq:Rbkup_upbk}
\end{equation}
\begin{equation}
{\cal{\slash{R}}}^{bu}({\bf k|q|p})+ {\cal{\slash{R}}}^{ub}({\bf q|k|p})=0 ,
\label{eq:Rbkuq_uqbk}
\end{equation}
\begin{equation}
{\cal{\slash{R}}}^{bu}({\bf p|q|k})+ {\cal{\slash{R}}}^{ub}({\bf q|p|k})=0 .
\label{eq:Rbpuq_uqbp}
\end{equation}

\end{enumerate}

The solutions of these equations are not unique and the expression for
the energy transfer between modes cannot be obtained from the above
equations alone. As in
sections~\ref{s:mode_to_mode_fluids} and \ref{subs:umode_umode},
we will now explore the solutions of the above equations.

We define the following quantities :
\begin{equation}
{\cal{\slash{S}}}^{ub}({\bf k|p|q}) \equiv - Re \left( i{\bf \left[k.b(q)\right]} %
{\bf \left[u(k).b(p)\right]} \right) ,
\label{eq:Sukbp_def}
\end{equation}    
\begin{equation}
{\cal{\slash{S}}}^{bu}({\bf k|p|q}) \equiv - Re \left( i{\bf \left[k.b(q)\right]} %
{\bf \left[b(k).u(p)\right]} \right) ,
\label{eq:bkup_def}
\end{equation}    
Replacing ${\cal{\slash{S}}}^{ub}$  and ${\cal{\slash{S}}}^{bu}$ 
for ${\cal{\slash{R}}}^{ub}$ and ${\cal{\slash{R}}}^{bu}$ respectively
in Eqs.~(\ref{eq:Rukbp_ukbq})-(\ref{eq:Rbpuq_uqbp}) we find that 
the ${\cal{\slash{S}}}$'s are a solution to those equations, i.e.,
\begin{equation}
{\cal{\slash{S}}}^{ub}({\bf k|p|q}) + {\cal{\slash{S}}}^{ub}({\bf k|q|p})
                                    = S^{ub}({\bf k|p,q}) ,
\label{eq:Sukbp_ukbq}
\end{equation}
\begin{equation}
{\cal{\slash{S}}}^{ub}({\bf p|k|q}) + {\cal{\slash{S}}}^{ub}({\bf p|q|k})
                                    = S^{ub}({\bf p|k,q}) ,
\label{eq:Supbk_upbq}
\end{equation}
\begin{equation}
{\cal{\slash{S}}}^{ub}({\bf q|k|p}) + {\cal{\slash{S}}}^{ub}({\bf q|p|k})
                                    = S^{ub}({\bf q|k,p}) .
\label{eq:Suqbk_uqbp}
\end{equation}
\begin{equation}
{\cal{\slash{S}}}^{bu}({\bf k|p|q}) + {\cal{\slash{S}}}^{bu}({\bf k|q|p})
                                    = S^{bu}({\bf k|p,q}) ,
\label{eq:Sbkup_bkuq}
\end{equation}
\begin{equation}
{\cal{\slash{S}}}^{bu}({\bf p|k|q}) + {\cal{\slash{S}}}^{bu}({\bf p|q|k})
                                    = S^{bu}({\bf p|k,q}) ,
\label{eq:Sbpuk_bpuq}
\end{equation}
\begin{equation}
{\cal{\slash{S}}}^{bu}({\bf q|k|p}) + {\cal{\slash{S}}}^{bu}({\bf q|p|k})
                                    = S^{bu}({\bf q|k,p}),
\label{eq:Sbquk_bqup}
\end{equation}
and
\begin{equation}
{\cal{\slash{S}}}^{ub}({\bf k|p|q})+ {\cal{\slash{S}}}^{bu}({\bf p|k|q})=0 ,
\label{eq:Sukbp_bpuk}
\end{equation}
\begin{equation}
{\cal{\slash{S}}}^{ub}({\bf k|q|p})+ {\cal{\slash{S}}}^{bu}({\bf q|k|p})=0 ,
\label{eq:Sukbq_bquk}
\end{equation}
\begin{equation}
{\cal{\slash{S}}}^{ub}({\bf p|q|k})+ {\cal{\slash{S}}}^{bu}({\bf q|p|k})=0 ,
\label{eq:Supbq_bqup}
\end{equation}
\begin{equation}
{\cal{\slash{S}}}^{bu}({\bf k|p|q})+ {\cal{\slash{S}}}^{ub}({\bf p|k|q})=0 ,
\label{eq:Sbkup_upbk}
\end{equation}
\begin{equation}
{\cal{\slash{S}}}^{bu}({\bf k|q|p})+ {\cal{\slash{S}}}^{ub}({\bf q|k|p})=0 ,
\label{eq:Sbkuq_uqbk}
\end{equation}
\begin{equation}
{\cal{\slash{S}}}^{bu}({\bf p|q|k})+ {\cal{\slash{S}}}^{ub}({\bf q|p|k})=0 .
\label{eq:Sbpuq_uqbp}
\end{equation}
The ${\cal{\slash{S}}}^{ub}$ 's are just a single instance of the
the ${\cal{\slash{R}}}^{ub}$'s. It can be seen by inspection 
that {\it all} solutions can be expressed in the form :
\begin{equation}
{\cal{\slash{R}}}^{bu}({\bf k|p|q}) = 
{\cal{\slash{S}}}^{bu}({\bf k|p|q})+Z_{\Delta},
\label{eq:Skpq_Zbu}
\end{equation}
\begin{equation}
{\cal{\slash{R}}}^{bu}({\bf k|q|p}) = 
{\cal{\slash{S}}}^{bu}({\bf k|q|p})-Z_{\Delta},
\label{eq:Skqp_Zbu}
\end{equation}
\begin{equation}
{\cal{\slash{R}}}^{bu}({\bf p|k|q}) = 
{\cal{\slash{S}}}^{bu}({\bf p|k|q})-Z_{\Delta},
\label{eq:Spkq_Zbu}
\end{equation}
\begin{equation}
{\cal{\slash{R}}}^{bu}({\bf p|q|k}) = 
{\cal{\slash{S}}}^{bu}({\bf p|q|k})+Z_{\Delta},
\label{eq:Spqk_Zbu}
\end{equation}
\begin{equation}
{\cal{\slash{R}}}^{bu}({\bf q|k|p}) = 
{\cal{\slash{S}}}^{bu}({\bf q|k|p})+Z_{\Delta},
\label{eq:Sqkp_Zbu}
\end{equation}
\begin{equation}
{\cal{\slash{R}}}^{bu}({\bf q|p|k}) = 
{\cal{\slash{S}}}^{bu}({\bf q|p|k})-Z_{\Delta},
\label{eq:Sqpk_Zbu}
\end{equation}
\begin{equation}
{\cal{\slash{R}}}^{ub}({\bf k|p|q}) = 
{\cal{\slash{S}}}^{ub}({\bf k|p|q})+Z_{\Delta},
\label{eq:Skpq_Zub}
\end{equation}
\begin{equation}
{\cal{\slash{R}}}^{ub}({\bf k|q|p}) = 
{\cal{\slash{S}}}^{ub}({\bf k|q|p})-Z_{\Delta},
\label{eq:Skqp_Zub}
\end{equation}
\begin{equation}
{\cal{\slash{R}}}^{ub}({\bf p|k|q}) = 
{\cal{\slash{S}}}^{ub}({\bf p|k|q})-Z_{\Delta},
\label{eq:Spkq_Zub}
\end{equation}
\begin{equation}
{\cal{\slash{R}}}^{ub}({\bf p|q|k}) = 
{\cal{\slash{S}}}^{ub}({\bf p|q|k})+Z_{\Delta},
\label{eq:Spqk_Zub}
\end{equation}
\begin{equation}
{\cal{\slash{R}}}^{ub}({\bf q|k|p}) = 
{\cal{\slash{S}}}^{ub}({\bf q|k|p})+Z_{\Delta},
\label{eq:Sqkp_Zub}
\end{equation}
\begin{equation}
{\cal{\slash{R}}}^{ub}({\bf q|p|k}) = 
{\cal{\slash{S}}}^{ub}({\bf q|p|k})-Z_{\Delta},
\label{eq:Sqpk_Zub}
\end{equation}
where $Z_\Delta$ is an arbitrary function dependent on
{\bf k, p, q, u(k), u(p), u(q), b(k), b(p), b(q)}.
These solutions are pictorially represented in Fig. 9
below.

We are already familiar with such solutions from the earlier
discussions.  From Fig. 9 we see that 
$Z_\Delta$ transfers energy from ${\bf u(p) \rightarrow b(k) \rightarrow
u(q) \rightarrow b(p) \rightarrow u(k) \rightarrow b(q) \rightarrow}$
and back to ${\bf u(p)}$ without resulting in a change in
modal energy. 
Hence, following the discussions in Section~\ref{subs:umode_umode} and 
Section~\ref{subs:bmode_bmode}, we 
can interpret the quantity $Z_\Delta$ as a circulating transfer.
The ${\cal{\slash{S}}}^{bu}$  can be interpreted as the effective 
mode-to-mode transfers. For example, 
${\cal{\slash{S}}}^{bu}({\bf k|p|q})$ is the effective transfer from
{\bf u(p)} to {\bf b(k)},i.e,
\begin{equation}
{\cal{\slash{R}}}^{bu}_{eff}({\bf k|p|q})=
{\cal{\slash{S}}}^{bu}({\bf k|p|q}),
\end{equation}
which is mediated by the {\it magnetic mode} {\bf b(q)}.

In the next section, we will use the effective mode-to-mode transfer
to define, (a) shell-to-shell transfers between the velocity and the
magnetic modes, and (b) cascade rates between the velocity and the magnetic
modes.
\section{Shell-to-Shell energy transfer and cascade rates in MHD turbulence}
\label{s:shell_to_shell_mhd}
In this section, we will extend the formulation of
Section~\ref{s:shell_redefn} to define shell-to-shell transfer rates and 
cascade rates in MHD turbulence.

We shall use the term {\it u}-shell and {\it u}-sphere respectively to
denote a shell and a sphere in {\it k}-space containing velocity modes,
and {\it b}-shell and {\it b}-sphere for the corresponding magnetic modes.

\subsection{Shell-to-Shell energy transfer rates}
In Section~\ref{s:shell_redefn} we had defined the 
effective shell-to-shell transfer rates between two {\it u}-shells in
Fourier space. The expression for the effective
mode-to-mode transfer between two velocity modes
is the same for both  Fluid and MHD flows 
[see Section~\ref{subs:mode_to_mode_soln} and \ref{subs:umode_umode}].
Hence, the effective shell-to-shell transfer rates between two 
{\it u}-shells in MHD will also be 
defined by Eq.~(\ref{eq:shell_eff_defn}) of Section~\ref{s:shell_redefn}, i.e.,
\begin{equation}
T^{uu}_{mn} = \sum_{{\bf k} {\large \epsilon}{\it m}} \sum_{{\bf p} 
              {\large \epsilon {\it n}}} 
                       {\cal{\slash{S}}}^{uu}({\bf k|p|q}).
\label{eq:ushell_ushell}
\end{equation}

Similarly, the quantity ${\cal{\slash{R}}}^{bb}({\bf k|p|q})$ 
gives the energy transfer rate from {\bf b(p)} to {\bf b(k)}, mediated by
{\bf q}. Hence the energy transfer rate from $n^{th}$ {\it b}-shell to
the $m^{th}$ {\it b}-shell can be obtained by summing 
${\cal{\slash{R}}}^{bb}({\bf k|p|q})$ over every {\bf p} in the 
the $n^{th}$ {\it b}-shell  and  over every {\bf k} in the
$m^{th}$ {\it b}-shell.
From Section~\ref{subs:bmode_bmode} we know that 
${\cal{\slash{R}}}^{bb}={\cal{\slash{S}}}^{bb}({\bf k|p|q})+Y_\Delta$,
where ${\cal{\slash{S}}}^{bb}({\bf k|p|q})$ is the effective mode-to-mode
transfer from {\bf b(p)} to {\bf b(k)}, and $Y_\Delta$ is a 
circulating transfer.
Thus, $Y_\Delta$ is transferred from {\bf b(p)} in
shell {\it n} to {\bf b(k)} in shell {\it m} and then back to
{\bf b(p)} via the mode {\bf b(q)}. Therefore, the 
{\it effective} shell-to-shell transfer transfer rate
from the $n^{th}$ to the $m^{th}$ {\it b}-shell is
\begin{equation}
T^{bb}_{mn} = \sum_{{\bf k} {\large \epsilon}{\it m}} \sum_{{\bf p} 
              {\large \epsilon {\it n}}} 
                       {\cal{\slash{S}}}^{bb}({\bf k|p|q}).
\label{eq:bshell_bshell}
\end{equation}

We can now also define shell-to-shell transfers rates between a {\it u}-shell 
and a {\it b}-shell. ${\cal{\slash{R}}}^{bu}({\bf k|p|q})$ is the
energy transfer rate from {\bf u(p)} to {\bf b(k)} mediated by mode {\bf q}.
Hence the energy transfer rate from the
the $n^{th}$ {\it u}-shell to the $m^{th}$ {\it b}-shell  can be
calculated by summing ${\cal{\slash{R}}}^{bu}({\bf k|p|q})$ 
over every {\bf p} in the $n^{th}$ {\it u}-shell and over every
{\bf k} in $m^{th}$ {\it b}-shell. 
We have shown in Section~\ref{subs:umode_bmode} that 
${\cal{\slash{R}}}^{bu}({\bf k|p|q})={\cal{\slash{S}}}^{bu}({\bf k|p|q})
+Z_{\Delta}$, where ${\cal{\slash{S}}}^{bu}({\bf k|p|q})$ is the effective
mode-to-mode transfer from the mode {\bf u(p)} to the mode {\bf b(k)} in
the triad {\bf k, p, q}, and $Z_\Delta$ is the circulating transfer which
is transferred along ${\bf u(p) \rightarrow b(k) \rightarrow u(q)
\rightarrow b(p) \rightarrow u(k) \rightarrow b(q) \rightarrow u(p)}$.
Hence, $Z_\Delta$ is transferred from the mode {\bf u(p)} in shell {\it n}
to {\bf b(k)} in shell {\it m} and then flows back to {\bf u(p)} via the
modes ${\bf u(q) \rightarrow b(p) \rightarrow u(k) \rightarrow b(q) 
\rightarrow u(p) \rightarrow u(p)}$. Therefore, we interpret the
quantity obtained by summing ${\cal{\slash{S}}}^{bu}({\bf k|p|q})$ over
every {\bf p} in the $n^{th}$ {\it u}-shell and every {\bf k} in the
$m^{th}$ {\it b}-shell as the {\it effective} shell-to-shell transfer.
That is,
\begin{equation}
T^{bu}_{mn} = \sum_{{\bf k} {\large \epsilon}{\it m}} \sum_{{\bf p} 
              {\large \epsilon {\it n}}} 
                       {\cal{\slash{S}}}^{bu}({\bf k|p|q})
\label{eq:bshell_ushell}
\end{equation}
is the effective transfer from the the $n^{th}$ {\it u}-shell to the
$m^{th}$ {\it b}-shell.
The energy transfer rate from the $n^{th}$ {\it b}-shell to the
$m^{th}$ {\it u}-shell, $T^{ub}_{mn}=-T^{bu}_{mn}$.

In Appendix A-C, we have argued that the circulating
transfer $X_{\Delta}=0$, $Y_{\Delta}=0$, and $Z_{\Delta}=0$ to the first order
in the expansion coefficients.
If $X_{\Delta}=0$, $Y_{\Delta}=0$, $Z_{\Delta}=0$, then the effective
shell-to-shell transfers will be the same as the total shell-to-shell
transfers. However, we had pointed out is 
Section~\ref{s:shell_redefn} that the circulating transfer, if nonzero,
does not contribute to a change in energy of any shell, and hence
from the physical point of view the effective shell-to-shell
transfers is of greater significance than the total shell-to-shell
transfers.  We have numerically computed
 the effective shell-to-shell transfers given by
Eqs.~(\ref{eq:ushell_ushell})-(\ref{eq:bshell_ushell}) in 
numerical simulations of 2-D MHD turbulence and gained important insights 
into the energy transfer process. The results of the simulations will
be discussed in a companion paper \cite{Dar2000_2}.

\subsection{Energy cascade rates}

There are various types of cascade rates (energy fluxes) in MHD
turbulence. We have schematically shown these transfers in 
Fig. 10. In this section we will derive the
formulae to calculate these cascade rates within the framework of
effective transfers.

In Section~\ref{subs:flux_ns_defn}, we derived an expression for
kinetic energy flux in terms of the mode-to-mode transfer
[Eq.~(\ref{eq:shell_new_defn})]. We showed 
that the circulating transfer $X_\Delta$ does not contribute to the
flux. Hence, the kinetic energy flux can be expressed by 
Eq.~(\ref{eq:deudt_fluid3}) in terms of the effective mode-to-mode transfer
${\cal{\slash{S}}}^{uu}({\bf k|p|q})$.
The kinetic energy flux $\Pi^{u<}_{u>}$ (the energy transfer rate
from the {\it u}-sphere to outside the same sphere)
in MHD will also be given by the same equation, i.e.,
\begin{equation}
\Pi^{u<}_{u>}(K) = \sum_{|{\bf p}|<K} \sum_{|{\bf k}|>K} 
            {\cal{\slash{S}}}^{uu}({\bf k|p|q}) .
\label{eq:flux_uinuout}
\end{equation}

The magnetic energy flux $\Pi^{b<}_{b>}(K)$ is defined as
the rate of energy lost by the {\it b}-sphere to the modes outside
the {\it b}-sphere. Since ${\cal{\slash{R}}}^{bb}({\bf k|p|q})$ is the
mode-to-mode transfer from {\bf b(p)} to {\bf b(k)}, $\Pi^{b<}_{b>}(K)$
can be obtained by summing ${\cal{\slash{R}}}^{bb}({\bf k|p|q})$ over
every mode {\bf p} inside the {\it b}-sphere and every mode
{\bf k}  outside the {\it b}-sphere, i.e.,
\begin{equation}
\Pi^{b<}_{b>}(K) =
   \sum_{|{\bf p}|<K} \sum_{|{\bf k}|>K} 
            {\cal{\slash{R}}}^{bb}({\bf k|p|q}) .
\label{eq:debdt_fluid2}
\end{equation}
As in the kinetic energy flux, the circulating transfer 
$Y_{\Delta}$  will not contribute to the magnetic energy flux in
Eq.~(\ref{eq:debdt_fluid2}). We briefly explain this as follows: 
let us take the mode {\bf b(q)} to be outside the {\it b}-sphere.
Then if the mode {\bf b(k)} loses $Y_\Delta$ to mode
{\bf b(p)} then it will also
gain $Y_\Delta$ from mode {\bf b(q)}, and hence $Y_\Delta$  
will  not contribute to the magnetic energy flux. Similar arguments apply
when ${\bf b(q)}$ is inside the {\it b}-sphere.
Therefore, we can write the magnetic energy flux in 
Eq.~(\ref{eq:debdt_fluid2}) as
\begin{equation}
\Pi^{b<}_{b>}(K) =
   \sum_{|{\bf p}|<K} \sum_{|{\bf k}|>K} 
            {\cal{\slash{S}}}^{bb}({\bf k|p|q}) .
\label{eq:flux_binbout}
\end{equation}
where ${\cal{\slash{S}}}^{bb}({\bf k|p|q})$ is given by
Eq.~(\ref{eq:Sbkbp_def}) of Section~\ref{subs:bmode_bmode}.

Similarly, 
the energy transfer between the kinetic energy and magnetic energy can
be described by four types of fluxes. We shall define these fluxes below.

There is a transfer of energy from a
{\it u}-sphere of radius $K$ to the {\it b}-sphere 
of the same radius. The rate of loss of energy from the {\it u}-sphere
to the corresponding {\it b}-sphere can be calculated by
by summing ${\cal{\slash{R}}}^{bu}({\bf k|p|q})$ over every mode
inside the {\it b}-sphere and the {\it u}-sphere. 
Following the definition of the effective shell-to-shell transfer 
$T^{bu}_{mn}$ we can define the effective flux from the {\it u}-sphere
to the {\it b}-sphere as
\begin{equation}
\Pi^{u<}_{b<}(K) =
   \sum_{|{\bf p}|<K} \sum_{|{\bf k}|<K} 
            {\cal{\slash{S}}}^{bu}({\bf k|p|q}) .
\label{eq:flux_uinbin}
\end{equation}
by summing the effective mode-to-mode transfer,
${\cal{\slash{S}}}^{bu}({\bf k|p|q})$, from {\bf u(p)} to {\bf b(k)}, 
over the modes inside the {\it u}-sphere and the {\it b}-sphere.

There is a transfer of energy from the {\it u}-sphere to the
magnetic modes outside the corresponding {\it b}-sphere. The rate of
loss of the
the energy from the {\it u}-sphere to the modes outside the {\it b}-sphere
can be obtained by
summing ${\cal{\slash{R}}}^{bu}({\bf k|p|q})$ over every mode
inside the {\it u}-sphere and every mode outside the {\it b}-sphere. 
The corresponding effective flux can be calculated using
\begin{equation}
\Pi^{u<}_{b>}(K) =
   \sum_{|{\bf p}|<K} \sum_{|{\bf k}|>K} 
            {\cal{\slash{S}}}^{bu}({\bf k|p|q}) .
\label{eq:flux_uinbout}
\end{equation}

Similarly, there is a transfer of magnetic energy from a {\it b}-sphere
to velocity modes outside the {\it u}-sphere. The rate of loss of energy from
the {\it b}-sphere to the modes outside the {\it u}-sphere can be calculated
by 
\begin{equation}
\Pi^{b<}_{u>}(K) =
   \sum_{|{\bf p}|<K} \sum_{|{\bf k}|>K} 
            {\cal{\slash{S}}}^{ub}({\bf k|p|q}) .
\label{eq:flux_binuout}
\end{equation}

There is a transfer of energy from modes outside the 
{\it u}-sphere to the modes outside the {\it b}-sphere. The rate of loss
of energy by the modes outside the
{\it u}-sphere to the modes outside the {\it b}-sphere can be obtained by
\begin{equation}
\Pi^{u>}_{b>}(K) =
   \sum_{|{\bf p}|>K} \sum_{|{\bf k}|>K} 
            {\cal{\slash{S}}}^{bu}({\bf k|p|q}) .
\label{eq:flux_uoutbout}
\end{equation}

The total effective flux is defined as the total energy (kinetic+magnetic) lost
by the 
$K$-sphere to the modes outside, i.e.,
\begin{equation}
\Pi_{tot}(K) = \Pi^{u<}_{u>}(K) + \Pi^{b<}_{b>}(K) + \Pi^{u<}_{b>}(K) +
                \Pi^{b<}_{u>}(K). 
\label{eq:flux_tot}
\end{equation}

A schematic illustration of the effective fluxes defined in
Eqs.~(\ref{eq:flux_uinuout})-(\ref{eq:flux_uoutbout}) was
given in Fig. 10.
If the circulating transfer is zero, then each of the effective fluxes
will be the same as the corresponding total fluxes. 
In a subsequent paper we will present the results of the detailed study of
these fluxes in numerical simulations of 2-D MHD turbulence. 

\section{Conclusion}
\label{s:conclusion_chap3}

In literature we find a description of energy transfer rates from two
modes in a triad to the third mode, i.e., from {\bf u(p)} and {\bf u(q)}
to {\bf u(k)}. Here we have constructed new formulae to describe
energy transfer rates between a pair of modes in the triad (mode-to-mode
transfer), say from
{\bf u(p)} to {\bf u(k)}. The third mode in the triad acts as a mediator
in the transfer process.

The mode-to-mode energy transfer in our formalism
can be expressed as a combination of an  ``effective transfer'' 
{\it and} a ``circulating transfer''. 
In Appendices A-C we have shown by imposing galilean
invariance, symmetry and some other constraints that the circulating
transfer should be zero. A completely general proof is not available at
present. However even if the circulating transfer happens to be nonzero,
it will not result in a change of modal energy, since the amount of circulating
transfer gained by mode {\bf k} from mode {\bf p} is also lost by mode 
{\bf k} to mode {\bf q} (see Figs. 3, 8, 9).
Only the effective transfer is
responsible for modal energy change. As the circulating transfer
does not have any observable effect on the energy 
of the modes, it may be correct to ignore it from the study of energy
transfer.

Using the notion of effective mode-to-mode transfer and the circulating
transfer we defined  effective shell-to-shell energy transfer rates and 
energy fluxes in fluid turbulence 
[see Eq.~(\ref{eq:shell_eff_defn}) and Eq.~(\ref{eq:deudt_fluid3})] 
and in MHD turbulence 
[see Eqs.~(\ref{eq:ushell_ushell})-(\ref{eq:bshell_ushell}) and
Eqs.~(\ref{eq:flux_uinuout})-(\ref{eq:flux_uoutbout})].
Some of these energy transfers can be obtained only using the
``mode-to-mode'' energy  transfer formulae.
In a companion paper \cite{Dar2000_2} we
 present the results of our numerical study of the 
effective shell-to-shell energy transfer
rates and energy fluxes in 2-D MHD turbulence.

\newpage

\appendix
\setcounter{section}{0}
\section{Derivation of the circulating transfer between the velocity modes 
in a triad}
\setcounter{equation}{0}
\seceqaa

In Section~\ref{s:mode_to_mode_fluids}
we showed that the mode-to-mode transfer is
given by ${\cal{\slash{R}}}^{uu}={\cal{\slash{S}}}^{uu}+X_\Delta$. We 
interpreted $X_\Delta$ as a circulating transfer and 
${\cal{\slash{S}}}^{uu}$ [Eq.~(\ref{eq:Sukup_def})]
 as an effective mode-to-mode transfer.
In this appendix we shall determine the circulating transfer $X_\Delta$
for fluid and MHD turbulence, using symmetry considerations.
In the first section we shall discuss $X_{\Delta}$ in fluid turbulence 
and in the next section we will discuss the MHD turbulence.

\subsection{$X_{\Delta}$ in fluid turbulence}
\label{s:Xfluid_app}
The energy transfer between pair of modes should be invariant under a
rotation (because it is a scalar), and  galilean
transformation, and  it should be finite for all values of {\bf k, p, q,
u(k), u(p), u(q)}. We will determine the form of the circulating 
transfer by imposing these and a few other symmetry constraints.

We construct a scalar expression for $X_{\Delta}$ dependent on
{\bf k, p, q, u(k), u(p), u(q)}, and which is cubic in {\bf u} 
and linear in {\bf k, p} or {\bf q}, to satisfy the for dimensional reasons.
We write the general expression for such a scalar as 
\begin{eqnarray}
X_{\Delta} = 
Re ( & {\bf \left[k.u(q)\right]} &
\{ {\bf \left[u(k).u(p)\right]} \alpha +
{\bf \left[u(k).u(q)\right]} \beta +
 {\bf \left[u(p).u(q)\right]} \gamma  +  
\nonumber \\
&  &  {\bf \left[u(k).u(k)\right]} \delta  +
{\bf \left[u(p).u(p)\right]} \nu  +
{\bf \left[u(q).u(q)\right]} \mu \} +  
\nonumber \\
&  {\bf \left[k.u(p)\right]} &
\{ {\bf \left[u(k).u(p)\right]} \eta +
{\bf \left[u(k).u(q)\right]} \zeta +
{\bf \left[u(p).u(q)\right]} \epsilon  +  
\nonumber \\
&   & {\bf \left[u(k).u(k)\right]} \kappa  +
{\bf \left[u(p).u(p)\right]} \lambda  +
{\bf \left[u(q).u(q)\right]} \theta \} +  
\nonumber \\
&   {\bf \left[p.u(k)\right]} &
\{ {\bf \left[u(k).u(p)\right]}\psi + ... +
{\bf \left[u(p).u(q)\right]} \omega + ... + 
{\bf \left[u(q).u(q)\right]} \chi \} +  
\nonumber \\
&  {\bf \left[p.u(q)\right]} &
\{ {\bf \left[u(k).u(p)\right]} \tau + ... +
{\bf \left[u(p).u(q)\right]} \phi + ... + 
{\bf \left[u(q).u(q)\right]} \varphi \} +  
\nonumber \\
&   {\bf \left[q.u(k)\right]} &
\{ {\bf \left[u(k).u(p)\right]} \sigma + ... +
{\bf \left[u(p).u(q)\right]} \xi + ... + 
{\bf \left[u(q).u(q)\right]} \varsigma \} +  
\nonumber \\
&   {\bf \left[q.u(p)\right]}  &
[ {\bf \left[u(k).u(p)\right]} \pi+ 
{\bf \left[u(k).u(q)\right]} \rho + ... + 
{\bf \left[u(q).u(q)\right]} \varrho \}  )
\label{eq:X_general}
\end{eqnarray}
where all the coefficients $\alpha, \beta$, etc. are atmost
non-dimensional functions of the wavenumbers {\bf k, p, q} and of
{\bf u(k), u(p), u(q)}. 
`$Re$' in the expression in Eq.~(\ref{eq:X_general}) indicates
the real part of the expression. We should have infact written 
$X_\Delta$ as a linear combination of the real and the imaginary parts of
the entire expression in the above equation.
However, we will only consider the real part explicitly. The imaginary
part can be treated in the same manner and the following
conclusions, although explicitly stated for the real part will also be
valid for the imaginary part. The most general expression should
also contain terms of the kind $Re[{\bf k.u(q)}] Re[{\bf u(k).u(p)}]$,
$Re[{\bf k.u(q)}] Im[{\bf u(k).u(p)}]$, and
$Im[{\bf k.u(q)}] Im[{\bf u(k).u(p)}]$. However, we will explain below 
that such terms should not be present in $X_\Delta$
on account of galilean invariance.
Hence, from the outset we do not include such terms.
The task now is to determine the coefficients in Eq.~(\ref{eq:X_general}).

The interactions involves all three modes of the triad, i.e.,
if the Fourier coefficients of any of the wavenumbers is zero,
interaction between the modes is turned off. The mode-to-mode transfer
should respect this fundamental feature of the triad interactions. 
Therefore, if {\bf u(k), u(p)} or {\bf u(q)} approach zero,
$X_{\Delta}$ should approach zero as well. We will argue below that 
$\alpha, \beta$, etc., in Eq.~(\ref{eq:X_general}) are independent of 
the magnitude of {\bf u(k), u(p), u(q)}. By inspecting Eq.~(\ref{eq:X_general})
for $X_{\Delta}$, hence we find that a few terms in the 
expression are independent of
the magnitude of atleast one of the Fourier components. For example, the term
${\bf [k.u(q)][u(k).u(q)]}\beta$ is independent of ${\bf u(p)}$.
Such terms should be dropped from the expression of circulating transfer. 
After dropping such terms, we obtain
\begin{eqnarray}
X_\Delta & =  
Re( & {\bf \left[k.u(q)\right]}{\bf \left[u(k).u(p)\right]} \alpha +
{\bf \left[k.u(p)\right]} [{\bf \left[u(k).u(q)\right]} \zeta + \nonumber \\
&   & {\bf \left[p.u(k)\right]} [ {\bf \left[u(p).u(q)\right]} \omega +
{\bf \left[p.u(q)\right]} [ {\bf \left[u(k).u(p)\right]} \tau  + \nonumber \\
&  & {\bf \left[q.u(k)\right]} [ {\bf \left[u(p).u(q)\right]} \xi + 
{\bf \left[q.u(p)\right]} [ {\bf \left[u(k).u(q)\right]} \rho )
\label{eq:X1}
\end{eqnarray}

The requirement of finiteness of the mode-to-mode transfer
imposes restrictions on the form of $X_\Delta$.
In Eq.~(\ref{eq:X_general}) the dimensional terms are
${\bf \left[k.u(q)\right]}[{\bf \left[u(k).u(p)\right]}$, 
 ${\bf \left[k.u(p)\right]}[{\bf \left[u(k).u(q)\right]}$, etc.
and they are finite
for all values of {\bf k, p, q, u(k), u(p), u(q)}. The coefficients
$\alpha, \beta$, etc. are `dimensionless scalars'. Therefore they
can be written as
\begin{equation}
\alpha = f[ \frac{{\bf k.p}}{kp}, \frac{{\bf k.p}}{kq},...,
\frac{{\bf k.u(p)}}{k|{\bf u(p)}|}, \frac{{\bf k.u(p)}}{k|{\bf u(q)}|},...,
\frac{{\bf u(k).u(p)}}{|{\bf u(k)}||{\bf u(p)}|},
\frac{{\bf u(k).u(p)}}{|{\bf u(k)}||{\bf u(q)}|},...]
\label{eq:alpha_general}
\end{equation}
where the arguments of the function have been non-dimensionalised. 
Consider triads in which one of the wavenumbers, say {\bf q}, tends to zero.
Then the terms like ${\bf k.p}/kq$ with {\bf q} in the denominator diverge.
Similarly if one of the Fourier coefficients, say {\bf u(q)},
is equal to zero, then the arguments 
$\frac{{\bf k.u(p)}}{k|{\bf u(q)}|}$ and
$\frac{{\bf u(k).u(p)}}{|{\bf u(k)}||{\bf u(q)}|}$, etc. diverge. 
Hence terms which depend on the magnitude of vectors must not appear
in the expression of the dimensionless scalars $\alpha, \beta$, etc.
Only the arguments which depend on the angles between vectors, e.g.,
$\frac{{\bf k.u(p)}}{k|{\bf u(p)}|}$, 
$\frac{{\bf u(k).u(p)}}{|{\bf u(k)}||{\bf u(p)}|}$, should be
included. Therefore, all the
dimensionless coefficients should be a function of only these arguments,
i.e., 
$\alpha = f[ \frac{{\bf k.p}}{kp},..., \frac{{\bf k.u(p)}}{k|{\bf u(p)}|},...,
\frac{{\bf u(k).u(p)}}{|{\bf u(k)}||{\bf u(p)}|},
...]$. 

We now write the coefficient $\alpha$, $\zeta$, etc. in the form
\begin{equation}
\alpha = \alpha^{(0)}+ \alpha^{(1)} + \alpha^{(2)} + \alpha^{(3)} + 
         \alpha^{(4)} + \alpha^{(5)} 
\label{eq:alpha_lnr_nlnr}
\end{equation}
where $\alpha^{(0)}$ is a constant, and
\begin{equation}
\alpha^{(1)} = \alpha^{(1)}_{1} \frac{{\bf k.p}}{kp} +
               \alpha^{(1)}_{2} \frac{{\bf k.q}}{kq} +
               \alpha^{(1)}_{3} \frac{{\bf p.q}}{pq} ,
\label{eq:alpha1}
\end{equation}
\begin{equation}
\alpha^{(2)} = \alpha^{(2)}_{1} \frac{{\bf k.u(p})}{k|{\bf u(p)}|} +
               \alpha^{(2)}_{2} \frac{{\bf k.u(q})}{k|{\bf u(q)}|} +
               \alpha^{(2)}_{3} \frac{{\bf p.u(k)}}{p|{\bf u(k)}|} +
               \alpha^{(2)}_{4} \frac{{\bf p.u(q)}}{p|{\bf u(q)}|} +
               \alpha^{(2)}_{5} \frac{{\bf q.u(k)}}{q|{\bf u(k)}|} +
               \alpha^{(2)}_{6} \frac{{\bf q.u(p)}}{q|{\bf u(p)}|} ,
\label{eq:alpha2}
\end{equation}
\begin{equation}
\alpha^{(3)} =  \alpha^{(3)}_{1} \frac{{\bf u(k).u(p})}{{\bf |u(k)||u(p)|}} +
   \alpha^{(3)}_{2} \frac{{\bf u(k).u(q})}{{\bf |u(k)||u(q)|}} +
   \alpha^{(3)}_{3} \frac{{\bf u(p).u(q)}}{{\bf |u(p)||u(q)|}} . 
\label{eq:alpha3}
\end{equation}
The quantities $\alpha^{(1)}, \alpha^{(2)}, \alpha^{(3)}$ are `linear'
in the scalars of the type ${\bf k.p}/kp$, ${\bf k.u(p)}/k|{\bf u(p)}|$, and
${\bf u(k).u(p)}/{\bf |u(k)||u(p)|}$ respectively.
The quantity $\alpha^{(4)}$ contains terms like
${\bf (k.p)^2}/k^2p^2$, ${\bf [k.u(p)]}^{2}/k^2({\bf [u(p).u(p)]}$, etc.
which are higher order in these scalars.
All the other coefficients $\zeta, \gamma, \omega,$ etc.  in
Eq.~(\ref{eq:X_general}) can be similarly divided into linear and non-linear 
terms.

We now study the consequences of galilean invariance of $X_\Delta$
on Eq.~(\ref{eq:X_general}).
Consider two frames of reference $S$ and $S'$ with $S$ moving with respect
to $S'$ with velocity ${\bf U_0}$. 
We denote the Fourier coefficient of velocity in frame $S$ and $S'$ by
${\bf u(k)}$ and ${\bf u'(k)}$ respectively.
The relationship between ${\bf u(k)}$ and ${\bf u'(k)}$ is given by
\begin{equation}
{\bf u'(k) = u(k)} \exp^{i ({\bf k.U_0})t} + {\bf U_0} \delta ({\bf k}) .
\label{eq:uk_galilean}
\end{equation}
The second term is non-zero only for the ${\bf k}=0$ modes. 
In the subsequent discussion relating to galilean invariance we
can drop this term for sake of simplicity without affecting the
derivation.
The transformation rule for scalars of the type 
{\bf k.p, k.u(p), u(k).u(p)} between the two frames of references 
$S$ and $S'$ can be deduced from the transformation
property of ${\bf u(k)}$. 
\begin{equation}
{\bf k'.u'(q')} =  {\bf k.u(q)} \exp^{i ({\bf q.U_0})t}
\label{eq:k_uq_galilean}
\end{equation}
\begin{equation}
{\bf u'(k').u'(p')} =  {\bf u(k).u(p)} \exp^{i {\bf (k+p).U_0}t}
\label{eq:uk_up_galilean}
\end{equation}
It follows from the above two transformation rules that
${\bf k.u(q)}$ and ${\bf u(k).u(p)}$ are not invariant under a change
of reference frame. 
However, a product of these quantities can be seen to
be invariant 
\begin{eqnarray}
{\bf [k'.u'(q')][u'(k').u'(p')]} & = & {\bf [k.u(q)][u(k).u(p)]} 
                                 \exp^{i {\bf (k+p+q).U_0}t} \nonumber \\
                             & = & {\bf [k.u(q)][u(k).u(p)]} 
\label{eq:kuq_ukup_galilean}
\end{eqnarray}
using {\bf k+p+q=0}. However,
\begin{eqnarray}
Re{\bf [k'.u'(q')]} Re{\bf [u'(k').u'(p')]} \neq
Re{\bf [k.u(q)]} Re{\bf [u(k).u(p)]},
\label{eq:rekuq_reukup_galilean}
\end{eqnarray}
\begin{eqnarray}
Re{\bf [k'.u'(q')]} Im{\bf [u'(k').u'(p')]} \neq
Re{\bf [k.u(q)]} Im{\bf [u(k).u(p)]},
\label{eq:rekuq_imukup_galilean}
\end{eqnarray}
and
\begin{eqnarray}
Im{\bf [k'.u'(q')]} Im{\bf [u'(k').u'(p')]} \neq
Im{\bf [k.u(q)]} Im{\bf [u(k).u(p)]} 
\label{eq:imkuq_imukup_galilean}
\end{eqnarray}
Thus, we have shown that the terms of the kind given in the above three
equations are not galilean invariant. 

Clearly, the invariance under Galilean transformation demands that all
of the vectors {\bf u(k), u(p), u(q)} appear either once, twice, or
$n$ times in the scalar.  A scalar of the type
${\bf [k.u(q)][u(k).u(q)]}$ having two or more repeated wavenumber indices
is not invariant since the term appearing in the exponential is not zero
for such a scalar. The transformation properties of any quantity formed
from the scalars like {\bf k.p, k.u(p), u(k).u(p)} can be obtained in a 
similar manner.
If the quantity is not galilean invariant then it should
not be included  in the expression for mode-to-mode transfer. 

We have shown above that scalars of the type 
${\bf k.u(q)}$ and ${\bf u(k).u(p)}$ are not  Galilean invariant
[Eqs.~(\ref{eq:k_uq_galilean}) and (\ref{eq:uk_up_galilean})].
It follows that $\alpha^{(2)}$,$\beta^{(2)}$, etc., 
$\alpha^{(3)}$, $\beta^{(3)}$, etc. are not Galilean invariant.
 To preserve the Galilean invariance of $X_\Delta$, the
coefficients $\alpha^{(2)}$ and  $\alpha^{(3)}$ should be dropped from 
Eq.~(\ref{eq:alpha_lnr_nlnr}).
However, the nonlinear quantity $\alpha^{(4)}$ can be constructed in a 
Galilean invariant form. 
Currently, we have dropped these higher order terms.
Our derivation of
$X_\Delta$ therefore has the limitation of being strictly valid only upto the
linear order. A complete proof containing all order of terms in
$X_\Delta$ is beyond the scope of this thesis and is left for future
work.
After dropping the terms $\alpha^{(2)}, \alpha^{(3)}, \alpha^{(4)}$, the
Eq.~(\ref{eq:alpha_lnr_nlnr}) reduces to
\begin{equation}
\alpha = \alpha^{(0)}+\alpha^{(1)}
\label{eq:alpha_eq_alpha1}
\end{equation}

Using equations (\ref{eq:alpha_lnr_nlnr}) and (\ref{eq:alpha_eq_alpha1}) 
we now write X in Eq.~(\ref{eq:X1}) as,
\begin{eqnarray}
X_\Delta & =  
Re( & {\bf \left[k.u(q)\right]}{\bf \left[u(k).u(p)\right]} \{
\alpha_{0} +
\alpha_{1} \frac{{\bf k.p}}{kp} +
\alpha_{2} \frac{{\bf k.q}}{kq} +
\alpha_{3} \frac{{\bf p.q}}{pq} \} + \nonumber \\
&  & {\bf \left[k.u(p)\right]}{\bf \left[u(k).u(q)\right]} \{
\zeta_{0} +
\zeta_{1} \frac{{\bf k.p}}{kp} +
\zeta_{2} \frac{{\bf k.q}}{kq} +
\zeta_{3} \frac{{\bf p.q}}{pq} \} + \nonumber \\
&  & {\bf \left[p.u(k)\right]}{\bf \left[u(p).u(q)\right]} \{
\omega_{0} +
\omega_{1} \frac{{\bf k.p}}{kp} +
\omega_{2} \frac{{\bf k.q}}{kq} +
\omega_{3} \frac{{\bf p.q}}{pq} \} + \nonumber \\
&  & {\bf \left[p.u(q)\right]}{\bf \left[u(k).u(p)\right]} \{
\tau_{0} +
\tau_{1} \frac{{\bf k.p}}{kp} +
\tau_{2} \frac{{\bf k.q}}{kq} +
\tau_{3} \frac{{\bf p.q}}{pq} \} + \nonumber \\
&  & {\bf \left[q.u(k)\right]}{\bf \left[u(p).u(q)\right]} \{
\xi_{0} +
\xi_{1} \frac{{\bf k.p}}{kp} +
\xi_{2} \frac{{\bf k.q}}{kq} +
\xi_{3} \frac{{\bf p.q}}{pq} \} + \nonumber \\
&  & {\bf \left[q.u(p)\right]}{\bf \left[u(k).u(q)\right]} \{
\rho_{0} +
\rho_{1} \frac{{\bf k.p}}{kp} +
\rho_{2} \frac{{\bf k.q}}{kq} +
\rho_{3} \frac{{\bf p.q}}{pq} \} )
\label{eq:X2}
\end{eqnarray}
We have dropped the superscript $'1'$ from $\alpha^{(1)}_{1},
\alpha^{(1)}_{2}, \alpha^{(1)}_{3}$. Using the incompressibility relationship
${\bf q.u(q)=0}$ we obtain ${\bf p.u(q) = -(k+q).u(q) = -k.u(q)}$. 
Therefore, fourth term in the above equation can be absorbed into 
the first term. 
Similarly, using the relationships ${\bf q.u(p) = -k.u(p)}$ and 
${\bf q.u(k) = -p.u(k)}$, the last term can be combined with the second term
and the fifth term combined with the third term. Now, we can write
\begin{eqnarray}
X_\Delta & =  
Re( & {\bf \left[k.u(q)\right]}{\bf \left[u(k).u(p)\right]} \{
\alpha_{0} +
\alpha_{1} \frac{{\bf k.p}}{kp} +
\alpha_{2} \frac{{\bf k.q}}{kq} +
\alpha_{3} \frac{{\bf p.q}}{pq} \} + \nonumber \\
&  & {\bf \left[k.u(p)\right]}{\bf \left[u(k).u(q)\right]} \{
\zeta_{0} +
\zeta_{1} \frac{{\bf k.p}}{kp} +
\zeta_{2} \frac{{\bf k.q}}{kq} +
\zeta_{3} \frac{{\bf p.q}}{pq} \} + \nonumber \\
&  & {\bf \left[p.u(k)\right]}{\bf \left[u(p).u(q)\right]} \{
\omega_{0} +
\omega_{1} \frac{{\bf k.p}}{kp} +
\omega_{2} \frac{{\bf k.q}}{kq} +
\omega_{3} \frac{{\bf p.q}}{pq} \} )
\label{eq:Xkpq}
\end{eqnarray}
by  making the replacements $(\alpha - \tau) \rightarrow \alpha$, 
$(\zeta - \rho) \rightarrow \zeta$ and $(\omega - \xi) \rightarrow \omega$.

The quantity $X_{\Delta}$ represents the energy 
circulating in the anti-clockwise
direction, i.e., $X_{\Delta}$ flows along ${\bf p} 
\rightarrow  {\bf k} \rightarrow {\bf q} \rightarrow {\bf p}$ 
[see Fig. 3 in  Section~\ref{subs:mode_to_mode_soln}]. 
We will now denote the circulating transfer along
${\bf p} \rightarrow {\bf k}$ by $X({\bf k|p|q})$, transfer
along ${\bf k} \rightarrow {\bf q}$ by $X({\bf q|k|p})$ and so on for the
circulating transfer between each pair of modes.
The above $X_{\Delta}$ is the circulating transfer
from {\bf p} to {\bf k}, i.e., $X({\bf k|p|q|}) \equiv X_\Delta$.
Then by symmetry the circulating transfer from 
{\bf q} to {\bf k} can be written as
\begin{eqnarray}
X({\bf k|q|p}) & =  
Re( & {\bf \left[k.u(p)\right]}{\bf \left[u(k).u(q)\right]} \{
\alpha_{0} +
\alpha_{1} \frac{{\bf k.q}}{kq} +
\alpha_{2} \frac{{\bf k.p}}{kp} +
\alpha_{3} \frac{{\bf p.q}}{pq} \} + \nonumber \\
&  & {\bf \left[k.u(q)\right]}{\bf \left[u(k).u(p)\right]} \{
\zeta_{0} +
\zeta_{1} \frac{{\bf k.q}}{kq} +
\zeta_{2} \frac{{\bf k.p}}{kp} +
\zeta_{3} \frac{{\bf p.q}}{pq} \} + \nonumber \\
&  & {\bf \left[q.u(k)\right]}{\bf \left[u(p).u(q)\right]} \{
\omega_{0} +
\omega_{1} \frac{{\bf k.q}}{kq} +
\omega_{2} \frac{{\bf k.p}}{kp} +
\omega_{3} \frac{{\bf p.q}}{pq} \} ),
\label{eq:Xkqp}
\end{eqnarray}
and the circulating transfer from {\bf p} to {\bf k} is
\begin{eqnarray}
X({\bf p|k|q}) & =  
Re( & {\bf \left[p.u(q)\right]}{\bf \left[u(k).u(p)\right]} \{
\alpha_{0} +
\alpha_{1} \frac{{\bf k.p}}{kp} +
\alpha_{2} \frac{{\bf p.q}}{pq} +
\alpha_{3} \frac{{\bf k.q}}{kq} \} + \nonumber \\
&  & {\bf \left[p.u(k)\right]}{\bf \left[u(p).u(q)\right]} \{
\zeta_{0} +
\zeta_{1} \frac{{\bf k.p}}{kp} +
\zeta_{2} \frac{{\bf p.q}}{pq} +
\zeta_{3} \frac{{\bf k.q}}{kq} \} + \nonumber \\
&  & {\bf \left[k.u(p)\right]}{\bf \left[u(k).u(q)\right]} \{
\omega_{0} +
\omega_{1} \frac{{\bf k.p}}{kp} +
\omega_{2} \frac{{\bf p.q}}{pq} +
\omega_{3} \frac{{\bf k.q}}{kq} \} )
\label{eq:Xpkq}
\end{eqnarray}
From Fig. 3, $X({\bf k|p|q})$ flows along the
 anti-clockwise direction while
$X({\bf k|q|p})$ and $X({\bf p|k|q})$ 
flows along the clockwise direction. The former should therefore
be equal and opposite to the latter two transfers, i.e., 
\begin{equation}
X({\bf k|p|q}) + X({\bf k|q|p}) = 0 ,
\end{equation}
\label{eq:Xkpq_Xkqp1}
and
\begin{equation}
X({\bf k|p|q}) + X({\bf p|k|q}) = 0 .
\label{eq:Xkpq_Xpkq1}
\end{equation}
Substitution of expressions 
$X({\bf k|p|q})$, $X({\bf k|q|p})$, $X({\bf p|k|q})$ 
[Eqs.~(\ref{eq:Xkpq})-(\ref{eq:Xpkq})] into the first of these 
two relationships yields
\begin{eqnarray}
{\bf \left[k.u(q)\right]}{\bf \left[u(k).u(p)\right]} \{
(\alpha_{0}+\zeta_{0}) +
(\alpha_{1}+\zeta_{2}) \frac{{\bf k.p}}{kp} +
(\alpha_{2}+\zeta_{1}) \frac{{\bf k.q}}{kq} +
(\alpha_{3}+\zeta_{3}) \frac{{\bf p.q}}{pq} \}  \nonumber \\
{\bf \left[k.u(p)\right]}{\bf \left[u(k).u(q)\right]} \{
(\zeta_{0}+\alpha_{0}) +
(\zeta_{1}+\alpha_{2}) \frac{{\bf k.p}}{kp} +
(\zeta_{2}+\alpha_{1}) \frac{{\bf k.q}}{kq} +
(\zeta_{3}+\alpha_{3}) \frac{{\bf p.q}}{pq} \}  \nonumber \\
{\bf \left[p.u(k)\right]}{\bf \left[u(p).u(q)\right]} \{
(\omega_{1}-\omega_{2}) \frac{{\bf k.p}}{kp} +
(\omega_{2}-\omega_{1}) \frac{{\bf k.q}}{kq} \} & = 0,
\nonumber \\
&
\label{eq:Xkpq_Xkqp2}
\end{eqnarray}
and 
\begin{eqnarray}
{\bf \left[k.u(q)\right]}{\bf \left[u(k).u(p)\right]} \{
(\alpha_{2}-\alpha_{3}) \frac{{\bf k.q}}{kq} +
(\alpha_{3}-\alpha_{2}) \frac{{\bf p.q}}{pq} \} &  & \nonumber \\
 + {\bf \left[k.u(p)\right]}{\bf \left[u(k).u(q)\right]} \{
(\zeta_{0}+\omega_{0}) +
(\zeta_{1}+\omega_{1}) \frac{{\bf k.p}}{kp} +
(\zeta_{2}+\omega_{3}) \frac{{\bf k.q}}{kq} + &
(\zeta_{3}+\omega_{2}) \frac{{\bf p.q}}{pq} \} & \nonumber \\
 +  {\bf \left[p.u(k)\right]}{\bf \left[u(p).u(q)\right]} \{
(\omega_{0}+\zeta_{0}) +
(\omega_{1}+\zeta_{1}) \frac{{\bf k.p}}{kp} +
(\omega_{2}+\zeta_{3}) \frac{{\bf k.q}}{kq} + &
(\omega_{3}+\zeta_{2}) \frac{{\bf p.q}}{pq} \} & = 0 .
\nonumber \\
& &
\label{eq:Xkpq_Xpkq2}
\end{eqnarray}
Each of the three terms in the two equations above has a different
functional dependence on {\bf u(k), u(p), u(q)}. If we only rotate
{\bf u(k)} about {\bf k} without changing its magnitude then each of
the terms will change by a different factor depending on the direction
of {\bf u(p)}, {\bf u(q)} which can be independently chosen.  
Since the sum of three terms should be zero and each term can change
a different factor, each of the terms should be individually zero. 
In addition, this should hold for all wavenumber triads. 
Therefore each term inside the
curly bracket should also be individually zero. 
This condition gives the 
the following relationships between the constants: 
$\alpha_{0}+\zeta_{0}=0, \alpha_1+\zeta_2=0, \alpha_2+\zeta_1=0,
\alpha_3+\zeta_3=0, \omega_1-\omega_2=0$ from Eq.~(\ref{eq:Xkpq_Xkqp2}), and 
$\omega_{0}+\zeta_{0}=0, \alpha_2-\alpha_3=0, 
\zeta_1+\omega_1=0, \zeta_2+\omega_3=0, 
\zeta_3+\omega_2=0$ from Eq.~(\ref{eq:Xkpq_Xpkq2}).

We now consider the following geometry -
the wavenumber triad form an isosceles triangle with ${\bf |p|=|q|}$,
and the Fourier modes ${\bf u(p)}$ and ${\bf u(q)}$ are perpendicular
to the plane of {\bf k, p, q} and are equal in magnitude. For this
geometry the clockwise and the anti-clockwise direction are equivalent.
There is no preferred direction along which the circulating transfer may
flow. Therefore, for the geometry shown the circulating transfer should
be zero. We may now write the circulating transfer [Eq.~(\ref{eq:Xkpq})]
for this geometry as
\begin{eqnarray}
X & =  
Re( &  {\bf \left[p.u(k)\right]}{\bf \left[u(p).u(q)\right]} \{
(\omega_{1} + \omega_{2}) \frac{{\bf k.p}}{kp} +
\omega_{3} \frac{{\bf p.q}}{pq} + a \} ) \nonumber \\
 & = & 0
\label{eq:X_isosceles}
\end{eqnarray}
With {\bf u(p)} and {\bf u(q)} fixed perpendicular to the plane of the
triad, X should be zero for all wavenumber triads which form an
isosceles triangle.
For $X$ to be zero for every such geometry,
each term inside the curly bracket should be equated to zero 
giving the additional relationships:
$\omega_{1}=-\omega_{2}, \omega_{3}=0,
\omega_{0}=0$. 

We have obtained 17 relationships between the constants
$\alpha, \zeta$, etc. These equations may be solved to get the values
of all these constants. It can be verified that the only solution of these
equations is $\alpha$'s$=0$, $\zeta$'s$=0$, $\omega$'s$=0$, $a=b=c=0$.

Therefore we get $X=0$. It follows that the mode-to-mode transfer
${\cal{\slash{R}}}^{uu}={\cal{\slash{S}}}^{uu}$.

We have thus determined the mode-to-mode transfer. The proof
used rotational invariance, galilean invariance, finiteness of
mode-to-mode transfer, and symmetry of X with respect to ${\bf k, p, q,
u(k), u(p), u(q)}$. However, in the series we have taken only linear
terms in ${\bf k.p, k.u(p), u(k).u(p)}$. A complete proof containing all
orders of terms is beyond the scope of this thesis.
It may be possible to obtain a completely general proof by finding some
symmetries which we have ignored in our proof. Otherwise a novel approach
may be required.

\subsection{$X_{\Delta}$ in MHD turbulence}
\label{s:mhd}

The circulating transfer between the velocity modes in MHD can also
be shown to be equal to zero to linear order terms in ${\bf k.p,
k.u(p), u(k).u(p)}$
. The arguments leading to this are nearly 
the same as that for the fluid case. However, we additionally 
need to show that the general expression for the circulating transfer 
$X_{\Delta}$ for MHD should not contain {\bf b(k), b(p), b(q)} and
can also be given by Eq.~(\ref{eq:X_general}) 

In the previous section we showed that each term in $X_\Delta$ 
should depend on {\bf u(k), u(p), u(q)} --- this followed from the 
fundamentally three-mode feature of the
interactions. The expression for the combined energy transfer to 
{\bf u(k)} from {\bf u(p)} and {\bf u(q)} is identical for the fluid and
for the MHD case. Hence, in the case of MHD also, each term in the 
circulating transfer between the 
velocity modes should be dependent on {\bf u(k), u(p)}, and {\bf u(q)}.
Thus, the general expression for the circulating transfer between the 
velocity modes in MHD is the same as that for Fluid turbulence. The
rest of the arguments leading to $X_{\Delta}=0$ in MHD are
identical to those in Section~\ref{s:Xfluid_app}.

\newpage

\section{ Derivation of circulating transfer between 
the magnetic modes in a triad}
\setcounter{equation}{0}
\seceqbb

In Section~\ref{subs:bmode_bmode} we had shown that the 
mode-to-mode transfer from
{\bf b(p)} to {\bf b(k)} can be written as
${\cal{\slash{R}}}^{bb}={\cal{\slash{S}}}^{bb}+Y_\Delta$, where
${\cal{\slash{S}}}^{bb}$ given by Eq.~(\ref{eq:Sbkbp_def}) is the effective
transfer between the modes and $Y_\Delta$ is the circulating transfer.
In this section we will derive the expression for the circulating
transfer $Y_\Delta$. The derivation for $Y_\Delta$ is
similar  to the derivation of  $X_\Delta$ for fluid turbulence given
in Appendix A.

Since the mode-to-mode transfer is a scalar, the circulating transfer 
$Y_\Delta$ will also be a scalar with a dependence  on 
the wavenumbers {\bf k, p, q} and the Fourier coefficients  
{\bf u(k), u(p), u(q), b(k), b(p), b(q)}. It 
should be linear in {\bf k, p} or {\bf q} and cubic in the Fourier
coefficients {\bf u(k), u(p), u(q), b(k), b(p), b(q)} for dimensional reasons. 

 The
most general form of $Y_\Delta$ that satisfies these properties can be written
as
\begin{equation}
Y_\Delta = Y_{\Delta}^{(1)}+ Y_{\Delta}^{(2)}+ Y_{\Delta}^{(3)}+
Y_{\Delta}^{(4)}
\label{eq:Y_general}
\end{equation}
where 
\begin{eqnarray}
Y_{\Delta}^{(1)} = 
Re ( & {\bf \left[k.u(q)\right]} &
\{ {\bf \left[u(k).b(k)\right]}... +
{\bf \left[u(k).b(p)\right]}...  +
 {\bf \left[u(k).b(q)\right]}...   + 
\nonumber \\
&  &  {\bf \left[u(p).b(k)\right]}...   +
{\bf \left[u(p).b(p)\right]}...   +
{\bf \left[u(p).b(q)\right]}...  + 
\nonumber \\
& & \{ {\bf \left[u(q).b(k)\right]}...  +
{\bf \left[u(q).b(p)\right]}...  +
{\bf \left[u(q).b(q)\right]}...  \} + 
\nonumber \\
& {\bf \left[k.u(p)\right]} &
\{ {\bf \left[u(k).b(k)\right]}...  + .... +
 {\bf \left[u(k).b(q)\right]}...  + .... +
 {\bf \left[u(q).b(k)\right]}...  + .... +
{\bf \left[u(q).b(q)\right]}...  \} + 
\nonumber \\
& {\bf \left[p.u(k)\right]} &
\{ {\bf \left[u(k).b(k)\right]}...  + .... +
 {\bf \left[u(p).b(q)\right]}...  + .... +
 {\bf \left[u(q).b(p)\right]}...  + .... +
{\bf \left[u(q).b(q)\right]}...  \}   )
\nonumber \\
\label{eq:Y1_general}
\end{eqnarray}

\begin{eqnarray}
Y_{\Delta}^{(2)} = 
Re ( &{\bf \left[k.u(q)\right]} &
\{ {\bf \left[b(k).b(k)\right]}  +
{\bf \left[b(k).b(p)\right]} \alpha +
 {\bf \left[b(k).b(q)\right]}...   + 
\nonumber \\
&  &  {\bf \left[b(p).b(p)\right]}   +
{\bf \left[b(p).b(q)\right]}...  +
{\bf \left[b(q).b(q)\right]}... \}+ 
\nonumber \\
& {\bf \left[k.u(p)\right]} &
\{ {\bf \left[b(k).b(k)\right]}...  + .... +
{\bf \left[b(k).b(q)\right]} \beta + .... +
{\bf \left[b(q).b(q)\right]}... \} + 
\nonumber \\
& {\bf \left[p.u(k)\right]} &
\{ {\bf \left[b(k).b(k)\right]}...  + ....+
{\bf \left[b(p).b(q)\right]}  \gamma + .... +
{\bf \left[b(q).b(q)\right]}... \}  )
\nonumber \\
\label{eq:Y2_general}
\end{eqnarray}

\begin{eqnarray}
Y_{\Delta}^{(3)} = 
Re ( & {\bf \left[k.b(q)\right]} &
\{ {\bf \left[u(k).b(k)\right]}...  +
{\bf \left[u(k).b(p)\right]} \delta +
 {\bf \left[u(k).b(q)\right]}...   + 
\nonumber \\
&  &  {\bf \left[u(p).b(k)\right]} \eta   +
{\bf \left[u(p).b(p)\right]}...   +
{\bf \left[u(p).b(q)\right]}...  + 
\nonumber \\
& & \{ {\bf \left[u(q).b(k)\right]}...  +
{\bf \left[u(q).b(p)\right]}...  +
{\bf \left[u(q).b(q)\right]}...  \} + 
\nonumber \\
& {\bf \left[k.b(p)\right]} &
\{ {\bf \left[u(k).b(k)\right]}...  + .... +
 {\bf \left[u(k).b(q)\right]} \mu  + .... +
 {\bf \left[u(q).b(k)\right]} \nu   + .... +
 {\bf \left[u(q).b(q)\right]}...  \} + 
\nonumber \\
& {\bf \left[p.b(k)\right]} &
\{ {\bf \left[u(k).b(k)\right]}...  + .... +
{\bf \left[u(p).b(q)\right]}  \psi + .... +
{\bf \left[u(q).b(p)\right]} \chi  + 
{\bf \left[u(q).b(q)\right]}...  \}  )
\nonumber \\
\label{eq:Y3_general}
\end{eqnarray}

\begin{eqnarray}
Y_{\Delta}^{(4)} = 
Re ( & {\bf \left[k.b(q)\right]} &
\{ {\bf \left[u(k).u(k)\right]}...  +
{\bf \left[u(k).u(p)\right]} \theta  +
 {\bf \left[u(k).u(q)\right]}...   + 
\nonumber \\
&  &  {\bf \left[u(p).u(p)\right]}...   +
{\bf \left[u(p).u(q)\right]}...  +
{\bf \left[u(q).u(q)\right]}...  \}+ 
\nonumber \\
& & \{ {\bf \left[u(q).b(k)\right]}...  +
{\bf \left[u(q).b(p)\right]}...  +
{\bf \left[u(q).b(q)\right]}...  \} + 
\nonumber \\
& {\bf \left[k.b(p)\right]} &
\{ {\bf \left[u(k).u(k)\right]}...  +
{\bf \left[u(k).u(q)\right]} \phi  + .... +
{\bf \left[u(q).u(q)\right]}...  \} + 
\nonumber \\
&   {\bf \left[p.b(k)\right]} &
\{ {\bf \left[u(k).u(k)\right]}... + 
{\bf \left[u(p).u(q)\right]}  \xi + .... + 
{\bf \left[u(q).u(q)\right]}... ]  )
\nonumber \\
\label{eq:Y4_general}
\end{eqnarray}
Following Appendix A we consider explicitly only the real part,
which is denoted by $Re$ in the above equations. If we take a
linear combination of the real and the imaginary parts instead,
the derivation and the conclusions regarding $Y_\Delta$ will not change.
In addition, terms of the 
$Re[{\bf k.u(q)}] Re[{\bf u(k).u(p)}]$,
$Re[{\bf k.u(q)}] Re[{\bf u(k).b(p)}]$, etc. are not included
as they violate galilean invariance - we had explained this in Appendix A
in context of $X_\Delta$.
The three dots `...' in the above equations
in front of some of the terms indicate that 
a coefficient multiplies each of these terms but which we have not
shown explicitly as they do not enter most of the future discussion.
The coefficients $\alpha, \beta$, etc. in the above expressions 
for $Y_{\Delta}^{(1)}$,..
$Y_{\Delta}^{(4)}$ are non-dimensional functions of the wavenumbers
{\bf k, p, q} and of {\bf u(k), u(p), u(q), b(k, b(p), b(q)}. To determine
$Y_{\Delta}$ we should now determine these coefficients. 
Note that besides the terms of the kind $[{\bf k.u(q)}]\{...\}, 
[{\bf k.u(p)}]\{...\}, [{\bf p.u(k)}]\{...\}$ 
in Eqs.~(\ref{eq:Y1_general})-(~\ref{eq:Y4_general}),
 there are also terms of the kind
$[{\bf p.u(q)}]\{...\}, [{\bf q.u(p)}]\{...\}, [{\bf q.u(k)}]\{...\}$. 
However, using the
incompressibility constraints ${\bf k.u(k)=0}$ and the constraint
${\bf k+p+q=0}$,  we have absorbed the latter three terms into the former
ones.
Now we will impose a few restrictions on the form of $Y_\Delta$.

The interactions between the magnetic modes fundamentally involve 
three-modes --- the combined energy transfer to a magnetic mode,
say {\bf b(k)}, depends on the Fourier coefficients
at all three wavenumbers ({\bf k, p, q}). For example, the combined
energy transfer to {\bf b(k)} from {\bf b(p)} and {\bf b(q)}, given
by 
$ S^{bb}({\bf k|p,q}) = 
                      -  Re \left( i{\bf \left [k.u(q)\right]} 
                                     {\bf \left [b(k).b(p)\right ]} + 
                                   i{\bf \left [k.u(p)\right]} 
                                    {\bf \left [b(k).b(q)\right]} \right)$,
depends on {\bf u(p)}, {\bf u(q)}, {\bf b(p)}, {\bf b(q)}, and 
{\bf b(k)}. The mode-to-mode transfer should not break the
fundamentally three-mode feature of these interactions, i.e., 
the mode-to-mode transfer should have Fourier coefficients from all
wavenumbers ${\bf k, p, q}$. Hence we will drop all those terms from
$Y_{\Delta}^{(1)}$ to $Y_{\Delta}^{(4)}$ which are independent of
atleast one of the wavenumbers ${\bf k, p, q}$. Then, we get
\begin{eqnarray}
Y_{\Delta}^{(1)} = 
Re ( & {\bf \left[k.u(q)\right]} 
\left\{ {\bf \left[u(k).b(p)\right]}...  +
 {\bf \left[u(p).b(k)\right]}... \right\} & +
 {\bf \left[k.u(p)\right]} 
\left\{ {\bf \left[u(k).b(q)\right]}... +
 {\bf \left[u(q).b(k)\right]}... \right\} +  
\nonumber \\
& {\bf \left[p.u(k)\right]} 
\left\{ {\bf \left[u(p).b(q)\right]}...  +
 {\bf \left[u(q).b(p)\right]}... \right\}  ) 
\label{eq:Y1_general2}
\end{eqnarray}
\begin{eqnarray}
Y_{\Delta}^{(2)} = 
Re ( & {\bf \left[k.u(q)\right]} 
{\bf \left[b(k).b(p)\right]} \alpha +
 {\bf \left[k.u(p)\right]} 
{\bf \left[b(k).b(q)\right]} \beta + 
 {\bf \left[p.u(k)\right]} 
{\bf \left[b(p).b(q)\right]}  \gamma  )
\nonumber \\
\label{eq:Y2_general2}
\end{eqnarray}
\begin{eqnarray}
Y_{\Delta}^{(3)} = 
Re ( & {\bf \left[k.b(q)\right]} &
\left\{ {\bf \left[u(k).b(p)\right]} \delta +
 {\bf \left[u(p).b(k)\right]} \eta \right\}  + 
\nonumber \\
& {\bf \left[k.b(p)\right]} &
\left\{  {\bf \left[u(k).b(q)\right]} \mu  +
 {\bf \left[u(q).b(k)\right]} \nu \right\}  + 
\nonumber \\
& {\bf \left[p.b(k)\right]} &
\left\{ {\bf \left[u(p).b(q)\right]}  \psi + 
{\bf \left[u(q).b(p)\right]} \chi \right\}   )
\label{eq:Y3_general2}
\end{eqnarray}
\begin{eqnarray}
Y_{\Delta}^{(4)} = 
Re ( & {\bf \left[k.b(q)\right]} 
{\bf \left[u(k).u(p)\right]} \theta  +
 {\bf \left[k.b(p)\right]} 
{\bf \left[u(k).u(q)\right]} \phi  + 
 {\bf \left[p.b(k)\right]} 
 {\bf \left[u(k).u(k)\right]}... & )
\nonumber \\
\label{eq:Y4_general2}
\end{eqnarray}

We put the following additional restriction on the form of $Y_\Delta$.
If none of the magnetic modes in the triad are gaining/losing energy
to the other two modes, i.e., if the combined energy transfer to each
one of them is equal to zero, then even the circulating transfer
should be equal to zero. 
We observe the following additional feature of the interactions.
The combined energy transfer to every magnetic mode in the triad 
from the other two magnetic modes is equal to zero, if any two of
${\bf b(k), b(p), b(q)}$ are zero (see Eq.~(\ref{eq:bkpq_def}).
The mode-to-mode transfer should also respect this
feature of the interactions, i.e., the circulating transfer should
be zero if \{{\bf b(k), b(p)=0}\}, or
\{{\bf b(k), b(q)=0}\}, or \{{\bf b(p), b(q)=0}\}.
Now, let us put 
{\bf b(k)=0, b(p)=0} in $Y_{\Delta}^{(1)}$ to $Y_{\Delta}^{(4)}$. It can
easily shown that $Y_{\Delta}$ will be zero, for arbitrary values
{\bf u}'s and {\bf b}'s, only if the all coefficients 
in $Y_{\Delta}^{(1)}$ and $Y_{\Delta}^{(4)}$ are
themselves equal to zero. Hence we drop $Y_{\Delta}^{(1)}$ and 
$Y_{\Delta}^{(4)}$ from $Y_\Delta$ [Eq.~(\ref{eq:Y_general})]. We get,
\begin{equation}
Y_{\Delta} =
Y_{\Delta}^{(2)} +
Y_{\Delta}^{(3)} 
\label{eq:Y_triad}
\end{equation}
and
\begin{eqnarray}
Y_{\Delta}^{(2)} = 
Re ( & {\bf \left[k.u(q)\right]} 
{\bf \left[b(k).b(p)\right]} \alpha +
 & {\bf \left[k.u(p)\right]} 
 {\bf \left[b(k).b(q)\right]} \beta  + 
\nonumber \\
& {\bf \left[p.u(k)\right]} 
{\bf \left[b(p).b(q)\right]} \gamma  & )
\label{eq:Y2_triad}
\end{eqnarray}
\begin{eqnarray}
Y_{\Delta}^{(3)} = 
Re( & {\bf \left[k.b(q)\right]} &
\{ {\bf \left[u(k).b(p)\right]} \delta +
{\bf \left[u(p).b(k)\right]} \eta \} + 
\nonumber \\
& {\bf \left[k.b(p)\right]} &
\{  {\bf \left[u(k).b(q)\right]} \mu  +
 {\bf \left[u(q).b(k)\right]} \nu \} + 
\nonumber \\
& {\bf \left[p.b(k)\right]} &
\{ {\bf \left[u(p).b(q)\right]} \psi +
{\bf \left[u(q).b(p)\right]} \chi \}  )
\label{eq:Y3_triad}
\end{eqnarray}

In Appendix A we had considered the derivation of
circulating transfer $X_\Delta$ between the velocity modes. 
By imposing the condition of finiteness of
the circulating transfer we had concluded that the
coefficients should be independent of the magnitude of the wavenumbers
and the Fourier coefficients {\bf u(k)}, etc.
On imposing the constraint that $Y_\Delta$ should be finite and
following the reasoning in Appendix A, we can conclude that the coefficients
$\alpha, \beta$, etc. should be independent of 
the magnitude of the wavenumbers and of the Fourier coefficients of the
velocity and the magnetic fields. Therefore, we may write the coefficient
$\alpha = f[\frac{{\bf k.p}}{kp}, ..., \frac{{\bf k.u(p)}}{k|u(p)|},
\frac{{\bf k.b(p)}}{k|b(p)|},..., \frac{{\bf u(k).u(p)}}{|u(k)||u(p)|},
\frac{{\bf b(k).b(p)}}{|b(k)||b(p)|},...,\frac{{\bf u(k).b(p)}}{|u(k)||b(p)|},
...]$. The other coefficients can also be written in the same manner.

We will now write the coefficients $\alpha, \beta$, etc. as
\begin{equation}
\alpha = \alpha^{(0)}+ \alpha^{(1)}+ \alpha^{(2)}+ \alpha^{(3)}+
 \alpha^{(4)},
\label{eq:alphabb_lnr_nlnr}
\end{equation}
where $\alpha^{(0)}$ is a constant, and
\begin{equation}
\alpha^{(1)} = \alpha^{(1)}_{1} \frac{{\bf k.p}}{kp} +
               \alpha^{(1)}_{2} \frac{{\bf k.q}}{kq} +
               \alpha^{(1)}_{3} \frac{{\bf p.q}}{pq} ,
\label{eq:alpha1_bb}
\end{equation}
\begin{eqnarray}
& \alpha^{(2)} = & \alpha^{(2)}_{1} \frac{{\bf k.u(p})}{k|{\bf u(p)}|} +
               \alpha^{(2)}_{2} \frac{{\bf k.u(q})}{k|{\bf u(q)}|} +
               \alpha^{(2)}_{3} \frac{{\bf p.u(k)}}{p|{\bf u(k)}|} +
               \alpha^{(2)}_{4} \frac{{\bf p.u(q)}}{p|{\bf u(q)}|} +
               \alpha^{(2)}_{5} \frac{{\bf q.u(k)}}{q|{\bf u(k)}|} +
               \alpha^{(2)}_{6} \frac{{\bf q.u(p)}}{q|{\bf u(p)}|} +
\nonumber \\
      & &         \alpha^{(2)}_{7} \frac{{\bf k.b(p})}{k|{\bf b(p)}|} +
               ... + 
               \alpha^{(2)}_{12} \frac{{\bf q.b(p)}}{q|{\bf b(p)}|} ,
\label{eq:alpha2_bb}
\end{eqnarray}
\begin{eqnarray}
\alpha^{(3)} & = & \alpha^{(3)}_{1} \frac{{\bf u(k).u(p})}{{\bf |u(k)||u(p)|}} +
   \alpha^{(3)}_{2} \frac{{\bf u(k).u(q})}{{\bf |u(k)||u(q)|}} +
   \alpha^{(3)}_{3} \frac{{\bf u(p).u(q)}}{{\bf |u(p)||u(q)|}}  +
\nonumber \\
&  & \alpha^{(3)}_{4} \frac{{\bf b(k).b(p})}{{\bf |b(k)||b(p)|}} +
   \alpha^{(3)}_{5} \frac{{\bf b(k).b(q})}{{\bf |b(k)||b(q)|}} +
   \alpha^{(3)}_{6} \frac{{\bf b(p).b(q)}}{{\bf |b(p)||b(q)|}} +
\nonumber \\
&  & \alpha^{(3)}_{7} \frac{{\bf u(k).b(k})}{{\bf |u(k)||b(k)|}} +
   \alpha^{(3)}_{8} \frac{{\bf u(k).b(p})}{{\bf |u(k)||b(p)|}} + ... +
   \alpha^{(3)}_{15} \frac{{\bf u(q).b(p)}}{{\bf |u(q)||b(p)|}} .
\label{eq:alpha3_bb}
\end{eqnarray}
where $\alpha^{(1)}_{1}, \alpha^{(1)}_{2}, \alpha^{(2)}_{1}$, etc.
are all unknown constants.
The quantities $\alpha^{(1)}, \alpha^{(2)}, \alpha^{(3)}$ are `linear'
in the scalars of the type ${\bf k.p}/kp$, ${\bf k.u(p)}/k|{\bf u(p)}|$,
${\bf k.b(p)}/k|{\bf b(p)}|$, ${\bf u(k).u(p)}/{\bf |u(k)||u(p)|}$,
${\bf b(k).b(p)}/{\bf |b(k)||b(p)|}$, and
${\bf u(k).b(p)}/{\bf |u(k)||b(p)|}$.
The quantity $\alpha^{(4)}$ contains higher order terms in these scalars
like ${\bf (k.p)^2}/k^2p^2$, ${\bf [k.u(p)]}^{2}/k^2({\bf [u(p).u(p)]}$, etc.
All the other coefficients $\zeta, \gamma, \omega,$ etc.  in
Eqs.~(\ref{eq:Y2_triad})-(\ref{eq:Y3_triad}) can be similarly divided into 
linear and non-linear terms.

The mode-to-mode transfer should be invariant under a 
galilean transformation. In Appendix A, we had given the transformation
rules for the scalars of the type {\bf k.p}, {\bf k.u(p)}, {\bf u(k).u(p)}
and for a combination of these like {\bf [k.u(p)][u(k).u(q)]}. For the
present case we also need to consider the transformation properties of
the magnetic field. MHD equations hold for low velocities and only 
quantities upto the order of $v/c$ are included in the equations. Under this
approximation the magnetic fields are equal in the two frames 
$S$ and $S'$ with a relative
velocity ${\bf U}_0$. Then, the Fourier coefficient
${\bf b(k)}$ in the frame  $S$ and ${\bf b'(k)}$ in  the frame $S'$ are
related as
\begin{equation}
{\bf b'(k) = b(k)} \exp^{i ({\bf k.U_0})t} 
\label{eq:bk_galilean}
\end{equation}
This differs from the  galilean transformation of {\bf u(k)} by the term
${\bf U_{0}} \delta({\bf k})$ only. Using the above relationship between
between ${\bf b(k)}$ and ${\bf b'(k)}$ the transformation rules for
{\bf k.b(q)}, {\bf b(k).b(p)}, {\bf u(k).b(p)} are 
\begin{equation}
{\bf k'.b'(q')} =  {\bf k.b(q)} \exp^{i ({\bf q.U_0})t}, 
\label{eq:k_bq_galilean}
\end{equation}
\begin{equation}
{\bf b'(k').b'(p')} =  {\bf b(k).b(p)} \exp^{i {\bf (k+p).U_0}t},
\label{eq:bk_bp_galilean}
\end{equation}
\begin{equation}
{\bf u'(k').b'(p')} =  {\bf u(k).b(p)} \exp^{i {\bf (k+p).U_0}t}.
\label{eq:uk_bp_galilean}
\end{equation}
These scalars are not galilean invariant (in Appendix A, we had likewise
shown that the scalars of the type {\bf k.u(p)}, {\bf u(k).u(p)}
are not galilean invariant). It follows that $\alpha^{(2)}, \beta^{(2)}$,
etc., $\alpha^{(3)}, \beta^{(3)}$, etc. are not invariant under a
galilean transformation. To preserve the galilean invariance of
$Y_\Delta$, the coefficients $\alpha^{(2)}$, $\alpha^{(3)}$,
$\beta^{(2)}$, $\beta^{(3)}$, and other such coefficients should be
dropped from Eq.~(\ref{eq:alphabb_lnr_nlnr}). The nonlinear quantity
$\alpha^{(4)}$, $\beta^{(4)}$, etc. can be constructed in a galilean
invariant form. Even then we will drop this term. We had also dropped this
term from the expression for $X_\Delta$ in Appendix A. Our derivation
for $Y_\Delta$, like the derivation of $X_\Delta$ in Appendix A, is 
is thus valid only upto the linear orders. After dropping the 
coefficients $\alpha^{(2)}$, $\alpha^{(3)}$, $\alpha^{(4)}$ and other
coefficients of this type, Eq.~(\ref{eq:alphabb_lnr_nlnr}) reduces to
\begin{equation}
\alpha = \alpha^{(0)}+\alpha^{(1)}
\label{eq:alphabb_lnr}
\end{equation}
Using Eqs.~(\ref{eq:alphabb_lnr}) and (\ref{eq:alpha1_bb}) we can now write 
$Y_{\Delta}^{(2)}$ and $Y_{\Delta}^{(3)}$ in 
Eqs.~(\ref{eq:Y2_triad})-(\ref{eq:Y3_triad}) as
\begin{eqnarray}
Y_{\Delta}^{(2)} = 
Re ( & {\bf \left[k.u(q)\right]} 
{\bf \left[b(k).b(p)\right]} 
\{\alpha^{(0)}+\alpha_{1}\frac{{\bf k.p}}{kp}+\alpha_{2}\frac{{\bf k.q}}{kq}+
\alpha_{3}\frac{{\bf p.q}}{pq} \} + &
\nonumber \\
&  {\bf \left[k.u(p)\right]} 
 {\bf \left[b(k).b(q)\right]} 
\{\beta^{(0)}+\beta_{1}\frac{{\bf k.p}}{kp}+\beta_{2}\frac{{\bf k.q}}{kq}+
\beta_{3}\frac{{\bf p.q}}{pq} \} + &
\nonumber \\
& {\bf \left[p.u(k)\right]} 
{\bf \left[b(p).b(q)\right]} 
\{\gamma^{(0)}+\gamma_{1}\frac{{\bf k.p}}{kp}+\gamma_{2}\frac{{\bf k.q}}{kq}+
\gamma_{3}\frac{{\bf p.q}}{pq} \}  & )
\label{eq:Y2_lnr}
\end{eqnarray}
\begin{eqnarray}
Y_{\Delta}^{(3)} = 
Re ( & {\bf \left[k.b(q)\right]} 
{\bf \left[u(k).b(p)\right]} 
\{\delta^{(0)}+\delta_{1}\frac{{\bf k.p}}{kp}+\delta_{2}\frac{{\bf k.q}}{kq}+
\delta_{3}\frac{{\bf p.q}}{pq} \} +  &
\nonumber \\
& {\bf \left[k.b(q)\right]} 
{\bf \left[u(p).b(k)\right]}   
\{\eta^{(0)}+\eta_{1}\frac{{\bf k.p}}{kp}+\eta_{2}\frac{{\bf k.q}}{kq}+
\eta_{3}\frac{{\bf p.q}}{pq} \} + &
\nonumber \\
& {\bf \left[k.b(p)\right]} 
  {\bf \left[u(k).b(q)\right]}  
\{\mu^{(0)}+\mu_{1}\frac{{\bf k.p}}{kp}+\mu_{2}\frac{{\bf k.q}}{kq}+
\mu_{3}\frac{{\bf p.q}}{pq} \} + &
\nonumber \\
& {\bf \left[k.b(p)\right]} 
 {\bf \left[u(q).b(k)\right]}  
\{\nu^{(0)}+\nu_{1}\frac{{\bf k.p}}{kp}+\nu_{2}\frac{{\bf k.q}}{kq}+
\nu_{3}\frac{{\bf p.q}}{pq} \} + &
\nonumber \\
& {\bf \left[p.b(k)\right]} 
 {\bf \left[u(p).b(q)\right]}  
\{\phi^{(0)}+\phi_{1}\frac{{\bf k.p}}{kp}+\phi_{2}\frac{{\bf k.q}}{kq}+
\phi_{3}\frac{{\bf p.q}}{pq} \} + &
\nonumber \\
& {\bf \left[p.b(k)\right]} 
{\bf \left[u(q).b(p)\right]} 
\{\chi^{(0)}+\chi_{1}\frac{{\bf k.p}}{kp}+\chi_{2}\frac{{\bf k.q}}{kq}+
\chi_{3}\frac{{\bf p.q}}{pq} \}  & )
\label{eq:Y3_lnr}
\end{eqnarray}
where the superscript `1' has been dropped from
$\alpha^{(1)}$, $\alpha^{(2)}$, etc.


The quantity 
$Y_{\Delta}= Y_{\Delta}^{(2)}+Y_{\Delta}^{(3)}$
 represents the energy circulating in the anti-clockwise
direction, i.e. $Y_\Delta$ flows along ${\bf p} 
\rightarrow  {\bf k} \rightarrow {\bf q} \rightarrow {\bf p}$ 
[see Fig. 8 in Section~\ref{subs:bmode_bmode}].
 We will now denote the circulating transfer along
${\bf p} \rightarrow {\bf k}$ by $Y_{\Delta}({\bf k|p|q})$, transfer
along ${\bf k} \rightarrow {\bf q}$ by $Y_{\Delta}({\bf q|k|p})$ 
and so on for the
circulating transfer between each pair of modes.
We take $Y_{\Delta}= Y_{\Delta}^{(2)}+Y_{\Delta}^{(3)}$ from
Eqs.~(\ref{eq:Y2_lnr})-(\ref{eq:Y3_lnr}) 
to be the circulating transfer
from {\bf p} to {\bf k}, i.e., $Y_{\Delta}({\bf k|p|q|}) \equiv Y_\Delta$.
In the same manner $Y_{\Delta}({\bf k|q|p}))= Y_{\Delta}^{(2)}({\bf k|q|p})+
Y_{\Delta}^{(3)}({\bf k|q|p})$ can be written as
\begin{eqnarray}
Y_{\Delta}^{(2)}({\bf k|q|p}) = 
Re( & {\bf \left[k.u(p)\right]} 
{\bf \left[b(k).b(q)\right]} 
\{\alpha^{(0)}+\alpha_{1}\frac{{\bf k.q}}{kq}+\alpha_{2}\frac{{\bf k.p}}{kp}+
\alpha_{3}\frac{{\bf p.q}}{pq} \} + &
\nonumber \\
&  {\bf \left[k.u(q)\right]} 
 {\bf \left[b(k).b(p)\right]} 
\{\beta^{(0)}+\beta_{1}\frac{{\bf k.q}}{kq}+\beta_{2}\frac{{\bf k.p}}{kp}+
\beta_{3}\frac{{\bf p.q}}{pq} \} + &
\nonumber \\
& {\bf \left[q.u(k)\right]} 
{\bf \left[b(p).b(q)\right]} 
\{\gamma^{(0)}+\gamma_{1}\frac{{\bf k.q}}{kq}+\gamma_{2}\frac{{\bf k.p}}{kp}+
\gamma_{3}\frac{{\bf p.q}}{pq} \} & )
\label{eq:Y2_kqp}
\end{eqnarray}
and 
\begin{eqnarray}
Y_{\Delta}^{(3)}({\bf k|q|p}) = 
Re( & {\bf \left[k.b(p)\right]} 
{\bf \left[u(k).b(q)\right]} 
\{\delta^{(0)}+\delta_{1}\frac{{\bf k.q}}{kq}+\delta_{2}\frac{{\bf k.p}}{kp}+
\delta_{3}\frac{{\bf p.q}}{pq} \} + &
\nonumber \\
& {\bf \left[k.b(p)\right]} 
{\bf \left[u(q).b(k)\right]}   
\{\eta^{(0)}+\eta_{1}\frac{{\bf k.q}}{kq}+\eta_{2}\frac{{\bf k.p}}{kp}+
\eta_{3}\frac{{\bf p.q}}{pq} \} + &
\nonumber \\
& {\bf \left[k.b(q)\right]} 
  {\bf \left[u(k).b(p)\right]}  
\{\mu^{(0)}+\mu_{1}\frac{{\bf k.q}}{kq}+\mu_{2}\frac{{\bf k.p}}{kp}+
\mu_{3}\frac{{\bf p.q}}{pq} \} + &
\nonumber \\
& {\bf \left[k.b(q)\right]} 
 {\bf \left[u(p).b(k)\right]}  
\{\nu^{(0)}+\nu_{1}\frac{{\bf k.q}}{kq}+\nu_{2}\frac{{\bf k.p}}{kp}+
\nu_{3}\frac{{\bf p.q}}{pq} \} + &
\nonumber \\
& {\bf \left[q.b(k)\right]} 
 {\bf \left[u(q).b(p)\right]}  
\{\phi^{(0)}+\phi_{1}\frac{{\bf k.q}}{kq}+\phi_{2}\frac{{\bf k.p}}{kp}+
\phi_{3}\frac{{\bf p.q}}{pq} \} + &
\nonumber \\
& {\bf \left[q.b(k)\right]} 
{\bf \left[u(p).b(q)\right]} 
\{\chi^{(0)}+\chi_{1}\frac{{\bf k.q}}{kq}+\chi_{2}\frac{{\bf k.p}}{kp}+
\chi_{3}\frac{{\bf p.q}}{pq} \} & )
\label{eq:Y3_kqp}
\end{eqnarray}
and the circulating transfer from 
{\bf p} to {\bf k},
 $Y_{\Delta}({\bf p|k|q}))= Y_{\Delta}^{(2)}({\bf p|k|q})+
Y_{\Delta}^{(3)}({\bf p|k|q})$ 
 can be written as
\begin{eqnarray}
Y_{\Delta}^{(2)}({\bf p|k|q}) = 
Re( & {\bf \left[p.u(q)\right]} 
{\bf \left[b(k).b(p)\right]} 
\{\alpha^{(0)}+\alpha_{1}\frac{{\bf k.p}}{kp}+
\alpha_{2}\frac{{\bf p.q}}{pq}+ \alpha_{3}\frac{{\bf k.q}}{kq} \} + &
\nonumber \\
&  {\bf \left[p.u(k)\right]} 
 {\bf \left[b(p).b(q)\right]} 
\{\beta^{(0)}+\beta_{1}\frac{{\bf k.p}}{kp}+\beta_{2}\frac{{\bf p.q}}{pq}+
\beta_{3}\frac{{\bf k.q}}{kq} \} + &
\nonumber \\
& {\bf \left[k.u(p)\right]} 
{\bf \left[b(k).b(q)\right]} 
\{\gamma^{(0)}+\gamma_{1}\frac{{\bf k.p}}{kp}+\gamma_{2}\frac{{\bf p.q}}{kq}+
\gamma_{3}\frac{{\bf k.q}}{kq} \} & )
\label{eq:Y2_pkq}
\end{eqnarray}
\begin{eqnarray}
Y_{\Delta}^{(3)}({\bf p|k|q}) = 
Re( & {\bf \left[p.b(q)\right]} 
{\bf \left[u(p).b(k)\right]} 
\{\delta^{(0)}+\delta_{1}\frac{{\bf k.p}}{kp}+\delta_{2}\frac{{\bf p.q}}{pq}+
\delta_{3}\frac{{\bf k.q}}{kq} \} + &
\nonumber \\
& {\bf \left[p.b(q)\right]} 
{\bf \left[u(k).b(p)\right]}   
\{\eta^{(0)}+\eta_{1}\frac{{\bf k.p}}{kp}+\eta_{2}\frac{{\bf p.q}}{pq}+
\eta_{3}\frac{{\bf k.q}}{kq} \} + &
\nonumber \\
& {\bf \left[p.b(k)\right]} 
  {\bf \left[u(p).b(q)\right]}  
\{\mu^{(0)}+\mu_{1}\frac{{\bf k.p}}{kp}+\mu_{2}\frac{{\bf p.q}}{pq}+
\mu_{3}\frac{{\bf k.q}}{kq} \} + &
\nonumber \\
& {\bf \left[p.b(k)\right]} 
 {\bf \left[u(q).b(p)\right]}  
\{\nu^{(0)}+\nu_{1}\frac{{\bf k.p}}{kp}+\nu_{2}\frac{{\bf p.q}}{pq}+
\nu_{3}\frac{{\bf k.q}}{kq} \} + &
\nonumber \\
& {\bf \left[k.b(p)\right]} 
 {\bf \left[u(k).b(q)\right]}  
\{\phi^{(0)}+\phi_{1}\frac{{\bf k.p}}{kp}+\phi_{2}\frac{{\bf p.q}}{pq}+
\phi_{3}\frac{{\bf k.q}}{kq} \} + &
\nonumber \\
& {\bf \left[k.b(p)\right]} 
{\bf \left[u(q).b(k)\right]} 
\{\chi^{(0)}+\chi_{1}\frac{{\bf k.p}}{kp}+\chi_{2}\frac{{\bf p.q}}{pq}+
\chi_{3}\frac{{\bf k.q}}{kq} \} & )
\label{eq:Y3_pkq}
\end{eqnarray}
$Y_{\Delta}({\bf k|p|q})$ flows along the anti-clockwise direction while
$Y_{\Delta}({\bf k|q|p})$ and $Y_{\Delta}({\bf p|k|q})$ flow along
the clockwise direction. The former should therefore be equal and opposite
to the latter two transfers, i.e.,
\begin{eqnarray}
Y_{\Delta}({\bf k|p|q}) + Y_{\Delta}({\bf k|q|p}) = 0 ,
\label{eq:Ykpq_kqp1}
\end{eqnarray}
and
\begin{equation}
Y_{\Delta}({\bf k|p|q}) + Y_{\Delta}({\bf p|k|q}) = 0 .
\label{eq:Ykpq_pkq1}
\end{equation}
Replacing the expression for $Y_{\Delta}({\bf k|p|q})$ from
Eqs.~(\ref{eq:Y2_lnr})-(\ref{eq:Y3_lnr}) and for
$Y_{\Delta}({\bf k|q|p})$ from
Eqs.~(\ref{eq:Y2_kqp})-(\ref{eq:Y3_kqp}) 
into  Eq. (B.28) we get
\begin{eqnarray}
{\bf \left[k.u(q)\right]} 
{\bf \left[b(k).b(p)\right]} 
\{(\alpha^{(0)}+\beta^{(0)})+
(\alpha_{1}+\beta_{2})\frac{{\bf k.p}}{kp}+
(\alpha_{2}+\beta_{1})\frac{{\bf k.q}}{kq}+
(\alpha_{3}+\beta_{3}\frac{{\bf p.q}}{pq} \} + &
\nonumber \\
  {\bf \left[k.u(p)\right]} 
 {\bf \left[b(k).b(q)\right]} 
\{(\beta^{(0)}+\alpha_{(0)})+
(\beta_{1}+\alpha_{2})\frac{{\bf k.p}}{kp}+
(\beta_{2}+\alpha_{1}\frac{{\bf k.q}}{kq}+
\beta_{3}+\alpha_{3}\frac{{\bf p.q}}{pq} \} + &
\nonumber \\
 {\bf \left[p.u(k)\right]} 
{\bf \left[b(p).b(q)\right]} 
\{(\gamma_{1}-\gamma_{2})\frac{{\bf k.p}}{kp}+
(\gamma_{2}-\gamma_{1})\frac{{\bf k.q}}{kq}+&
\nonumber \\
{\bf \left[k.b(q)\right]} 
{\bf \left[u(k).b(p)\right]} 
\{(\delta^{(0)}+\nu^{(0)})+
(\delta_{1}+\mu_{2})\frac{{\bf k.p}}{kp}+
(\delta_{2}+\mu_{1})\frac{{\bf k.q}}{kq}+
(\delta_{3}+\mu_{3})\frac{{\bf p.q}}{pq} \} + &
\nonumber \\
 {\bf \left[k.b(q)\right]} 
{\bf \left[u(p).b(k)\right]}   
\{(\eta^{(0)}+\nu^{(0)})
(\eta_{1}+\nu_{2})\frac{{\bf k.p}}{kp}+
(\eta_{2}+\nu_{1})\frac{{\bf k.q}}{kq}+
(\eta_{3}+\nu_{3})\frac{{\bf p.q}}{pq} \} + &
\nonumber \\
 {\bf \left[k.b(p)\right]} 
  {\bf \left[u(k).b(q)\right]}  
\{(\mu^{(0)}+\delta^{(0)})+
(\mu_{1}+\delta_{2})\frac{{\bf k.p}}{kp}+
(\mu_{2}+\delta_{1})\frac{{\bf k.q}}{kq}+
(\mu_{3}+\delta_{3})\frac{{\bf p.q}}{pq} \} + &
\nonumber \\
 {\bf \left[k.b(p)\right]} 
 {\bf \left[u(q).b(k)\right]}  
\{(\nu^{(0)}+\eta^{(0)})+
(\nu_{1}+\eta_{2})\frac{{\bf k.p}}{kp}+
(\nu_{2}+\eta_{1})\frac{{\bf k.q}}{kq}+
(\nu_{3}+\eta_{3})\frac{{\bf p.q}}{pq} \} + &
\nonumber \\
 {\bf \left[p.b(k)\right]} 
 {\bf \left[u(p).b(q)\right]}  
\{(\phi^{(0)}-\chi^{(0)})+
(\phi_{1}-\chi_{2})\frac{{\bf k.p}}{kp}+
(\phi_{2}-\chi_{1})\frac{{\bf k.q}}{kq}+
(\phi_{3}-\chi_{3})\frac{{\bf p.q}}{pq} \} + &
\nonumber \\
 {\bf \left[p.b(k)\right]} 
{\bf \left[u(q).b(p)\right]} 
\{(\chi^{(0)}-\phi^{(0)})+
(\chi_{1}-\phi_{2})\frac{{\bf k.p}}{kp}+
(\chi_{2}-\phi_{1})\frac{{\bf k.q}}{kq}+
(\chi_{3}-\phi_{3})\frac{{\bf p.q}}{pq} \} & = 0
\nonumber \\
 &
\label{eq:junk2}
\end{eqnarray}
and replacing the expression for $Y_{\Delta}({\bf k|p|q})$ 
and $Y_{\Delta}({\bf p|k|q})$ from
Eqs.~(\ref{eq:Y2_pkq})-(\ref{eq:Y3_pkq}) into 
 Eq.(B.29) we get
\begin{eqnarray}
{\bf \left[k.u(q)\right]} 
{\bf \left[b(k).b(p)\right]} 
\{(\alpha_{2}-\alpha_{3})\frac{{\bf k.q}}{kq}+
(\alpha_{3}-\alpha_{2})\frac{{\bf p.q}}{pq} \} + &
\nonumber \\
  {\bf \left[k.u(p)\right]} 
 {\bf \left[b(k).b(q)\right]} 
\{(\beta^{(0)}+\gamma^{(0)})+
(\beta_{1}+\gamma_{2})\frac{{\bf k.p}}{kp}+
(\beta_{2}+\gamma_{3}\frac{{\bf k.q}}{kq}+
\beta_{3}+\gamma_{2}\frac{{\bf p.q}}{pq} \} + &
\nonumber \\
 {\bf \left[p.u(k)\right]} 
{\bf \left[b(p).b(q)\right]} 
\{(\gamma^{(0)}+\beta^{(0)})+
(\gamma_{1}+\beta_{1})\frac{{\bf k.p}}{kp}+
(\gamma_{2}+\beta_{3})\frac{{\bf k.q}}{kq}+
(\gamma_{3}+\beta_{3})\frac{{\bf p.q}}{pq}+&
\nonumber \\
{\bf \left[k.b(q)\right]} 
{\bf \left[u(k).b(p)\right]} 
\{(\delta^{(0)}-\eta^{(0)})+
(\delta_{1}-\eta_{1})\frac{{\bf k.p}}{kp}+
(\delta_{2}-\eta_{3})\frac{{\bf k.q}}{kq}+
(\delta_{3}-\eta_{2})\frac{{\bf p.q}}{pq} \} + &
\nonumber \\
 {\bf \left[k.b(q)\right]} 
{\bf \left[u(p).b(k)\right]}   
\{(\eta^{(0)}-\delta^{(0)})+
(\eta_{1}-\delta_{1})\frac{{\bf k.p}}{kp}+
(\eta_{2}-\delta_{3})\frac{{\bf k.q}}{kq}+
(\eta_{3}-\delta_{2})\frac{{\bf p.q}}{pq} \} + &
\nonumber \\
 {\bf \left[k.b(p)\right]} 
  {\bf \left[u(k).b(q)\right]}  
\{(\mu^{(0)}+\phi^{(0)})+
(\mu_{1}+\phi_{1})\frac{{\bf k.p}}{kp}+
(\mu_{2}+\phi_{3})\frac{{\bf k.q}}{kq}+
(\mu_{3}+\phi_{2})\frac{{\bf p.q}}{pq} \} + &
\nonumber \\
 {\bf \left[k.b(p)\right]} 
 {\bf \left[u(q).b(k)\right]}  
\{(\nu^{(0)}+\chi^{(0)})+
(\nu_{1}+\chi_{1})\frac{{\bf k.p}}{kp}+
(\nu_{2}+\chi_{3})\frac{{\bf k.q}}{kq}+
(\nu_{3}+\chi_{2})\frac{{\bf p.q}}{pq} \} + &
\nonumber \\
 {\bf \left[p.b(k)\right]} 
 {\bf \left[u(p).b(q)\right]}  
\{(\phi^{(0)}+\mu^{(0)})+
(\phi_{1}+\mu_{1})\frac{{\bf k.p}}{kp}+
(\phi_{2}+\mu_{3})\frac{{\bf k.q}}{kq}+
(\phi_{3}+\mu_{2})\frac{{\bf p.q}}{pq} \} + &
\nonumber \\
 {\bf \left[p.b(k)\right]} 
{\bf \left[u(q).b(p)\right]} 
\{(\chi^{(0)}+\nu^{(0)})+
(\chi_{1}+\nu_{1})\frac{{\bf k.p}}{kp}+
(\chi_{2}+\nu_{3})\frac{{\bf k.q}}{kq}+
(\chi_{3}+\nu_{2})\frac{{\bf p.q}}{pq} \} & = 0
\nonumber \\
 &
\label{eq:Ykpq_Ypkq2}
\end{eqnarray}
Since each term in the above equation has a different functional
dependence on {\bf u(k), u(p), u(q), b(k), b(p), b(q)}, each term
should be individually zero. It follows that the constants in the
above equations should satisfy the following relationships :
from Eq. (B.30) we get
$\alpha^{(0)}+\beta^{(0)}=0, 
\alpha_{1}+\beta_{2}=0, 
\alpha_{2}+\beta_{1}=0,
 \alpha_{3}+\beta_{3}=0,  
\gamma_{1}-\gamma_{2}=0
 \delta^{(0)}+\mu^{(0)}=0, 
\delta_{1}+\mu_{2}=0,
\delta_{2}+\mu_{1}=0,
\delta_{3}+\mu_{3}=0,
\eta^{(0)}+\nu^{(0)}=0,
\eta_{1}+\nu_{2}=0,
\eta_{2}+\nu_{1}=0,
\eta_{3}+\nu_{3}=0,
\phi^{(0)}-\chi^{(0)}=0,
\phi_{1}-\chi_{2}=0,
\phi_{2}-\chi_{1}=0, 
\phi_{3}-\chi_{3}=0,$
and from Eq. (B.31) we get
$\alpha_{2}-\alpha_{3}=0,
\beta^{(0)}+\gamma^{(0)}=0,
\beta_{1}+\gamma_{1}=0,
\beta_{2}+\gamma_{3}=0,
\beta_{3}+\gamma_{2}=0,
\delta^{(0)}-\eta^{(0)}=0,
\delta_{1}-\eta_{1}=0,
\delta_{2}-\eta_{3}=0,
\delta_{3}-\eta_{2}=0,
\mu^{(0)}+\phi^{(0)}=0,
\mu_{1}+\phi_{1}=0,
\mu_{2}+\phi_{3}=0,
\mu_{3}+\phi_{2}=0,
\nu^{(0)}+\chi^{(0)}=0,
\nu_{2}+\chi_{3}=0,
\mu_{3}+\phi_{2}=0.$

The above relationships can be arranged into the following groups:
\{$\alpha^{(0)} = -\beta^{(0)} = \gamma^{(0)}$\},
\{$\delta^{(0)} = -\mu^{(0)} = \phi^{(0)} = \eta^{(0)} = -\nu^{(0)} = 
\chi^{(0)}$\},
\{$\alpha_{1} = -\beta_{2} = \gamma_{3}$\},
\{$\alpha_{2} = -\beta_{1} = \gamma_{1} = \gamma_{2} = \alpha_{3} = 
-\beta_{3}$\},
\{$\delta_{1} = -\mu_{2} = \phi_{3} = \chi_{3} = -\nu_{2} = \eta_{1}$\},
\{$\delta_{2} = -\mu_{1} = \phi_{1} = \chi_{2} = -\nu_{3} = \eta_{3}$\},
\{$\delta_{3} = -\mu_{3} = \phi_{2} = \chi_{1} = -\nu_{1} = \eta_{2}$\}.

Let us now consider a triad with 
$|{\bf p}|=|{\bf q}|$, ${\bf k.p=k.q}$ and with
the Fourier modes ${\bf u(p), u(q), b(p), b(q)}$ perpendicular to the
plane formed by this triad. We argued in Appendix A that the circulating
transfer should vanish for such a geometry. For this geometry we can write
$Y_\Delta$ from Eq.~(\ref{eq:Y2_lnr}) and (~\ref{eq:Y3_lnr}) as
\begin{eqnarray}
Y_{\Delta} = 
Re( &  {\bf \left[p.u(k)\right]} 
{\bf \left[b(p).b(q)\right]} 
\{\gamma^{(0)}+(\gamma_{1}+\gamma_{2})\frac{{\bf k.p}}{kp}+
\gamma_{3}\frac{{\bf p.q}}{pq} \} + &
\nonumber \\
& {\bf \left[p.b(k)\right]} 
 {\bf \left[u(p).b(q)\right]}  
\{\phi^{(0)}+(\phi_{1}+\phi_{2})\frac{{\bf k.p}}{kp}+
\phi_{3}\frac{{\bf p.q}}{pq} \} + &
\nonumber \\
& {\bf \left[p.b(k)\right]} 
{\bf \left[u(q).b(p)\right]} 
\{\chi^{(0)}+(\chi_{1}+\chi_{2})\frac{{\bf k.p}}{kp}+
\chi_{3}\frac{{\bf p.q}}{pq} \} & )
\nonumber \\
&=0 
\label{eq:Y_isosceles}
\end{eqnarray}
The above equality should hold for all triads with ${\bf |p|=|q|}$ and
{\bf k.p=k.q}. Therefore each term inside the curly brackets should be
individually zero. This leads us to the following additional relationships
between the various constants in the above equation:
$\gamma^{(0)}=0,
\gamma_{1}+\gamma_{2}=0,
\gamma_{3}=0,
\phi^{(0)}=0,
\phi_{2}+\phi_{3}=0,
\phi_{1}=0,
\chi^{(0)}=0,
\chi_{1}+\chi_{3}=0,
\chi_{2}=0,$
Similarly, by choosing ${\bf |k|=|p|}$, and
${\bf u(k), u(p), b(k), b(p)}$ perpendicular to the triad plane we get
the relationships
$\alpha^{(0)}=0,
\alpha_{1}=0,
\alpha_{2}+\alpha_{3}=0,
\delta^{(0)}=0,
\delta_{1}=0,
\delta_{2}+\delta_{3}=0,
\eta^{(0)}=0,
\eta_{1}=0,
\eta_{2}+\eta_{3}=0.$
and by choosing ${\bf |k|=|q|}$, and
${\bf u(k), u(q), b(k), b(q)}$ perpendicular to the triad plane we get
$\beta^{(0)}=0,
\beta_{2}=0,
\beta_{1}+\beta_{3}=0,
\mu^{(0)}=0,
\mu_{2}=0,
\mu_{1}+\mu_{3}=0,
\nu^{(0)}=0,
\nu_{2}=0,
\nu_{1}+\nu_{3}=0.$

Using these relationships the above groups take the following values:
\{ $\alpha^{(0)} = -\beta^{(0)} = \gamma^{(0)} = 0$\},
\{$\delta^{(0)} = -\mu^{(0)} = \phi^{(0)} = \eta^{(0)} = -\nu^{(0)} = 
\chi^{(0)} = 0$\},
\{$\alpha_{1} = -\beta_{2} = \gamma_{3} = 0$\},
\{$\alpha_{2} = -\beta_{1} = \gamma_{1} = \gamma_{2} = \alpha_{3} 
= -\beta_{3} = 0$\},
\{$\delta_{1} = -\mu_{2} = \phi_{3} = \chi_{3} = -\nu_{2} = \eta_{1} = 0$\},
\{$\delta_{2} = -\mu_{1} = \phi_{1} = \chi_{2} = -\nu_{3} = \eta_{3} =
-\delta_{3} = \mu_{3} = -\phi_{2} = -\chi_{1} = \nu_{1} = -\eta_{2}$\}.
Hence, some of the constants are equal to zero, and all the non-zero constants
are related to each other by just one undetermined
constant which we will denote by $'C'$, i.e.,
\{$\delta_{2} = -\mu_{1} = ... = C$\}

Putting the values of these constants in the expressions for
$Y_{\Delta}^{(2)}$ and $Y_{\Delta}^{(3)}$ in
 Eqs.~(\ref{eq:Y2_lnr})-(\ref{eq:Y3_lnr}) we get
\begin{eqnarray}
Y_{\Delta}^{(2)} = 0
\label{eq:Y2_end}
\end{eqnarray}
\begin{eqnarray}
Y_{\Delta}^{(3)} = 
Re( &   {\bf \left[k.b(q)\right]} 
{\bf \left[u(k).b(p)\right]} 
  \{C\frac{{\bf k.q}}{kq}-
C\frac{{\bf p.q}}{pq} \} + 
 {\bf \left[k.b(q)\right]} 
{\bf \left[u(p).b(k)\right]}   
\{-C\frac{{\bf k.q}}{kq}+
C\frac{{\bf p.q}}{pq} \} + &
\nonumber \\
& {\bf \left[k.b(p)\right]} 
  {\bf \left[u(k).b(q)\right]}  
\{-C\frac{{\bf k.p}}{kp}+
C\frac{{\bf p.q}}{pq} \} + 
 {\bf \left[k.b(p)\right]} 
 {\bf \left[u(q).b(k)\right]}  
\{C\frac{{\bf k.p}}{kp}-
C\frac{{\bf p.q}}{pq} \} +  &
\nonumber \\
& {\bf \left[p.b(k)\right]} 
 {\bf \left[u(p).b(q)\right]}  
\{C\frac{{\bf k.p}}{kp}-C\frac{{\bf k.q}}{kq}\} + 
 {\bf \left[p.b(k)\right]} 
{\bf \left[u(q).b(p)\right]} 
\{-C\frac{{\bf k.p}}{kp}+C\frac{{\bf k.q}}{kq}\} & )
\nonumber \\
\label{eq:Y3_end}
\end{eqnarray}
with a single undetermined constant. 

For the determination of $X_\Delta$ in Appendix A, the above considerations
were sufficient to determine $X_\Delta$. However, in the present case we
are still left with one undetermined constant. We will now impose the
constraint of locality which we state as follows:
In the limit ${\bf k, q \rightarrow \infty >> p}$, $Y_\Delta$ should tend to
zero. If 
$Z_\Delta$ remains finite then the noise, which is always inevitably
present in any flow at the smallest scales, would interact directly with
the large scales. Moreover, the locality constraint is also necessary in 
order that the circulating transfer is consistent with the
effective transfer between {\bf b(k)} and {\bf b(q)} which goes to zero for
in the above limit. 
Let us take {\bf u(k), u(q)} to be perpendicular
to the plane of the triad. Then we can write 
$Y_\Delta$ as
\begin{eqnarray}
Y_{\Delta}^{(3)} = 
Re( &  {\bf \left[k.b(p)\right]} 
  {\bf \left[u(k).b(q)\right]}  
\{-C\frac{{\bf k.p}}{kp}+
C\frac{{\bf p.q}}{pq} \} + 
 {\bf \left[k.b(p)\right]} 
 {\bf \left[u(q).b(k)\right]}  
\{C\frac{{\bf k.p}}{kp}
-C\frac{{\bf p.q}}{pq} \} &)
\nonumber \\
&
\label{eq:Y3_local}
\end{eqnarray}
which may be written in terms of the angles of the triangle formed
by {\bf k, p, q} [Fig. B.1] as
\begin{eqnarray}
Y_{\Delta}^{(3)} = 
Re( &  {\bf \left[\hat{k}.b(p)\right]} 
  {\bf \left[u(k).b(q)\right]}  
\{-C k \cos(\psi)-C k \cos(\theta)\} + &
\nonumber \\
& {\bf \left[\hat{k}.b(p)\right]} 
 {\bf \left[u(q).b(k)\right]}  
\{C k \cos(\psi)
+C k \cos(\theta) \} & )
\nonumber \\
&
\label{eq:Y3_angle}
\end{eqnarray}

We now take the limit ${\bf k, q \rightarrow  \infty}$, 
keeping $\psi$ and {\bf p}
constant.  In this limit $\phi \rightarrow 0$. Making use of the
relationships $\theta=\pi-\psi-\phi$, $k/\sin\theta=p/\sin\phi$, and
taking the limit $\phi \rightarrow 0$, we get
for $Y_{\Delta} \rightarrow 
{\bf \left[\hat{k}.b(p)\right]} 
  {\bf \left[u(k).b(q)\right]}  
\{- C p \sin^{2}(\psi)\} + 
 {\bf \left[\hat{k}.b(p)\right]} 
 {\bf \left[u(q).b(k)\right]}  
\{ C p \sin^{2}(\psi)\}$. However, in this limit $Y_\Delta$ should tend to
zero in order to satisfy locality. From this constraint we get $C=0$.
Putting the value of C in Eq.~(\ref{eq:Y3_end}) gives $Y_\Delta^{(3)}=0$.

Therefore we get $Y_\Delta=0$. It follows that the mode-to-mode transfer
${\cal{\slash{R}}}^{bb}={\cal{\slash{S}}}^{bb}$.

In the above proof we used rotational invariance, galilean invariance,
finiteness of mode-to-mode transfer, the symmetry of $Y_\Delta$ with
respect to {\bf k, p, q, u(k), u(p), u(q), b(k), b(p), b(q)}, and the
assumption that the circulating transfer should vanish if two of the
wavenumbers approach infinity. This proof is valid only upto linear
orders, as explained earlier in this appendix. A completely general proof
without taking recourse to the linearity assumption is lacking at present.

\newpage

\section{ Derivation of the circulating transfer between 
velocity modes and magnetic modes in a triad}
\setcounter{equation}{0}
\seceqcc

In Section~\ref{subs:umode_bmode} we had shown that the energy transfer from
{\bf u(p)} to {\bf b(k)} in the triad ({\bf k, p, q}) can be expressed
as
${\cal{\slash{R}}}^{bu}({\bf k|p|q})={\cal{\slash{S}}}^{bu}({\bf k|p|q})
+Z_{\Delta}$. 
${\cal{\slash{S}}}^{bu}({\bf k|p|q})$ is given by 
Eq.~(\ref{eq:bkup_def})
of Section~\ref{subs:umode_bmode} was interpreted as 
the effective transfer from 
{\bf u(p)} to {\bf b(k)} and
$Z_{\Delta}$ was interpreted as a circulating transfer between the
velocity and magnetic modes belonging to a triad. We will follow the
approach of Appendix A and Appendix B to determine
$Z_{\Delta}$ in this Appendix. We will impose the following conditions
on $Z_{\Delta}$. It should not violate the fundamentally triad nature
of the interactions, it should be finite, rotationally invariant,
galilean invariant, and should vanish if $k/p \rightarrow \infty$.

The circulating transfer $Z_{\Delta}$ is a scalar with a dependence on
atmost {\bf k, p, q, u(k), u(p), u(q), b(k), b(p), b(q)}. We know from
Appendix B that such a function can be expressed in the
form given by Eqs.~(\ref{eq:Y_general})-(\ref{eq:Y4_general}) of Appendix B.
The interactions
between the velocity and the magnetic modes are fundamentally three-mode, i.e.,
each term in the expression for combined energy transfer to a velocity
or a magnetic mode depends on all the three wavenumbers ({\bf k, p, q}).
the combined energy transfer to any of the velocity or the magnetic
modes in the triad. $Z_\Delta$ should not violate this fundamental nature of
the interactions. Hence we should drop all those terms from
Eqs.~(\ref{eq:Y_general})-(\ref{eq:Y4_general}) which are independent
of atleast one of the modes and then we get 
Eqs.~(\ref{eq:Y1_general2})-(\ref{eq:Y4_general2}) for $Z_\Delta$. 
The combined energy
transfer to velocity and magnetic modes in the triad are zero if
\{{\bf b(k), b(p) = 0}\}, \{{\bf b(k), b(q) = 0}\}, or 
\{{\bf b(p), b(q) = 0}\}. 
The terms which are non-vanishing if one of the above set 
of Fourier coefficients are zero will be dropped from  
Eqs.~(\ref{eq:Y1_general2})-(\ref{eq:Y4_general2}), giving 
Eqs.~(\ref{eq:Y_triad})-(\ref{eq:Y3_triad})
In Appendix B also we had imposed identical constraints on 
Eqs.~(\ref{eq:Y_general})-(\ref{eq:Y4_general}) to get
Eqs.~(\ref{eq:Y_triad})-(\ref{eq:Y3_triad}).
The dimensionless coefficients in these equations are
{\bf k, p, q, u(k), u(p), u(q), b(k),
 b(p), b(q)} dependent and we will now attempt to determine these.

In Appendix B we had considered the consequences of finiteness and
galilean invariance on Eqs.~(\ref{eq:Y_triad})-(\ref{eq:Y3_triad})
of Appendix B. We will briefly summarise the conclusions now. To ensure
the finiteness and galilean invariance of
Eqs.~(\ref{eq:Y_triad})-(\ref{eq:Y3_triad})
the dimensionless coefficients in these equations
should be independent of the magnitudes of the wavenumbers and the Fourier
coefficients.  Then each coefficient in the equations 
can be written as 
$\alpha=\alpha^{(0)}+\alpha^{(1)}+\alpha^{(2)}+\alpha^{(3)}+\alpha^{(4)}$ where
$\alpha^{(0)}$ is a constant,
$\alpha^{(1)}, \alpha^{(2)}, \alpha^{(3)}$ defined in
Eqs.~(\ref{eq:alpha1_bb})-(\ref{eq:alpha3_bb}) of Appendix B.
are linear in terms of the kind
{\bf k.p}, {\bf k.u(p)}, {\bf u(k).u(p)} respectively, and
$\alpha^{(4)}$ is nonlinear in these terms.
$\alpha^{(2)}$, and $\alpha^{(3)}$ are not galilean invariant and 
would therefore lead to the non-invariance of 
Eqs.~(\ref{eq:alpha1_bb})-(\ref{eq:alpha3_bb}) of Appendix B.
Hence, $\alpha^{(2)}$ and $\alpha^{(3)}$ should not be included in
$\alpha$. In Appendix B we had also dropped $\alpha^{(4)}$.
Therefore, we get $\alpha=\alpha^{(0)}+\alpha^{(1)}$. 
Following Appendix B we will now write
$Z_{\Delta}$ [given by Eqs.~(\ref{eq:Y_triad})-(\ref{eq:Y3_triad}) in
Appendix B] 
\begin{equation}
Z_{\Delta}^{bu} = Z_{2\Delta}^{bu}+ Z_{3\Delta}^{bu}
\label{eq:Z_lnr}
\end{equation}
where
\begin{eqnarray}
Z_{2\Delta}^{bu}({\bf k|p|q}) = 
Re( & {\bf \left[k.u(q)\right]} 
{\bf \left[b(k).b(p)\right]} 
\{\alpha^{(0)}+\alpha_{1}^{(1)}\frac{{\bf k.p}}{kp}+
\alpha_{2}^{(1)}\frac{{\bf k.q}}{kq}+
\alpha_{3}^{(1)}\frac{{\bf p.q}}{pq} \} + &
\nonumber \\
&  {\bf \left[k.u(p)\right]} 
 {\bf \left[b(k).b(q)\right]} 
\{\beta^{(0)}+\beta_{1}^{(1)}\frac{{\bf k.p}}{kp}+
\beta_{2}^{(1)}\frac{{\bf k.q}}{kq}+
\beta_{3}^{(1)}\frac{{\bf p.q}}{pq} \} + &
\nonumber \\
& {\bf \left[p.u(k)\right]} 
{\bf \left[b(p).b(q)\right]} 
\{\gamma^{(0)}+\gamma_{1}^{(1)}\frac{{\bf k.p}}{kp}+
\gamma_{2}^{(1)}\frac{{\bf k.q}}{kq}+
\gamma_{3}^{(1)}\frac{{\bf p.q}}{pq} \}  &)
\label{eq:Z2_lnr}
\end{eqnarray}
\begin{eqnarray}
Z_{3\Delta}^{bu}({\bf k|p|q}) = 
Re( & {\bf \left[k.b(q)\right]} 
{\bf \left[u(k).b(p)\right]} 
\{\delta^{(0)}+\delta_{1}^{(1)}\frac{{\bf k.p}}{kp}+
\delta_{2}^{(1)}\frac{{\bf k.q}}{kq}+
\delta_{3}^{(1)}\frac{{\bf p.q}}{pq} \} + &
\nonumber \\
& {\bf \left[k.b(q)\right]} 
{\bf \left[u(p).b(k)\right]}   
\{\eta^{(0)}+\eta_{1}^{(1)}\frac{{\bf k.p}}{kp}+
\eta_{2}^{(1)}\frac{{\bf k.q}}{kq}+
\eta_{3}^{(1)}\frac{{\bf p.q}}{pq} \} + &
\nonumber \\
& {\bf \left[k.b(p)\right]} 
  {\bf \left[u(k).b(q)\right]}  
\{\mu^{(0)}+\mu_{1}^{(1)}\frac{{\bf k.p}}{kp}+
\mu_{2}^{(1)}\frac{{\bf k.q}}{kq}+
\mu_{3}^{(1)}\frac{{\bf p.q}}{pq} \} + &
\nonumber \\
& {\bf \left[k.b(p)\right]} 
 {\bf \left[u(q).b(k)\right]}  
\{\nu^{(0)}+\nu_{1}^{(1)}\frac{{\bf k.p}}{kp}+
\nu_{2}^{(1)}\frac{{\bf k.q}}{kq}+
\nu_{3}^{(1)}\frac{{\bf p.q}}{pq} \} + &
\nonumber \\
& {\bf \left[p.b(k)\right]} 
 {\bf \left[u(p).b(q)\right]}  
\{\phi^{(0)}+\phi_{1}^{(1)}\frac{{\bf k.p}}{kp}+
\phi_{2}^{(1)}\frac{{\bf k.q}}{kq}+
\phi_{3}^{(1)}\frac{{\bf p.q}}{pq} \} + &
\nonumber \\
& {\bf \left[p.b(k)\right]} 
{\bf \left[u(q).b(p)\right]} 
\{\chi^{(0)}+\chi_{1}^{(1)}\frac{{\bf k.p}}{kp}+
\chi_{2}^{(1)}\frac{{\bf k.q}}{kq}+
\chi_{3}^{(1)}\frac{{\bf p.q}}{pq} \} &)
\label{eq:Z3_lnr}
\end{eqnarray}
The quantities $Z_{\Delta}^{bu} = Z_{2\Delta}^{bu}+ Z_{3\Delta}^{bu}$
 represent the
energy circulating from ${\bf u(p) \rightarrow b(k) \rightarrow u(q)}$
${\bf \rightarrow b(p) ... \rightarrow u(p)}$ 
[see Fig. 9 in Section~\ref{subs:umode_bmode}]. We will
denote the circulating transfer from ${\bf u(p) \rightarrow b(k)}$
by  $Z_{\Delta}^{bu}({\bf k|p|q})$, from ${\bf b(p) \rightarrow u(k)}$
by  $Z_{\Delta}^{ub}({\bf k|p|q})$ and so on for the circulating transfer
between other modes. We take  
$Z_{\Delta}^{bu} = Z_{2\Delta}+^{bu} Z_{3\Delta}^{bu}$
in Eqs.~(\ref{eq:Z2_lnr})-(~\ref{eq:Z3_lnr}) to be the circulating transfer
from ${\bf u(p) \rightarrow b(k)}$, i.e., 
$Z_{\Delta}^{bu}({\bf k|p|q}) \equiv Z_{\Delta}$. Then by symmetry the
circulating transfer from {\bf u(q)} to {\bf b(k)} can be written as
\begin{eqnarray}
Z_{\Delta}^{bu}({\bf k|q|p})=Z_{2\Delta}^{bu}({\bf k|q|p})+
Z_{3\Delta}^{bu}({\bf k|q|p})
\label{eq:Z_bkuq}
\end{eqnarray}
\begin{eqnarray}
Z_{2\Delta}^{bu}({\bf k|q|p}) = 
Re( & {\bf \left[k.u(p)\right]} 
{\bf \left[b(k).b(q)\right]} 
\{\alpha^{(0)}+\alpha_{1}\frac{{\bf k.q}}{kq}+\alpha_{2}\frac{{\bf k.p}}{kp}+
\alpha_{3}\frac{{\bf p.q}}{pq} \} + &
\nonumber \\
&  {\bf \left[k.u(q)\right]} 
 {\bf \left[b(k).b(p)\right]} 
\{\beta^{(0)}+\beta_{1}\frac{{\bf k.q}}{kq}+\beta_{2}\frac{{\bf k.p}}{kp}+
\beta_{3}\frac{{\bf p.q}}{pq} \} + &
\nonumber \\
& {\bf \left[q.u(k)\right]} 
{\bf \left[b(p).b(q)\right]} 
\{\gamma^{(0)}+\gamma_{1}\frac{{\bf k.q}}{kq}+\gamma_{2}\frac{{\bf k.p}}{kp}+
\gamma_{3}\frac{{\bf p.q}}{pq} \}  &)
\label{eq:Z2_bkuq}
\end{eqnarray}
\begin{eqnarray}
Z_{3\Delta}^{bu}({\bf k|q|p}) = 
Re( & {\bf \left[k.b(p)\right]} 
{\bf \left[u(k).b(q)\right]} 
\{\delta^{(0)}+\delta_{1}\frac{{\bf k.q}}{kq}+\delta_{2}\frac{{\bf k.p}}{kp}+
\delta_{3}\frac{{\bf p.q}}{pq} \} + &
\nonumber \\
& {\bf \left[k.b(p)\right]} 
{\bf \left[u(q).b(k)\right]}   
\{\eta^{(0)}+\eta_{1}\frac{{\bf k.q}}{kq}+\eta_{2}\frac{{\bf k.p}}{kp}+
\eta_{3}\frac{{\bf p.q}}{pq} \} + &
\nonumber \\
& {\bf \left[k.b(q)\right]} 
  {\bf \left[u(q).b(k)\right]}  
\{\mu^{(0)}+\mu_{1}\frac{{\bf k.q}}{kq}+\mu_{2}\frac{{\bf k.p}}{kp}+
\mu_{3}\frac{{\bf p.q}}{pq} \} + &
\nonumber \\
& {\bf \left[k.b(q)\right]} 
 {\bf \left[u(k).b(q)\right]}  
\{\nu^{(0)}+\nu_{1}\frac{{\bf k.q}}{kq}+\nu_{2}\frac{{\bf k.p}}{kp}+
\nu_{3}\frac{{\bf p.q}}{pq} \} + &
\nonumber \\
& {\bf \left[q.b(k)\right]} 
 {\bf \left[u(q).b(p)\right]}  
\{\phi^{(0)}+\phi_{1}\frac{{\bf k.q}}{kq}+\phi_{2}\frac{{\bf k.p}}{kp}+
\phi_{3}\frac{{\bf p.q}}{pq} \} + &
\nonumber \\
& {\bf \left[q.b(k)\right]} 
{\bf \left[u(p).b(q)\right]} 
\{\chi^{(0)}+\chi_{1}\frac{{\bf k.q}}{kq}+\chi_{2}\frac{{\bf k.p}}{kp}+
\chi_{3}\frac{{\bf p.q}}{pq} \} &)
\label{eq:Z3_bkuq}
\end{eqnarray}
We have dropped the superscript `1' in the above equations.
$Z_{\Delta}^{bu}({\bf k|p|q})$ and
$Z_{\Delta}^{ub}({\bf p|k|q})$ are physically the same transfers, but
the former represents the transfer from {\bf u(p)} to {\bf b(k)}
and the latter represents the same transfer from {\bf b(k)} to {\bf u(p)}
The former should therefore be equal and opposite in sign to the latter, i.e.,
\begin{equation}
Z_{\Delta}^{bu}({\bf k|p|q})+Z_{\Delta}^{ub}({\bf p|k|q})=0
\label{eq:Zbkup_upbk}
\end{equation}
$Z_{\Delta}^{ub}({\bf p|k|q})$ is thus given by the negative of 
Eqs.~(\ref{eq:Z_lnr})-(\ref{eq:Z3_lnr}).
Using the expression for $Z_{\Delta}^{ub}({\bf p|k|q})$ we can write
$Z_{\Delta}^{ub}({\bf k|p|q})$  to be 
\begin{eqnarray}
Z_{\Delta}^{ub}({\bf k|p|q}) = Z_{2\Delta}^{ub}({\bf k|p|q})+
Z_{3\Delta}^{ub}({\bf k|p|q}) 
\label{eq:Z_ukbp}
\end{eqnarray}
\begin{eqnarray}
Z_{2\Delta}^{ub}({\bf k|p|q}) = 
Re( & - {\bf \left[k.u(p)\right]} 
{\bf \left[b(k).b(q)\right]} 
\{\gamma^{(0)}+\gamma_{1}\frac{{\bf k.p}}{kp}+\gamma_{2}\frac{{\bf p.q}}{pq}+
\gamma_{3}\frac{{\bf k.q}}{kq} \}   &
\nonumber \\
& + {\bf \left[k.u(q)\right]} 
{\bf \left[b(k).b(p)\right]} 
\{\alpha^{(0)}+\alpha_{1}\frac{{\bf k.p}}{kp}+\alpha_{2}\frac{{\bf p.q}}{pq}+
\alpha_{3}\frac{{\bf k.q}}{kq} \}  &
\nonumber \\
&  - {\bf \left[p.u(k)\right]} 
 {\bf \left[b(p).b(q)\right]} 
\{\beta^{(0)}+\beta_{1}\frac{{\bf k.p}}{kp}+\beta_{2}\frac{{\bf p.q}}{pq}+
\beta_{3}\frac{{\bf k.q}}{kq} \} &)
\nonumber \\
\label{eq:Z2_ukbp}
\end{eqnarray}
\begin{eqnarray}
Z_{3\Delta}^{ub}({\bf k|p|q}) = 
Re( & - {\bf \left[k.b(p)\right]} 
 {\bf \left[u(k).b(q)\right]}  
\{\phi^{(0)}+\phi_{1}\frac{{\bf k.p}}{kp}+\phi_{2}\frac{{\bf p.q}}{pq}+
\phi_{3}\frac{{\bf k.q}}{kq} \}  & 
\nonumber \\
&  - {\bf \left[k.b(p)\right]} 
{\bf \left[u(q).b(k)\right]} 
\{\chi^{(0)}+\chi_{1}\frac{{\bf k.p}}{kp}+\chi_{2}\frac{{\bf p.q}}{pq}+
\chi_{3}\frac{{\bf k.q}}{kq} \}  &
\nonumber \\
&  + {\bf \left[k.b(q)\right]} 
{\bf \left[u(p).b(k)\right]} 
\{\delta^{(0)}+\delta_{1}\frac{{\bf k.p}}{kp}+\delta_{2}\frac{{\bf p.q}}{pq}+
\delta_{3}\frac{{\bf k.q}}{kq} \}  &
\nonumber \\
&  + {\bf \left[k.b(q)\right]} 
{\bf \left[u(k).b(p)\right]}   
\{\eta^{(0)}+\eta_{1}\frac{{\bf k.p}}{kp}+\eta_{2}\frac{{\bf p.q}}{pq}+
\eta_{3}\frac{{\bf k.q}}{kq} \}  &
\nonumber \\
&  - {\bf \left[p.b(k)\right]} 
  {\bf \left[u(p).b(q)\right]}  
\{\mu^{(0)}+\mu_{1}\frac{{\bf k.p}}{kp}+\mu_{2}\frac{{\bf p.q}}{pq}+
\mu_{3}\frac{{\bf k.q}}{kq} \}  &
\nonumber \\
&  - {\bf \left[p.b(k)\right]} 
 {\bf \left[u(q).b(p)\right]}  
\{\nu^{(0)}+\nu_{1}\frac{{\bf k.p}}{kp}+\nu_{2}\frac{{\bf p.q}}{pq}+
\nu_{3}\frac{{\bf k.q}}{kq} \}  &)
\nonumber \\
\label{eq:Z3_ukbp}
\end{eqnarray}
and $Z_{\Delta}^{ub}({\bf k|q|p})$ is given by
\begin{eqnarray}
Z_{\Delta}^{ub}({\bf k|q|p}) = Z_{2\Delta}^{ub}({\bf k|q|p})+
Z_{3\Delta}^{ub}({\bf k|q|p}) 
\label{eq:Z_ukbq}
\end{eqnarray}
\begin{eqnarray}
Z_{2\Delta}^{ub}({\bf k|q|p}) = 
Re( & - {\bf \left[k.u(q)\right]} 
{\bf \left[b(k).b(p)\right]} 
\{\gamma^{(0)}+\gamma_{1}\frac{{\bf k.q}}{kq}+\gamma_{2}\frac{{\bf p.q}}{pq}+
\gamma_{3}\frac{{\bf k.p}}{kp} \}   &
\nonumber \\
& + {\bf \left[k.u(p)\right]} 
{\bf \left[b(k).b(q)\right]} 
\{\alpha^{(0)}+\alpha_{1}\frac{{\bf k.q}}{kq}+\alpha_{2}\frac{{\bf p.q}}{pq}+
\alpha_{3}\frac{{\bf k.p}}{kp} \}  &
\nonumber \\
& -  {\bf \left[q.u(k)\right]} 
 {\bf \left[b(p).b(q)\right]} 
\{\beta^{(0)}+\beta_{1}\frac{{\bf k.q}}{kq}+\beta_{2}\frac{{\bf p.q}}{pq}+
\beta_{3}\frac{{\bf k.p}}{kp} \} & )
\nonumber \\
\label{eq:Z2_ukbq}
\end{eqnarray}
\begin{eqnarray}
Z_{3\Delta}^{ub}({\bf k|q|p}) = 
Re( & - {\bf \left[k.b(q)\right]} 
 {\bf \left[u(k).b(p)\right]}  
\{\phi^{(0)}+\phi_{1}\frac{{\bf k.q}}{kq}+\phi_{2}\frac{{\bf p.q}}{pq}+
\phi_{3}\frac{{\bf k.p}}{kp} \}   &
\nonumber \\
& - {\bf \left[k.b(q)\right]} 
{\bf \left[u(p).b(k)\right]} 
\{\chi^{(0)}+\chi_{1}\frac{{\bf k.q}}{kq}+\chi_{2}\frac{{\bf p.q}}{pq}+
\chi_{3}\frac{{\bf k.p}}{kp} \}  &
\nonumber \\
& + {\bf \left[k.b(p)\right]} 
{\bf \left[u(q).b(k)\right]} 
\{\delta^{(0)}+\delta_{1}\frac{{\bf k.q}}{kq}+\delta_{2}\frac{{\bf p.q}}{pq}+
\delta_{3}\frac{{\bf k.p}}{kp} \}  &
\nonumber \\
& + {\bf \left[k.b(p)\right]} 
{\bf \left[u(k).b(q)\right]}   
\{\eta^{(0)}+\eta_{1}\frac{{\bf k.q}}{kq}+\eta_{2}\frac{{\bf p.q}}{pq}+
\eta_{3}\frac{{\bf k.p}}{kp} \}   &
\nonumber \\
& - {\bf \left[q.b(k)\right]} 
  {\bf \left[u(q).b(p)\right]}  
\{\mu^{(0)}+\mu_{1}\frac{{\bf k.q}}{kq}+\mu_{2}\frac{{\bf p.q}}{pq}+
\mu_{3}\frac{{\bf k.p}}{kp} \}  &
\nonumber \\
& - {\bf \left[q.b(k)\right]} 
 {\bf \left[u(p).b(q)\right]}  
\{\nu^{(0)}+\nu_{1}\frac{{\bf k.q}}{kq}+\nu_{2}\frac{{\bf p.q}}{pq}+
\nu_{3}\frac{{\bf k.p}}{kp} \}  &)
\nonumber \\
\label{eq:Z3_ukbq}
\end{eqnarray}

The sum of the circulating transfer to {\bf b(k)} from {\bf u(p)} and
from {\bf u(q)} should be zero, i.e.,
\begin{equation}
Z_{\Delta}^{bu}({\bf k|p|q})+Z_{\Delta}^{bu}({\bf k|q|p})=0
\label{eq:Zbkup_bkuq1}
\end{equation}
and similarly the sum of the circulating transfer to {\bf u(k)} from {\bf b(p)} and
from {\bf b(q)} should also be equal to zero, i.e.,
\begin{equation}
Z_{\Delta}^{ub}({\bf k|p|q})+Z_{\Delta}^{ub}({\bf k|q|p})=0
\label{eq:Zukbp_ukbq1}
\end{equation}

Replacing $Z_{\Delta}^{bu}({\bf k|p|q})$ and 
$Z_{\Delta}^{bu}({\bf k|q|p})$ from Eqs.~(\ref{eq:Z_lnr})-(\ref{eq:Z3_lnr})
and
Eqs.~(\ref{eq:Z_bkuq})-(\ref{eq:Z3_bkuq})
into Eq.~(\ref{eq:Zbkup_bkuq1}) we get the following relationships between
the constants: $\alpha, \beta$, etc.
$\alpha^{(0)}+\beta^{(0)}=0, \alpha_{1}+\beta_{2}=0,
\alpha_{2}+\beta_{1}=0,
\alpha_{3}+\beta_{3}=0,
\gamma_{1}-\gamma_{2}=0,
\delta^{(0)}+\mu^{(0)}=0,
\delta_{1}+\mu_{2}=0,
\delta_{2}+\mu_{1}=0,
\delta_{3}+\mu_{3}=0,
\eta^{(0)}+\nu^{(0)}=0,
\eta_{1}+\nu_{2}=0,
\eta_{2}+\nu_{1}=0,
\eta_{3}+\nu_{3}=0,
\phi^{(0)}-\chi^{(0)}=0,
\phi_{1}-\chi_{2}=0,
\phi_{2}-\chi_{1}=0, 
\phi_{3}-\chi_{3}=0.$
After replacing 
$Z_{\Delta}^{ub}({\bf k|p|q})$ and
$Z_{\Delta}^{ub}({\bf k|q|p})$ from Eqs.~(\ref{eq:Z_lnr})-(\ref{eq:Z3_lnr})
and Eqs.~(\ref{eq:Z_ukbq})-(\ref{eq:Z3_ukbq})
into Eq.~(\ref{eq:Zukbp_ukbq1}) we get the following relationships:
$\alpha_{2}-\alpha_{3}=0,
\beta^{(0)}+\gamma^{(0)}=0,
\beta_{1}+\gamma_{1}=0,
\beta_{2}+\gamma_{3}=0,
\beta_{3}+\gamma_{2}=0,
\delta^{(0)}-\eta^{(0)}=0,
\delta_{1}-\eta_{1}=0,
\delta_{2}-\eta_{3}=0,
\delta_{3}-\eta_{2}=0,
\mu^{(0)}+\phi^{(0)}=0,
\mu_{1}+\phi_{1}=0,
\mu_{2}+\phi_{3}=0,
\mu_{3}+\phi_{2}=0,
\nu^{(0)}+\chi^{(0)}=0,
\nu_{2}+\chi_{3}=0,
\mu_{3}+\phi_{2}=0.$

Let us now consider the following geometry: the three wavenumbers in the
triad form an isosceles triangle, and we choose
Fourier coefficients at the two wavenumbers of equal lengths
to be perpendicular to {\bf k, p, q}.
We had considered this kind of geometry in Appendix A and
Appendix B.  We had argued that because of the symmetry between the
two wavenumbers of equal length, the circulating transfer should vanish
for this geometry. This constraint had given us an additional set of
relationships between the constants in $Y_{\Delta}$.
$Z^{bu}_{\Delta}$  should also be equal to zero 
for this geometry.  Since $Z^{bu}_{\Delta}$  and $Y_{\Delta}$  
have the same functional form, given by 
Eq.~(\ref{eq:Y2_lnr})-(\ref{eq:Y3_lnr}) of Appendix B 
and  Eq.~(\ref{eq:Z_lnr})-(\ref{eq:Z3_lnr})
 of this appendix respectively, hence imposition
of this constraint on $Z^{bu}_{\Delta}$  would give the same set of
relationships, i.e.,
$\gamma^{(0)}=0,
\gamma_{1}+\gamma_{2}=0,
\gamma_{3}=0,
\phi^{(0)}=0,
\phi_{2}+\phi_{3}=0,
\phi_{1}=0,
\chi^{(0)}=0,
\chi_{1}+\chi_{3}=0,
\chi_{2}=0,
\alpha^{(0)}=0,
\alpha_{1}=0,
\alpha_{2}+\alpha_{3}=0,
\delta^{(0)}=0,
\delta_{1}=0,
\delta_{2}+\delta_{3}=0,
\eta^{(0)}=0,
\eta_{1}=0,
\eta_{2}+\eta_{3}=0,
\beta^{(0)}=0,
\beta_{2}=0,
\beta_{1}+\beta_{3}=0,
\mu^{(0)}=0,
\mu_{2}=0,
\mu_{1}+\mu_{3}=0,
\nu^{(0)}=0,
\nu_{2}=0,
\nu_{1}+\nu_{3}=0.$

The above relationships between the constants in
Eqs.~(\ref{eq:Z2_lnr})-(\ref{eq:Z3_lnr}) are identical to the 
relationships between the constants in 
Eqs.~(\ref{eq:Y2_lnr})-(\ref{eq:Y3_lnr}) of Appendix B. 
We know from Appendix B that these relationships can be expressed
more clearly as follows:
\{ $\alpha^{(0)} = -\beta^{(0)} = \gamma^{(0)} = 0$\},
\{$\delta^{(0)} = -\mu^{(0)} = \phi^{(0)} = \eta^{(0)} = -\nu^{(0)} = 
\chi^{(0)} = 0$\},
\{$\alpha_{1} = -\beta_{2} = \gamma_{3} = 0$\},
\{$\alpha_{2} = -\beta_{1} = \gamma_{1} = \gamma_{2} = \alpha_{3} 
= -\beta_{3} = 0$\},
\{$\delta_{1} = -\mu_{2} = \phi_{3} = \chi_{3} = -\nu_{2} = \eta_{1} = 0$\},
\{$\delta_{2} = -\mu_{1} = \phi_{1} = \chi_{2} = -\nu_{3} = \eta_{3} =
-\delta_{3} = \mu_{3} = -\phi_{2} = -\chi_{1} = \nu_{1} = -\eta_{2}$\}.
There is just one undetermined
constant which we will denote by `C', i.e.,
\{$\delta_{2} = -\mu_{1} = ... = C$\}. Expressing
$Z_{\Delta}^{bu}$ [Eqs.~(\ref{eq:Z2_lnr})-(\ref{eq:Z3_lnr})]
in terms of this single constant, we get
\begin{eqnarray}
Z_{2\Delta} = 0
\label{eq:Z2_end}
\end{eqnarray}
\begin{eqnarray}
Z_{3\Delta} = 
Re( &  {\bf \left[k.b(q)\right]} 
{\bf \left[u(k).b(p)\right]} 
  \{C\frac{{\bf k.q}}{kq}-
C\frac{{\bf p.q}}{pq} \} + 
 {\bf \left[k.b(q)\right]} 
{\bf \left[u(p).b(k)\right]}   
\{-C\frac{{\bf k.q}}{kq}+
C\frac{{\bf p.q}}{pq} \} + &
\nonumber \\
& {\bf \left[k.b(p)\right]} 
  {\bf \left[u(k).b(q)\right]}  
\{-C\frac{{\bf k.p}}{kp}+
C\frac{{\bf p.q}}{pq} \} + 
 {\bf \left[k.b(p)\right]} 
 {\bf \left[u(q).b(k)\right]}  
\{C\frac{{\bf k.p}}{kp}+
-C\frac{{\bf p.q}}{pq} \} + &
\nonumber \\
& {\bf \left[p.b(k)\right]} 
 {\bf \left[u(p).b(q)\right]}  
\{C\frac{{\bf k.p}}{kp}-C\frac{{\bf k.q}}{kq}\} + 
 {\bf \left[p.b(k)\right]} 
{\bf \left[u(q).b(p)\right]} 
\{-C\frac{{\bf k.p}}{kp}+C\frac{{\bf k.q}}{kq}\} &)
\nonumber \\
\label{eq:Z3_end}
\end{eqnarray}
which is the same as the expression for
for $Y_\Delta$ in Eqs.~(\ref{eq:Y2_end})-(\ref{eq:Y3_end}) of Appendix B.

If $k/p$ and $q/p \rightarrow \infty$, while {\bf p} is kept constant the 
circulating transfer $Z_{\Delta}$ should tend to zero. This constraint
was also applied on the circulating transfer between magnetic modes,
$Y_\Delta$ in Appendix B. 
We had shown that the constant $C=0$ for this
constraint to be satisfied. Putting $C=0$ in 
Eqs.~(\ref{eq:Z2_end})-(\ref{eq:Z3_end}) we get
 $Z_{\Delta}^{bu}=Z_{2\Delta}^{bu}+Z_{3\Delta}^{bu}$ equal to zero.

The circulating transfer between the velocity and the magnetic modes
in the triad is equal to zero and the mode-to-mode transfer
${\cal{\slash{R}}}^{bu}({\bf k|p|q})={\cal{\slash{S}}}^{bu}({\bf k|p|q})$.

The derivation of $Z_{\Delta}^{bu}$, however took the coefficients to be
linear. A completely general derivation for 
$Z_{\Delta}^{bu}$ will require imposition of additional symmetry and
invariance or some other constraints, which we have ignored here. 

\newpage

\section{Equivalence of Kraichnan's formula and our formula
for the Kinetic energy flux}
\setcounter{equation}{0}
\seceqdd

In this appendix we will show that the flux formula ~(\ref{eq:deudt_fluid3})
based on the effective mode-to-mode transfer is equivalent to 
Eq.~(\ref{eq:deudt_fluid1}) derived by Kraichnan.

Kraichnan \cite{Kraichnan59} showed that the kinetic energy flux in 
terms of the combined energy transfer $S^{uu}$ given by
\begin{equation}
\Pi^{u<}_{u>}(K) =
- \sum_{{\bf |k|<K}} \sum_{{\bf p}}^{\Delta} \frac{1}{2}S^{uu}({\bf k|p,q}) 
\label{eq:deudt_fluid1_app}
\end{equation}
Our formula for flux in terms of the effective mode-to-mode transfer
${\cal{\slash{S}}}^{uu}$  is given by
\begin{equation}
\Pi^{u<}_{u>}(K) =
 - \sum_{|{\bf k}|<K} \sum_{|{\bf p}|>K}^{\Delta} 
            {\cal{\slash{S}}}^{uu}({\bf k|p|q}) .
\label{eq:deudt_fluid2_app}
\end{equation}

The equation (~\ref{eq:deudt_fluid1_app})  can be written in terms of
${\cal{\slash{S}}}$'s as follows:
\begin{eqnarray}
\Pi^{u<}_{u>}(K) =
\sum_{{\bf |k|<K}} \sum_{{\bf p}}^{\Delta} \frac{1}{2}
 \left( {\cal{\slash{S}}}^{uu}({\bf k|p|q}) +
 {\cal{\slash{S}}}^{uu}({\bf k|q|p}) \right) 
\end{eqnarray}
which can be further split up as
\begin{eqnarray}
\Pi^{u<}_{u>}(K) &  = &
\nonumber \\
 &   & \sum_{{\bf |k|<K}} \sum_{{\bf |p|,|q|}<K}^{\Delta} 
               \frac{1}{2} \left( {\cal{\slash{S}}}^{uu}({\bf k|p|q}) +
                       {\cal{\slash{S}}}^{uu}({\bf k|q|p}) \right) 
  \nonumber \\
  & + &      \sum_{{\bf |k|<K}} \sum_{{\bf |p|,|q|}>K}^{\Delta} 
              \frac{1}{2} \left( {\cal{\slash{S}}}^{uu}({\bf k|p|q}) +
                     {\cal{\slash{S}}}^{uu}({\bf k|q|p}) \right)
  \nonumber \\
 & +  & \sum_{{\bf |k|<K}} \sum_{{\bf |p|}<K, {\bf |q|}>K}^{\Delta} 
              \frac{1}{2} \left( {\cal{\slash{S}}}^{uu}({\bf k|p|q}) +
                    {\cal{\slash{S}}}^{uu}({\bf k|q|p}) \right) 
  \nonumber \\
  & + &      \sum_{{\bf |k|<K}} \sum_{{\bf |p|}>K, {\bf |q|}<K}^{\Delta} 
              \frac{1}{2}  \left( {\cal{\slash{S}}}^{uu}({\bf k|p|q}) +
                    {\cal{\slash{S}}}^{uu}({\bf k|q|p}) \right) 
\label{eq:flux_split}
\end{eqnarray}
These transfers are shown in Fig. D.1.
Making use of symmetries between ${\bf k,p,q}$ in the summations we get
the following relationships.
Since {\bf p, q} are interchangeable we get
\begin{equation}
        \sum_{{\bf |k|<K}} \sum_{{\bf |p|,|q|}>K}^{\Delta} 
                      {\cal{\slash{S}}}^{uu}({\bf k|p|q}) =
       \sum_{{\bf |k|<K}} \sum_{{\bf |p|,|q|}>K}^{\Delta} 
{\cal{\slash{S}}}^{uu}({\bf k|q|p}) , 
\end{equation}
since {\bf k, p} are interchangeable,
\begin{equation}
  \sum_{{\bf |k|<K}} \sum_{{\bf |p|}<K, {\bf |q|}>K}^{\Delta} 
                      {\cal{\slash{S}}}^{uu}({\bf k|p|q}) =
  \sum_{{\bf |k|<K}} \sum_{{\bf |p|}<K, {\bf |q|}>K}^{\Delta} 
 {\cal{\slash{S}}}^{uu}({\bf p|k|q}),
\end{equation}
since {\bf k, q} are interchangeable,
\begin{equation}
  \sum_{{\bf |k|<K}} \sum_{{\bf |q|}<K, {\bf |p|}>K}^{\Delta} 
                      {\cal{\slash{S}}}^{uu}({\bf k|q|p}) =
  \sum_{{\bf |k|<K}} \sum_{{\bf |q|}<K, {\bf |p|}>K}^{\Delta} 
{\cal{\slash{S}}}^{uu}({\bf q|k|p}) .  \nonumber
\end{equation}
The sum
$\sum_{{\bf |k|<K}} \sum_{{\bf |p|}<K,{\bf |q|}>K}^{\Delta} 
\frac{1}{2} {\cal{\slash{S}}}^{uu}({\bf k|p|q})$ involves
transfers from ${\bf k \rightarrow p}$ and also from
${\bf p \rightarrow k}$, which are equal and opposite. Therefore
\begin{equation}
\sum_{{\bf |k|<K}} \sum_{{\bf |p|<K,|q|}>K}^{\Delta} 
\frac{1}{2} {\cal{\slash{S}}}^{uu}({\bf k|p|q}) = 0.
\end{equation}
Similarly,
\begin{equation}
\sum_{{\bf |k|<K}} \sum_{{\bf |p|>K,|q|}<K}^{\Delta} 
\frac{1}{2} {\cal{\slash{S}}}^{uu}({\bf k|q|p}) = 0,
\end{equation}
and,
\begin{equation}
\sum_{{\bf |k|<K}} \sum_{{\bf |p|<K,|q|}<K}^{\Delta} 
\frac{1}{2} {\cal{\slash{S}}}^{uu}({\bf k|q|p}) = 0,
\end{equation}

In other words all terms except both legs of Fig. D.1,
${\bf p \rightarrow k}$ legs of Fig. D.1, and
${\bf q \rightarrow k}$ legs of Fig. D.1 survive.
We use these results to rewrite $\Pi^{u<}_{u>}(K)$ as
\begin{eqnarray}
\Pi^{u<}_{u>}(K) & = & 
  \sum_{{\bf |k|<K}} \sum_{{\bf |p|,|q|}>K}^{\Delta} 
                     {\cal{\slash{S}}}^{uu}({\bf k|p|q})
  +  \sum_{{\bf |k|<K}} \sum_{{\bf |p|}<K,{\bf |q|}>K}^{\Delta} 
  \frac{1}{2} {\cal{\slash{S}}}^{uu}({\bf k|q|p})    
\nonumber \\
  & + & \sum_{{\bf |k|<K}} \sum_{{\bf |q|}<K,{\bf |p|}>K}^{\Delta} 
  \frac{1}{2} {\cal{\slash{S}}}^{uu}({\bf k|p|q})  
\end{eqnarray}
Using symmetry between {\bf p} and {\bf q} we get,
\begin{eqnarray}
 \sum_{{\bf |k|<K}} \sum_{{\bf |p|<K,|q|}>K}^{\Delta}  
 \frac{1}{2}{\cal{\slash{S}}}^{uu}({\bf k|q|p})  =
 \sum_{{\bf |k|<K}} \sum_{{\bf |q|<K,|p|}>K}^{\Delta}  
 \frac{1}{2}{\cal{\slash{S}}}^{uu}({\bf k|p|q})  
\end{eqnarray}
Hence, we can write,
\begin{eqnarray}
\Pi^{u<}_{u>} & = &
   \sum_{{\bf |k|<K}} \sum_{{\bf |p|,|q|}>K}^{\Delta}  
 {\cal{\slash{S}}}^{uu}({\bf k|p|q})  +
   \sum_{{\bf |k|<K}} \sum_{{\bf |q|}<K,{\bf |p|}<K}^{\Delta}  
 {\cal{\slash{S}}}^{uu}({\bf k|p|q})   \nonumber \\
 & = &  \sum_{{\bf |k|<K}} \sum_{{\bf |p|}>K}^{\Delta}  
 {\cal{\slash{S}}}^{uu}({\bf k|p|q}) 
\end{eqnarray}
This is the same as $\Pi^{u<}_{u>}$ of Eq.~(\ref{eq:deudt_fluid1}).
Thus, we have shown that Kraichnan's formula for flux is the same as
our formula based on the mode-to-mode transfer.

\newpage

\bibliography{ref}

\begin{thebibliography}{10}

\bibitem{Domaradzki90}
J.~A. Domaradzki and R.~S. Rogallo.
\newblock {\em Phys. Fluids A}, 2:413, 1990.

\bibitem{Domaradzki88}
J.~A. Domaradzki.
\newblock {\em Phys. Fluids}, 31:2747, 1988.

\bibitem{Zhou93a}
Y.~Zhou.
\newblock {\em Phys. Fluids A}, 5:1092, 1993.

\bibitem{Zhou93b}
Y.~Zhou.
\newblock {\em Phys. Fluids A}, 5:2511, 1993.

\bibitem{Kraichnan76}
R.~H. Kraichnan.
\newblock {\em J. Atmos. Sci.}, 33:1521, 1976.

\bibitem{Domaradzki87}
J.~A. Domaradzki, R.~W. Metcalfe, R.~S. Rogallo, and J.~J. Riley.
\newblock {\em Phys. Rev. Lett.}, 58:547, 1987.

\bibitem{Batchelor50}
G.~K. Batchelor.
\newblock {\em Proc. Roy. Soc. (London)}, 2:413, 1950.

\bibitem{MKV96}
M.~K. Verma, D.~A. Roberts, M.~L. Golstein, S.~Ghosh, and W.~T. Stribling.
\newblock {\em J. Geophys. Res.}, 101:21619, 1996.

\bibitem{Pouquet76}
A.~Pouquet, U.~Frisch, and J.~Leorat.
\newblock {\em J. Fluid Mech.}, 77:321, 1976.

\bibitem{Pouquet78}
A.~Pouquet.
\newblock {\em J. Fluid Mech.}, 88:1, 1978.

\bibitem{Ishizawa298}
A.~Ishizawa and Y.~Hattori.
\newblock {\em chao-dyn/9810036}, 1998.

\bibitem{Ishizawa198}
A.~Ishizawa and Y.~Hattori.
\newblock {\em J. Phys. Soc. Jpn.}, 67:441, 1998.

\bibitem{Lesieur}
M.~Lesieur.
\newblock {\em Turbulence in Fluids - Stochastic and Numerical Modelling}.
\newblock Kluwer Academic Publishers, 1990.

\bibitem{Leslie}
D.~C. Leslie.
\newblock {\em Developments in the theory of turbulence}.
\newblock Oxford University Press, 1973.

\bibitem{Kraichnan59}
R.~H. Kraichnan.
\newblock {\em J. Fluid Mech.}, 5:497, 1959.

\bibitem{Stanisic}
M.~M. Stanisic.
\newblock {\em The Mathematical Theory of Turbulence}.
\newblock Springer-Verlag, 1985.

\bibitem{Dar2000_2}
G.~Dar, M.~K. Verma, and V.~Eswaran,  (submitted to Phys.~Plasma) 

\end{thebibliography}

\newpage

\begin{center}
FIGURE CAPTIONS
\end{center}

\noindent Fig. 1
two types of triads involved in transfer between shell {\it m}
and shell {\it n}. Mode {\bf q} could be either inside a shell or could
be located outside the shells.

\vspace{1.0cm}

\noindent Fig. 2
The mode-to-mode energy transfers 
${\cal{\slash{R}}}^{uu}$'s sought to be determined.

\vspace{1.0cm}

\noindent Fig. 3
Mode-to-mode transfer can be expressed as a sum of
circulating transfer and effective mode-to-mode transfer.

\vspace{1.0cm}

\noindent Fig. 4
The circulating transfer and mode-to-mode transfer from shell
{\it n} to shell {\it m} with the arrows pointing towards the direction
of energy transfer.

\vspace{1.0cm}

\noindent Fig. 5
The circulating transfer $X_\Delta$ does not contribute to the
energy flux out of the sphere K.

\vspace{1.0cm}

\noindent Fig. 6
The combined energy transfers in a triad. 
$S^{uu}({\bf k|p,q})$: ${\bf u(p)}$ and ${\bf u(q)} \rightarrow {\bf u(k)}$;
$S^{bu}({\bf q|k,p})$: ${\bf u(k)}$ and ${\bf u(p)} \rightarrow {\bf b(q)}$;
$S^{bb}({\bf k|p,q})$: ${\bf b(p)}$ and ${\bf b(q)} \rightarrow {\bf b(k)}$.

\vspace{1.0cm}

\noindent Fig. 7
The mode-to-mode energy transfers sought to be determined.

\vspace{1.0cm}

\noindent Fig. 8
Energy transfer between a pair of magnetic modes in a triad
can be expressed as a sum of
circulating transfer and effective mode-to-mode transfer.

\vspace{1.0cm}

\noindent Fig. 9
Energy transfer between a velocity and a magnetic
mode in a triad can be expressed as a sum of
circulating transfer and effective mode-to-mode transfer.

\vspace{1.0cm}

\noindent Fig. 10
The various energy fluxes defined in this section are shown.
The arrows indicate the directions of energy transfers corresponding to
a positive flux. These fluxes are computed in numerical simulations and
the results are given in a subsequent paper.

\vspace{1.0cm}

\noindent Fig. B.1
In the wavenumber triad ${\bf k, q \rightarrow \infty >> p}$
the circulating transfer, $Y_{\Delta}=0$ to avoid non-local transfer
from very large wavenumbers to very small wavenumbers in
comparison. 

\vspace{1.0cm}

\noindent Fig. D.1
he various triads involved in the terms in 
Eq.~(\ref{eq:flux_split}).
Triad of type A does not contribute to energy
flux. Triad of type C and D are equivalent and contribute the same amount.

\newpage

\begin{figure}[h]
\centerline{\mbox{\psfig{file=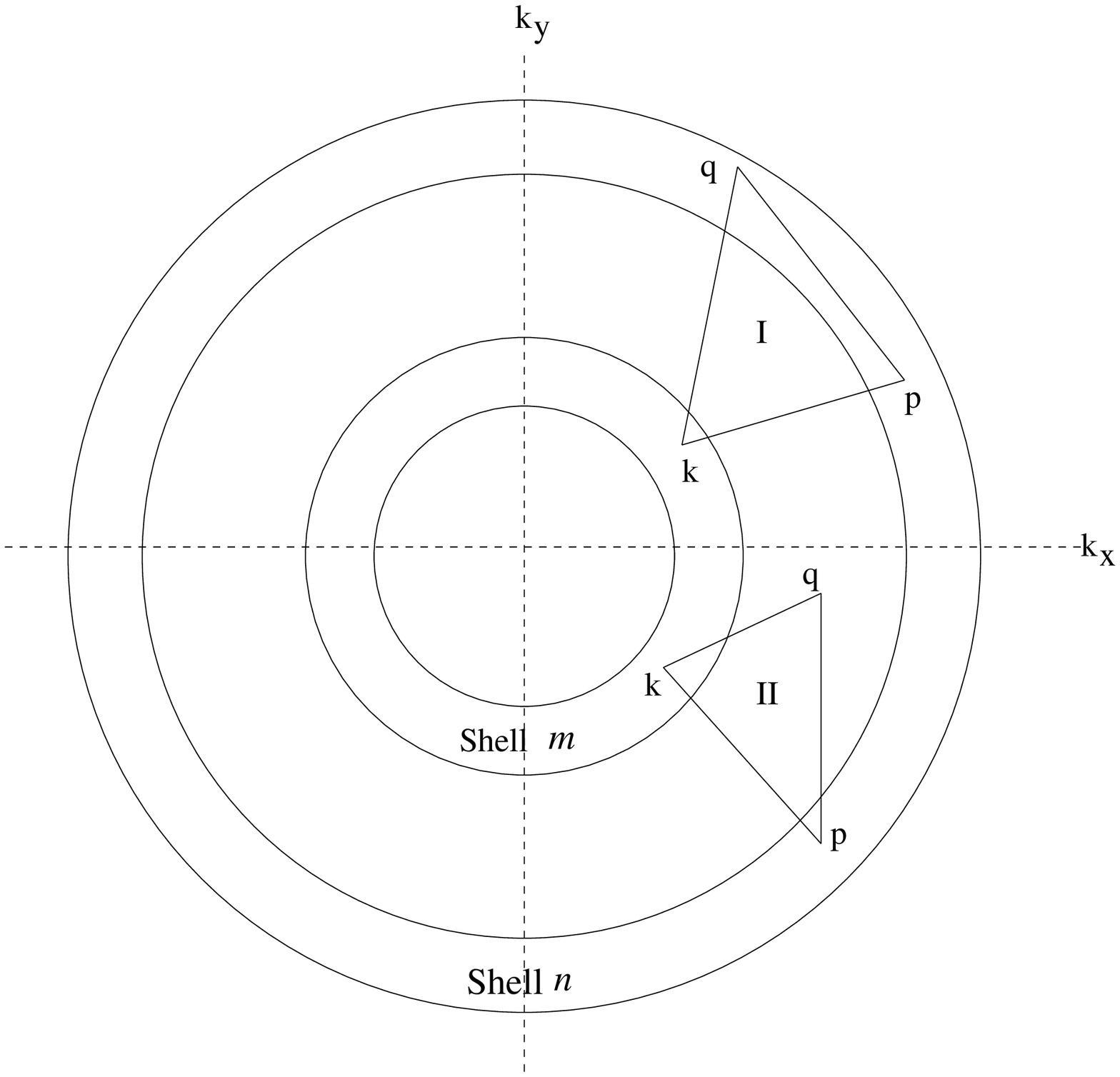,width=0.7\textwidth}}}
\label{fig:shell_old_defn}
\vskip 3.0cm
\caption{}
\end{figure}

\newpage

\begin{figure}[h]
\centerline{\mbox{\psfig{file=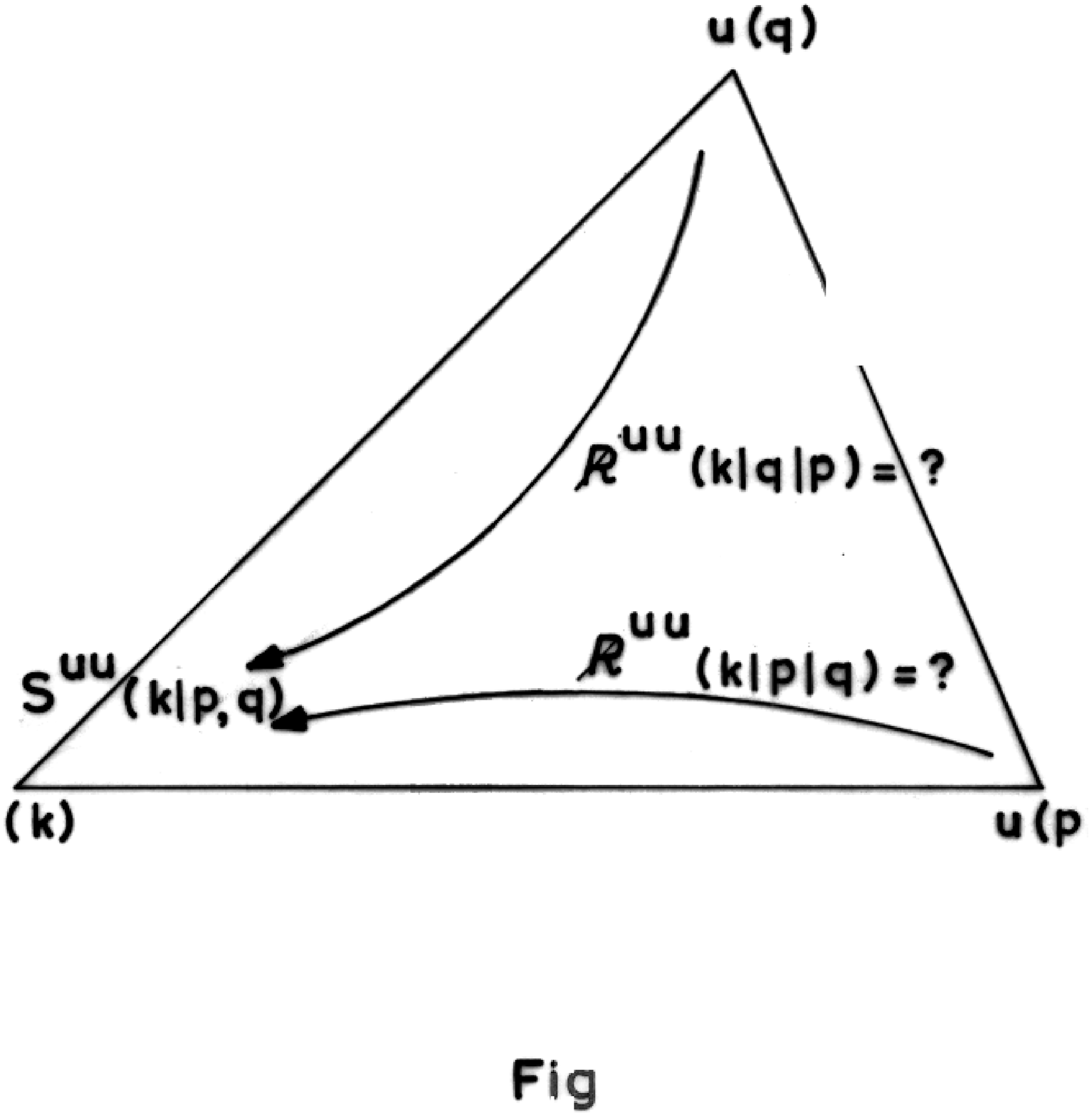,width=0.6\textwidth}}}
\label{fig:mode_uu_intro}
\vskip 3.0cm
\caption{}
\end{figure}

\newpage

\begin{figure}[h]
\centerline{\mbox{\psfig{file=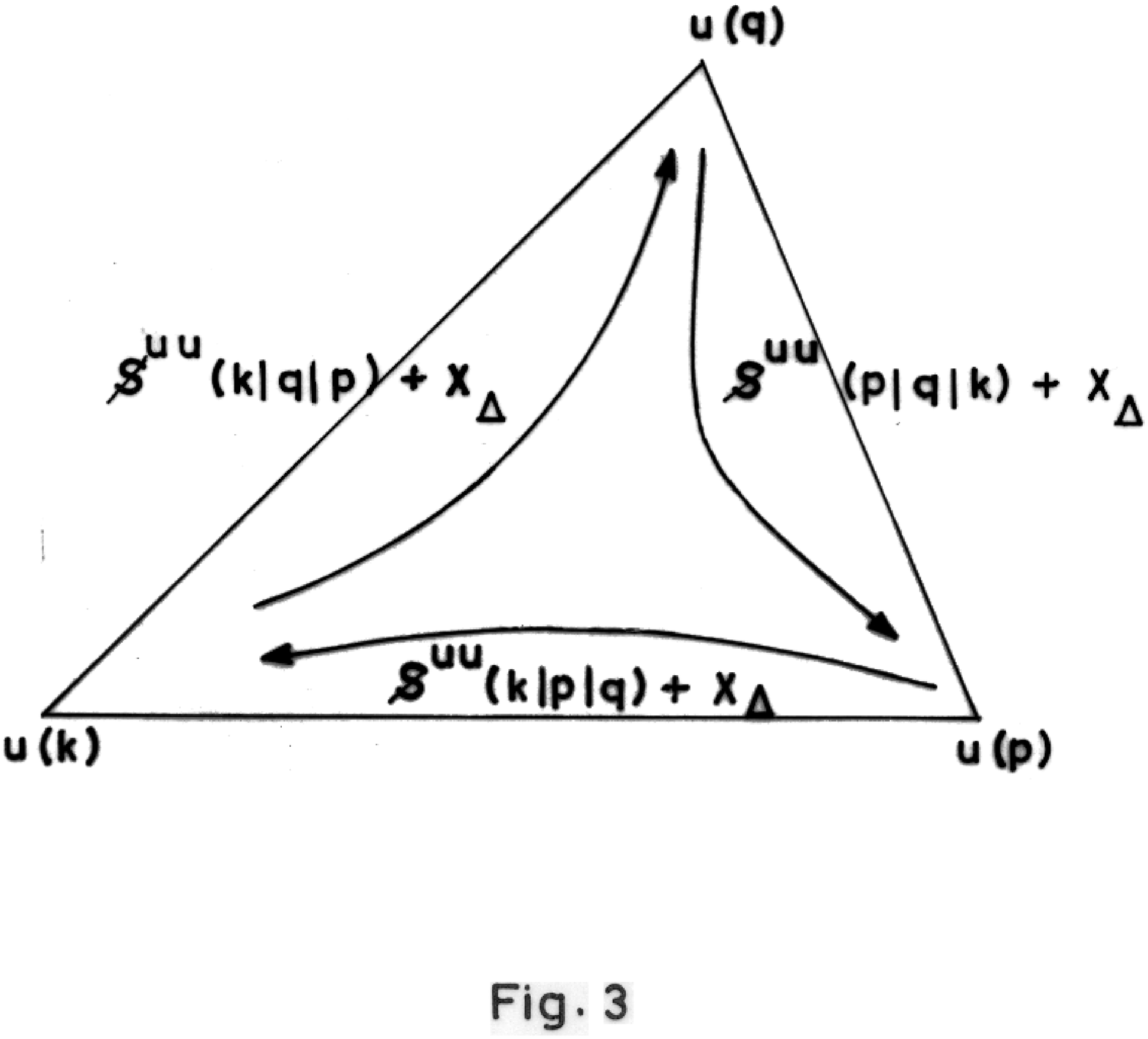,width=0.7\textwidth}}}
\label{fig:mode_uu_eff}
\vskip 3.0cm
\caption{}
\end{figure}

\newpage

\begin{figure}[h]
\centerline{\mbox{\psfig{file=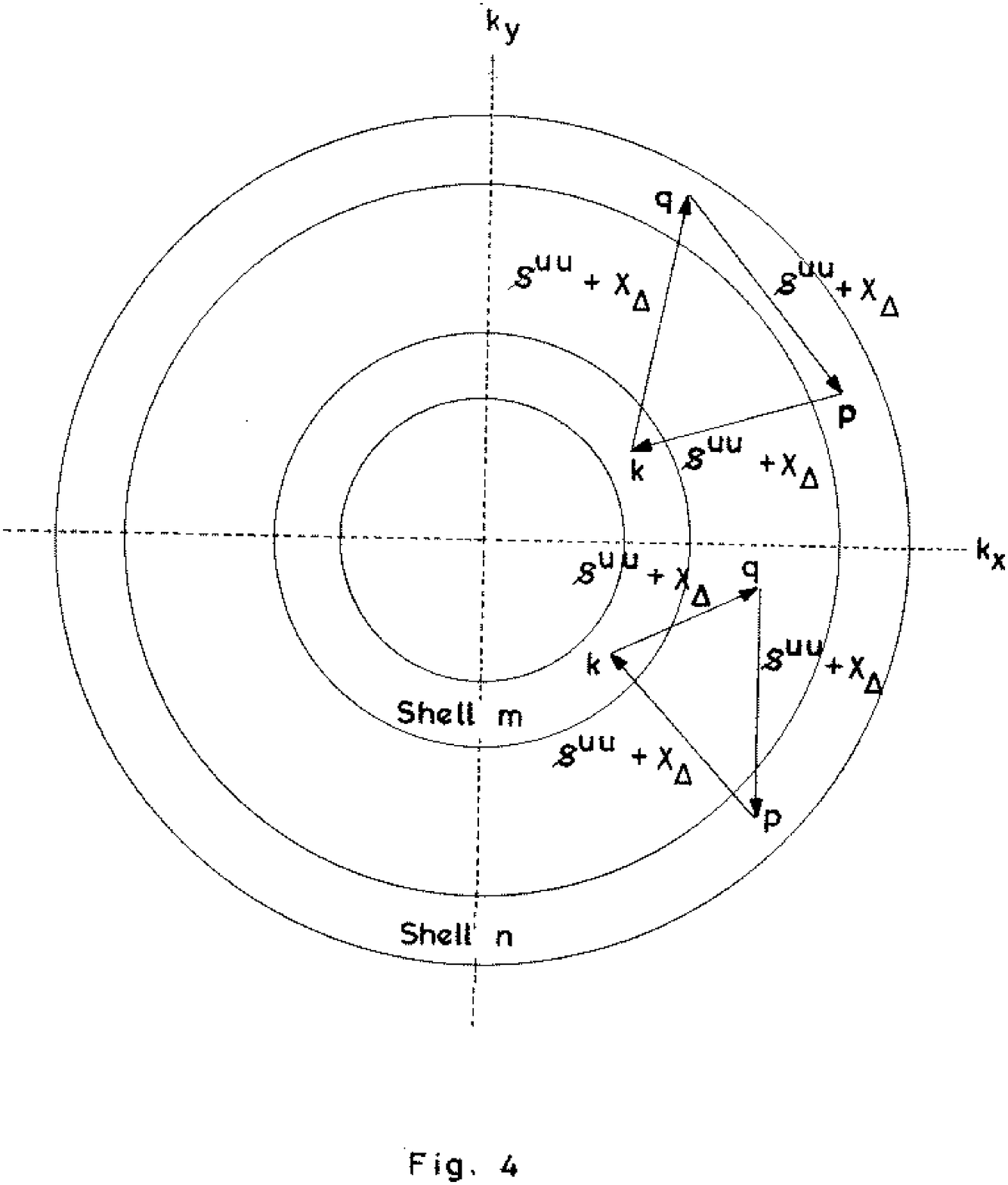,width=0.7\textwidth}}}
\label{fig:shell_eff}
\vskip 3.0cm
\caption{}
\end{figure}

\newpage

\begin{figure}[h]
\centerline{\mbox{\psfig{file=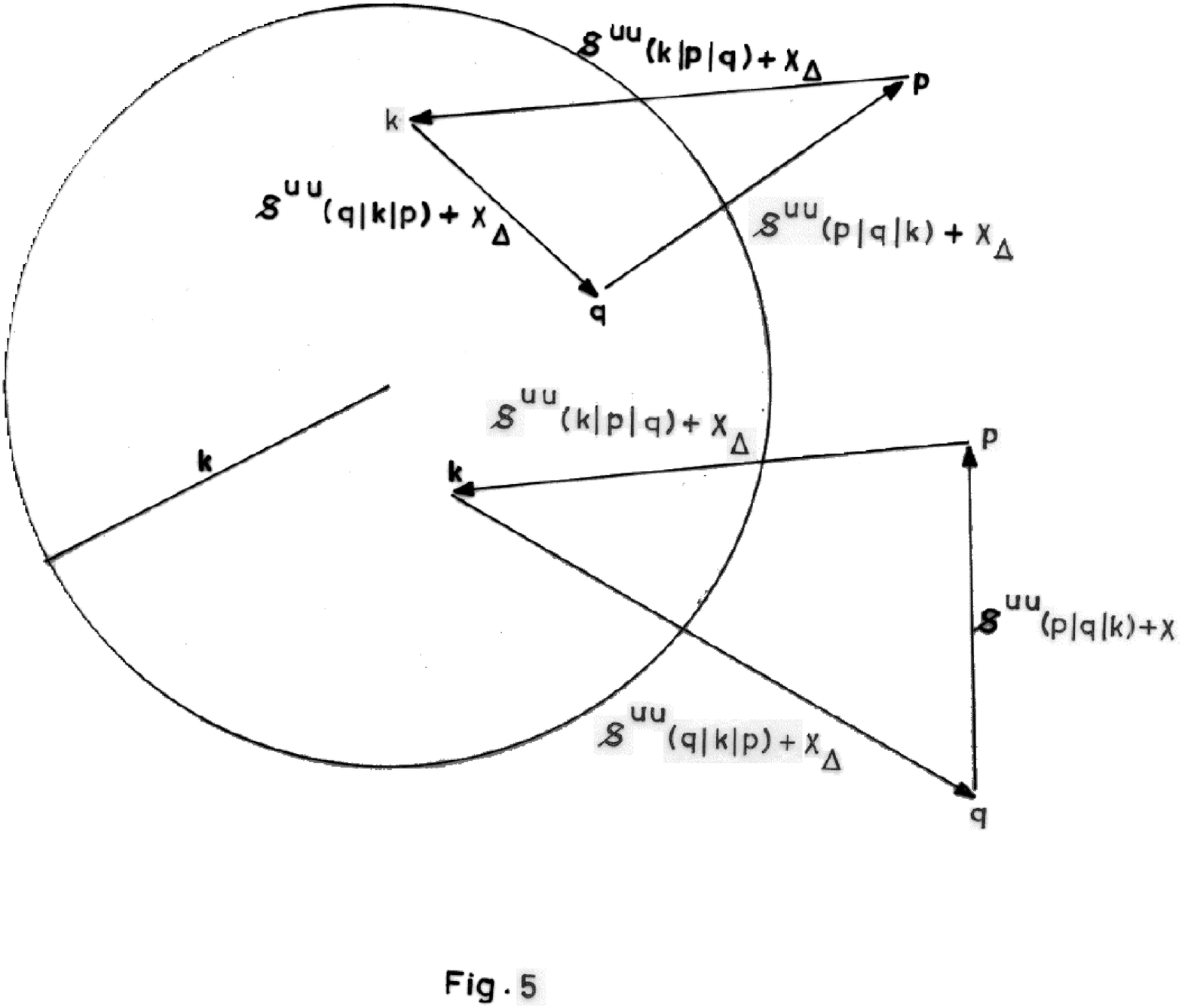,width=0.8\textwidth}}}
\label{fig:flux_uu_eff}
\vskip 3.0cm
\caption{}
\end{figure}

\newpage

\begin{figure}[h]
\centerline{\mbox{\psfig{file=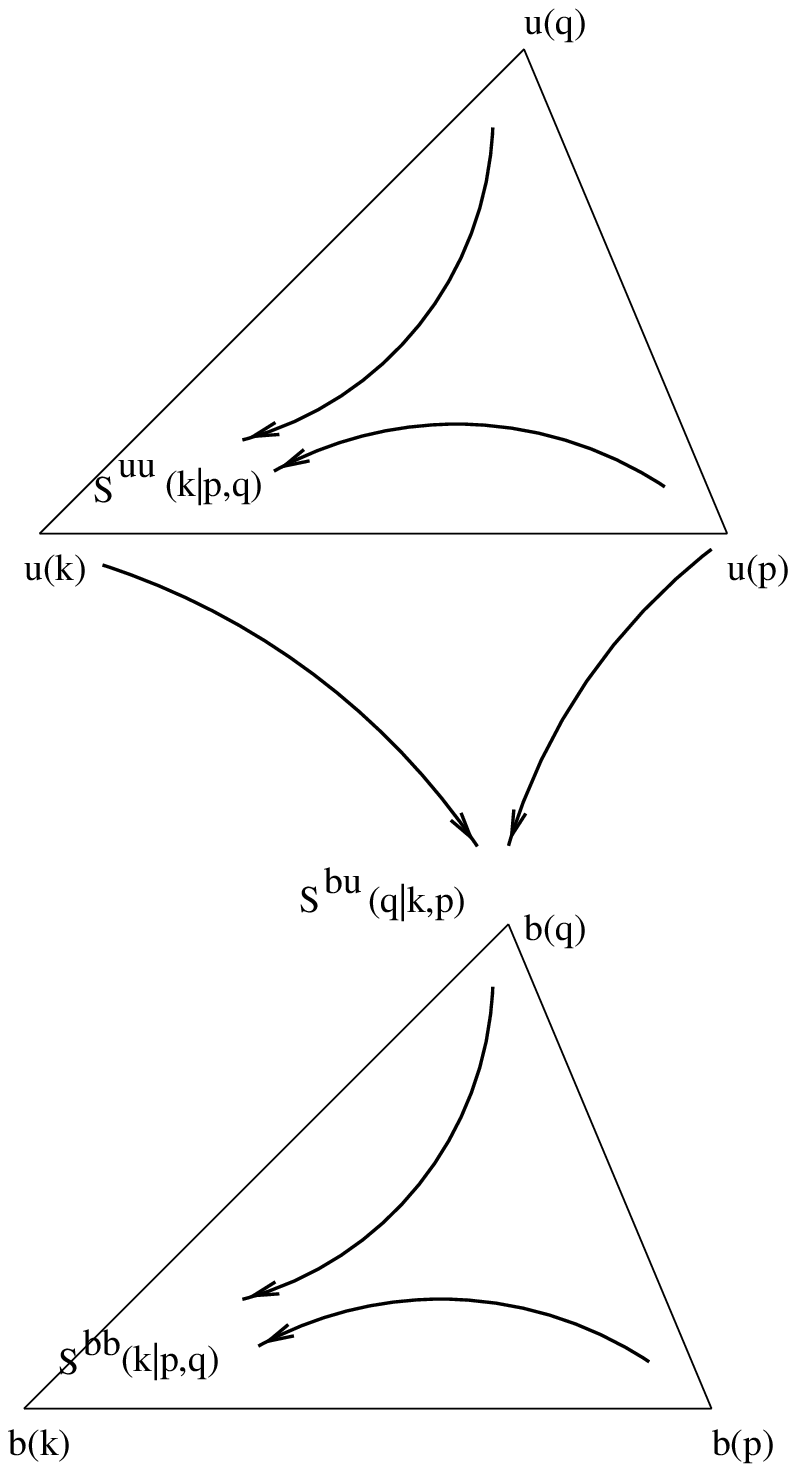,width=0.45\textwidth}}}
\label{fig:triad}
\vskip 3.0cm
\caption{}
\end{figure}

\newpage

\begin{figure}[h]
\centerline{\mbox{\psfig{file=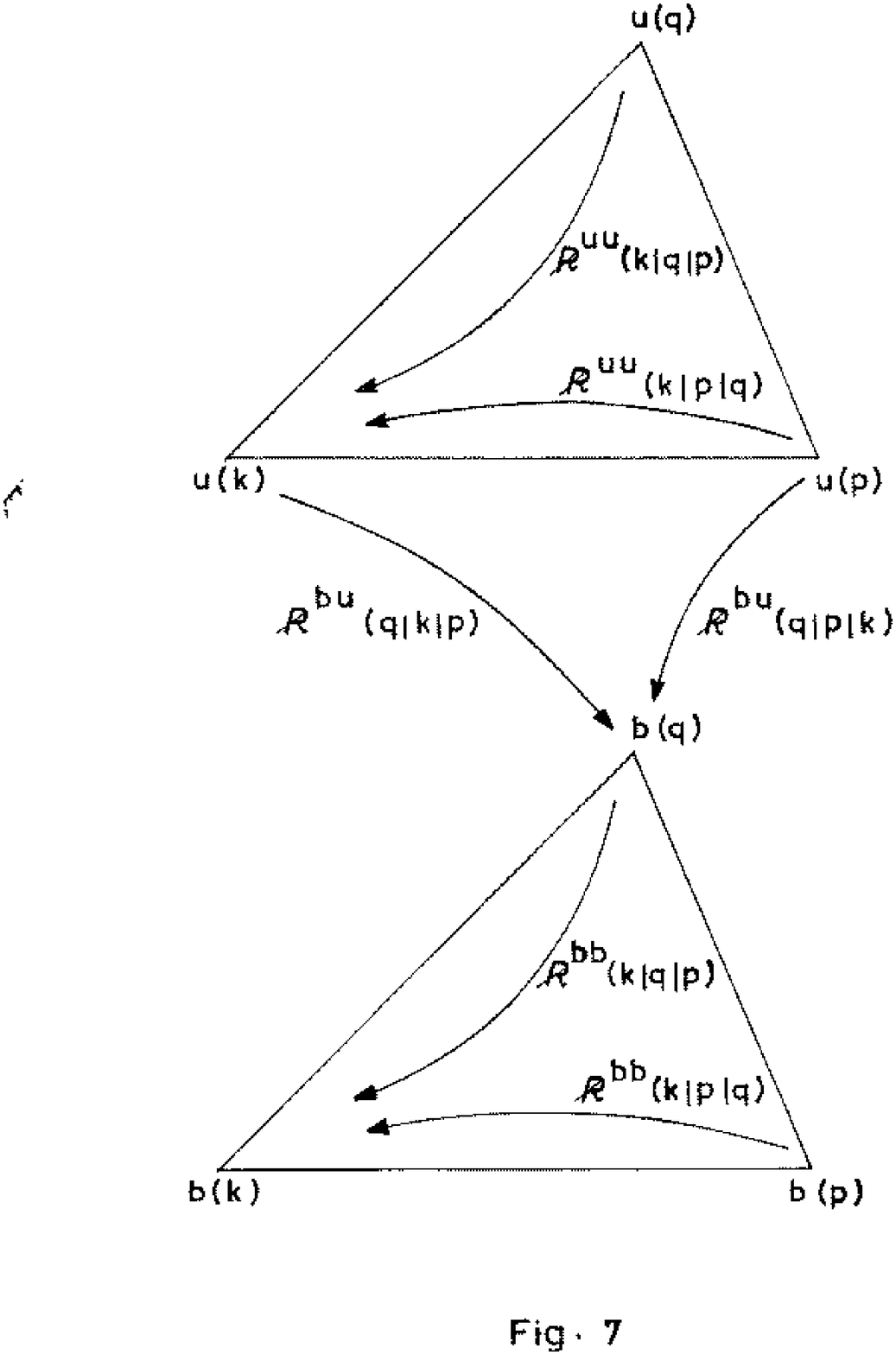,width=0.45\textwidth}}}
\vskip 3.0cm
\caption{}
\end{figure}

\newpage

\begin{figure}[h!]
\centerline{\mbox{\psfig{file=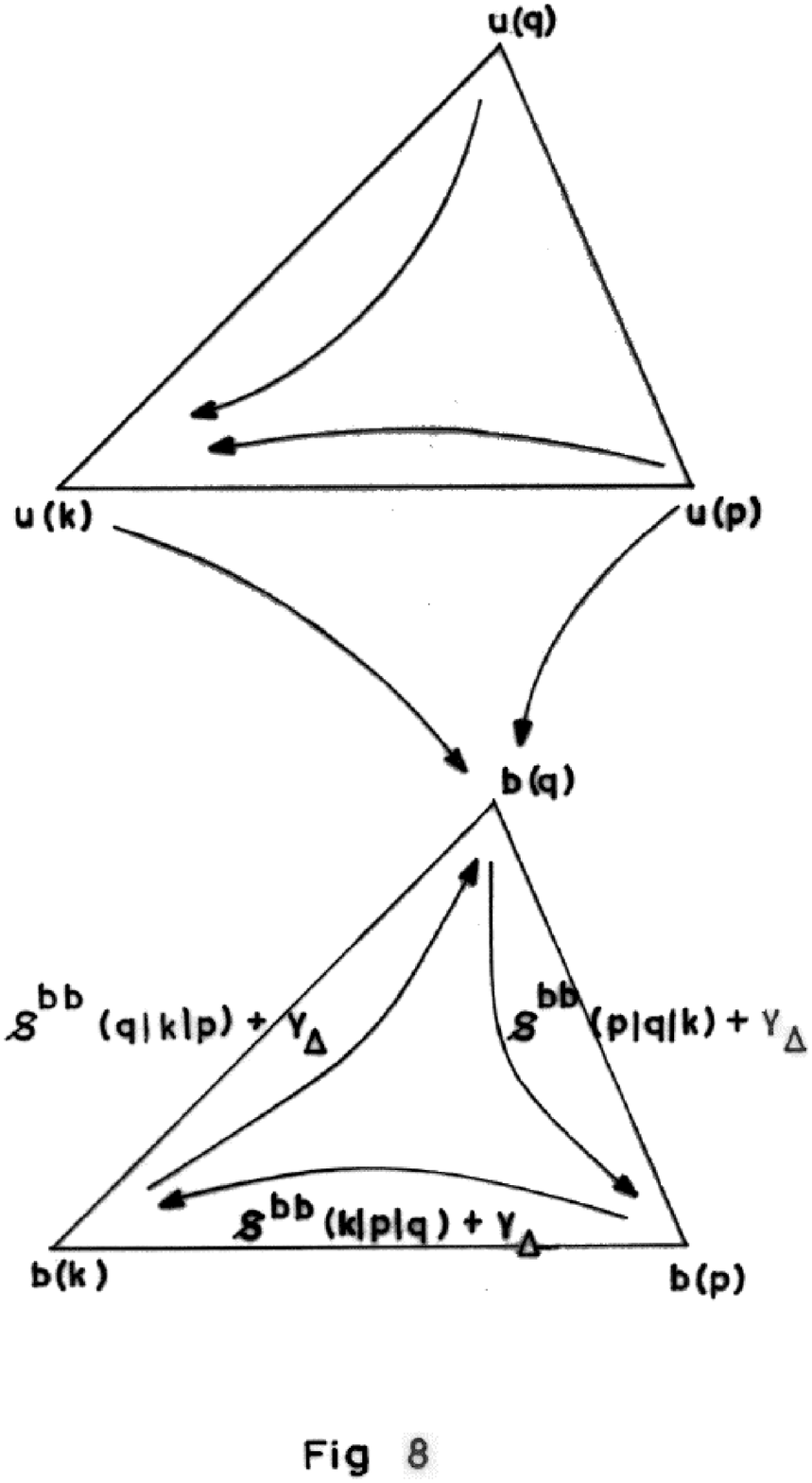,width=0.5\textwidth}}}
\label{fig:mode_bb_eff}
\vskip 3.0cm
\caption{}
\end{figure}

\newpage

\begin{figure}[h!]
\centerline{\mbox{\psfig{file=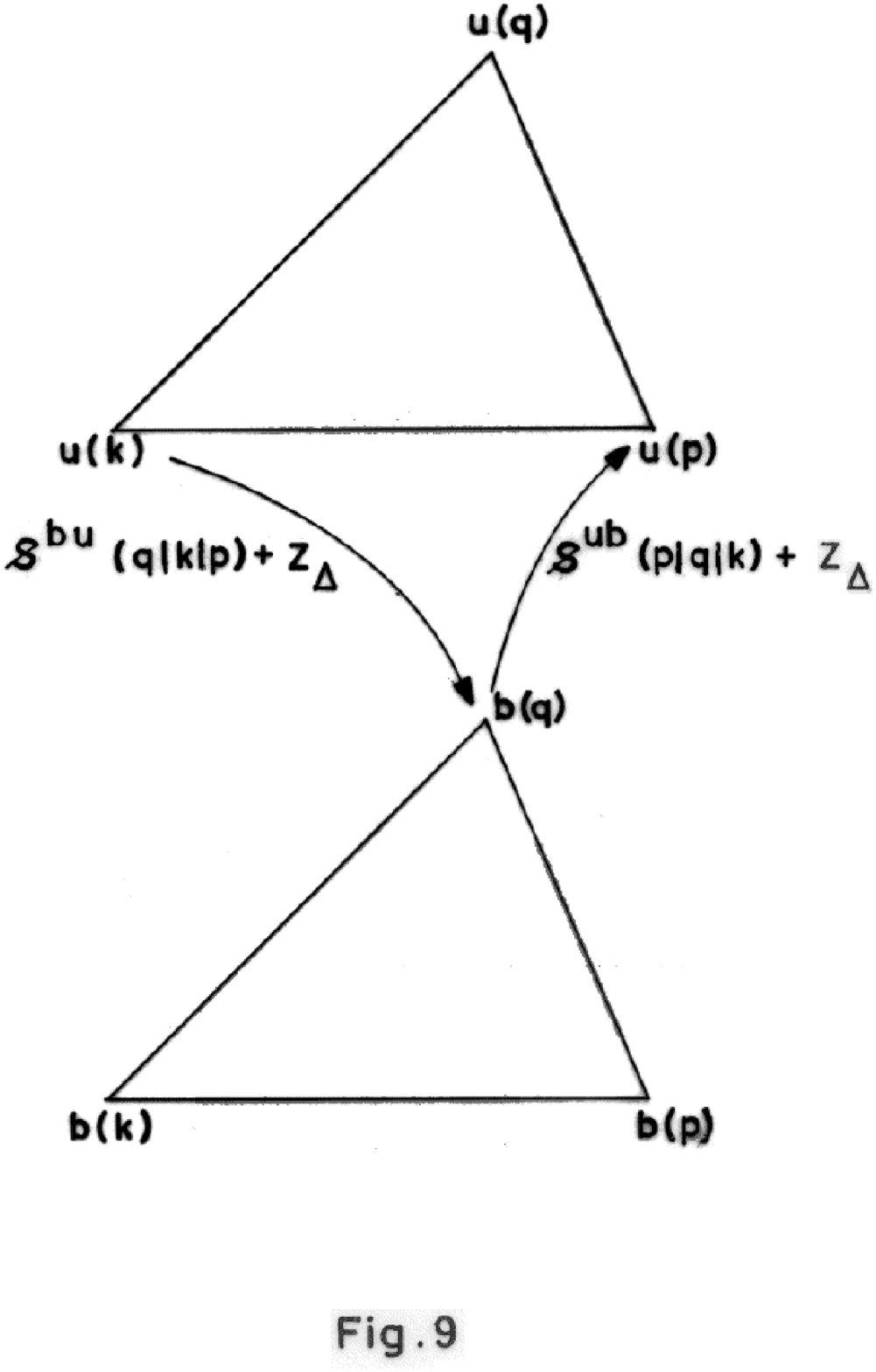,width=0.5\textwidth}}}
\label{fig:mode_ub_eff}
\vskip 3.0cm
\caption{}
\end{figure}

\newpage

\begin{figure}[h!]
\centerline{\mbox{\psfig{file=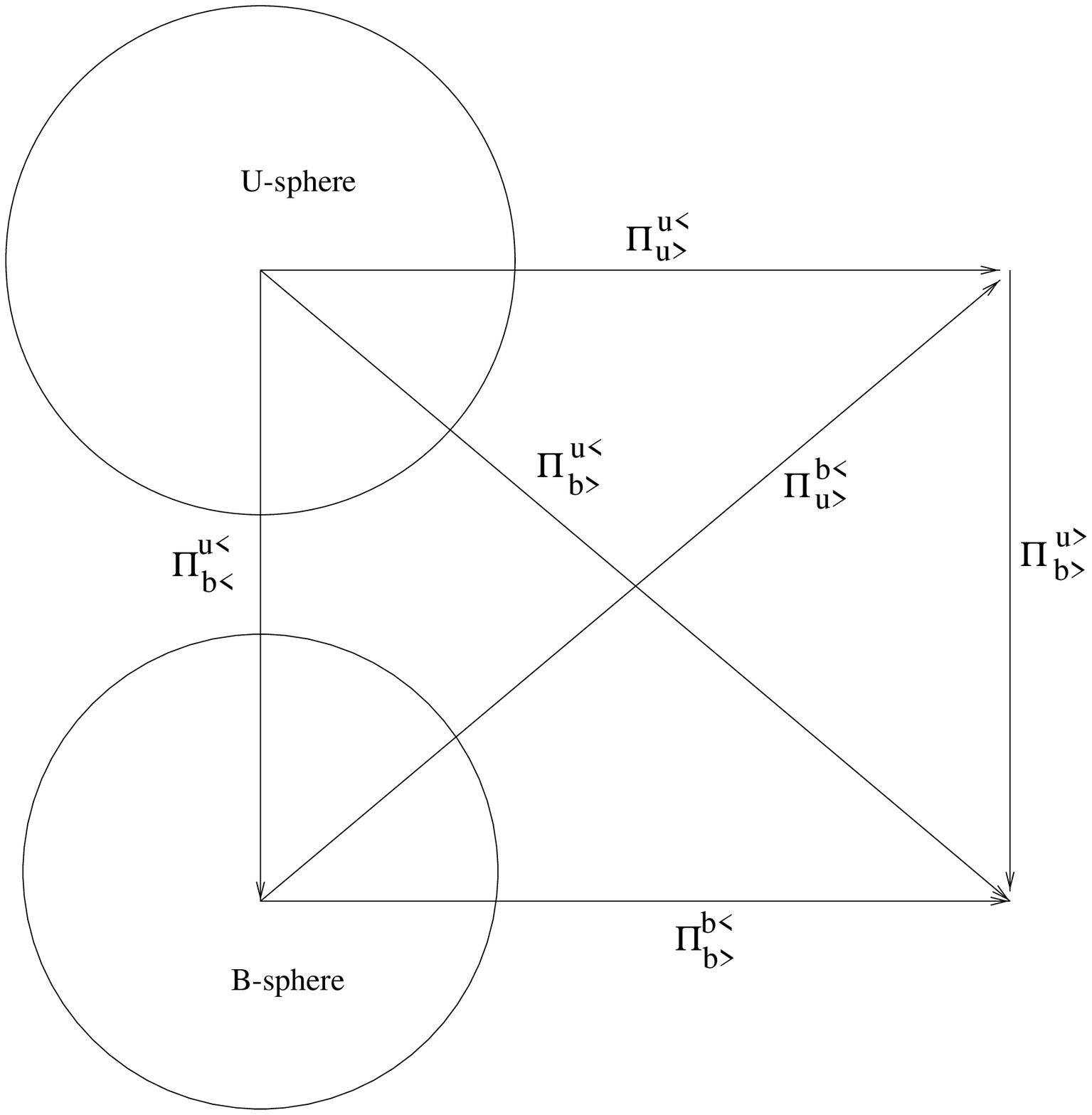,width=0.8\textwidth}}}
\label{fig:flux_eff_defn}
\vskip 3.0cm
\caption{}
\end{figure}

\newpage

\begin{figure}[h]
\centerline{\mbox{\psfig{file=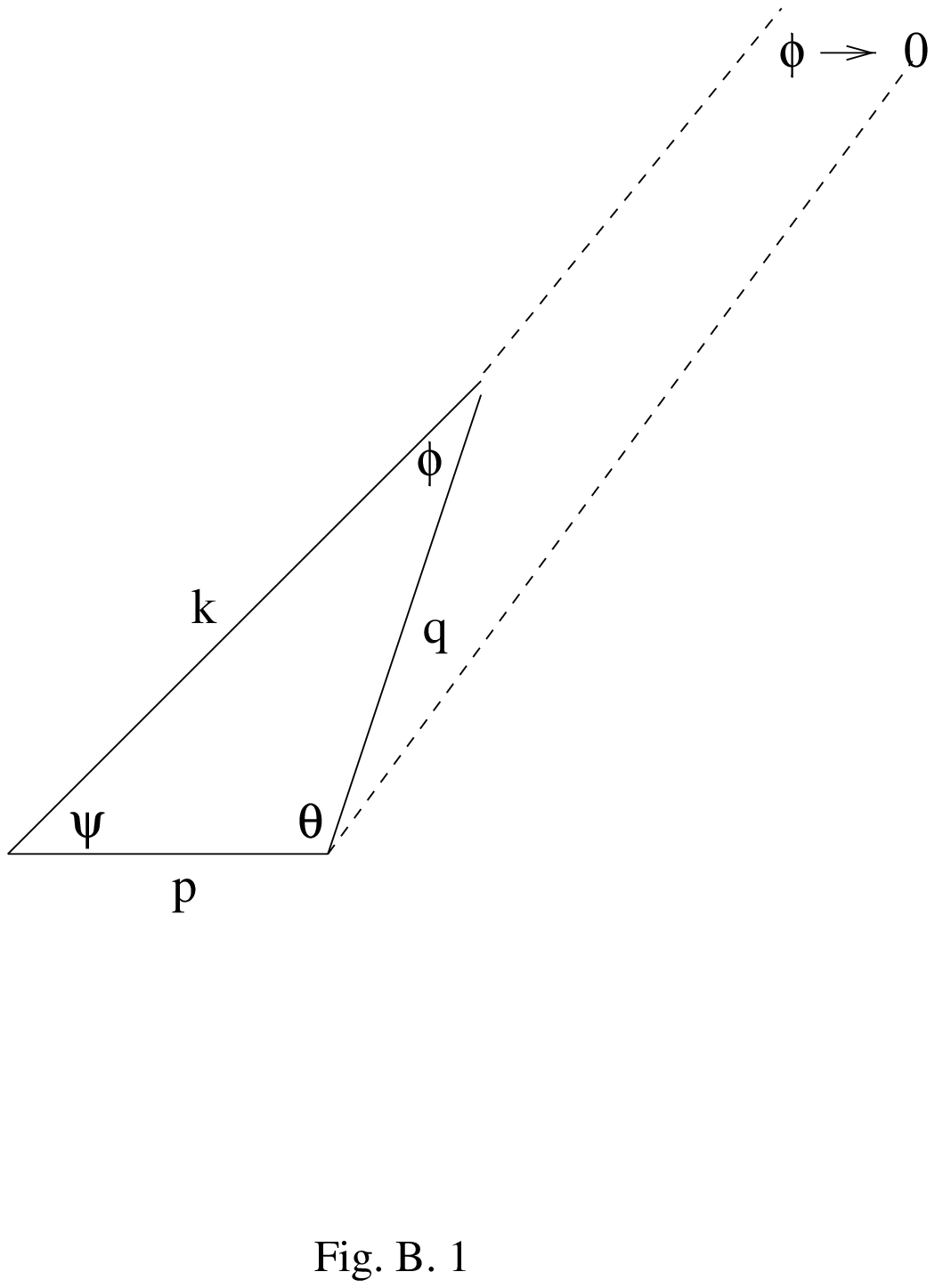,width=0.5\textwidth}}}
\label{fig:mode_local}
\end{figure}

\newpage

\begin{figure}[h]
\centerline{\mbox{\psfig{file=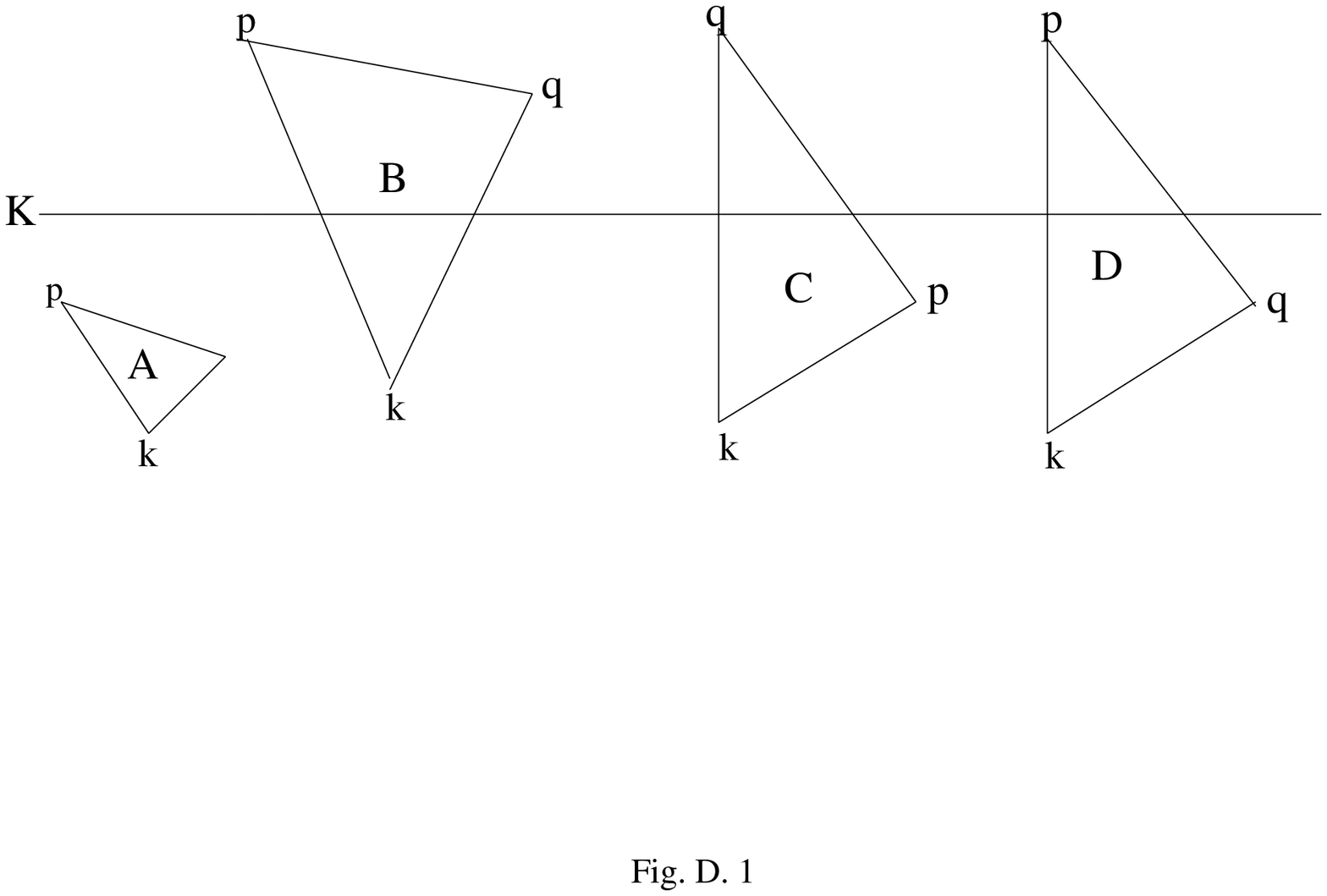,width=0.8\textwidth}}}
\label{fig:flux_triad}
\end{figure}

\end{document}